\documentclass[a4paper,11pt]{article}
\pdfoutput=1

\usepackage[bottom]{footmisc}

\usepackage{jheppub}

\usepackage[dvipsnames]{xcolor}
\usepackage{lmodern}
\usepackage{amsmath}
\usepackage{amssymb}
\usepackage{enumerate}
\usepackage{mathrsfs}
\usepackage{float}
\usepackage{graphicx}
\usepackage{subcaption}

\title{\boldmath Thermodynamic ensembles with cosmological horizons}
\author[]{Batoul Banihashemi}
\author[]{and Ted Jacobson}
\affiliation[]{Maryland Center for Fundamental Physics,\\  University of Maryland, College Park, MD 20742, USA}

\emailAdd{baniha@umd.edu}
\emailAdd{jacobson@umd.edu}

\def\D{\Delta}

\newcommand{\be}{\begin{equation}}
\newcommand{\ee}{\end{equation}}
\newcommand{\ben}{\begin{enumerate}}
\newcommand{\een}{\end{enumerate}}
\newcommand{\mc}{\mathcal}

\newcommand{\mce}{\varepsilon}
\newcommand{\im}{I_{\text{mic}}}

\newcommand{\euc}{Euclidean } 
\newcommand{\sds}{Schwarzschild-de Sitter }

\makeatletter 
\def\@fpheader{{\color{white}jhep}}
\makeatother

\abstract{
The entropy of a de Sitter horizon was derived long ago by Gibbons and Hawking via a gravitational partition function. Since there is no boundary at which to define the temperature or energy of the ensemble, the statistical foundation of their approach has remained obscure. To place the statistical ensemble on a firm footing we
introduce an artificial ``York boundary",
with either canonical or microcanonical boundary conditions, as has been done
previously for black hole ensembles.
The partition function and the density of states are expressed as
integrals over paths in the constrained, spherically reduced phase space
of pure 3+1 dimensional gravity with a positive cosmological constant.
Issues related to the domain and contour of integration are analyzed,
and the adopted choices for those are justified as far as possible.
The canonical ensemble includes a patch of spacetime without horizon, 
as well as configurations containing a black hole or a cosmological horizon. 
We study thermodynamic phases and (in)stability, and discuss 
an evolving reservoir model that can stabilize the cosmological horizon in the
canonical ensemble. Finally, we explain 
how the Gibbons-Hawking partition function on the 4-sphere
can be derived as a limit of well-defined thermodynamic ensembles 
and, from this viewpoint,  why it computes the dimension of the 
Hilbert space of states within a cosmological horizon.
}

\begin{document}
\maketitle
\flushbottom

\section{Introduction}

The nature of the static patch of de Sitter space as a thermodynamic ensemble presents some
persistent puzzles.
Most work on the subject relies on the pioneering study of Gibbons and Hawking (GH) \cite{GH}, 
who introduced the path integral formulation of the quantum gravitational partition function.
Their main focus was on the grand
 canonical ensemble containing
asymptotically flat, charged rotating black hole spacetimes. 
They
obtained a semiclassical approximation 
for the partition function $Z=e^{-I}$, 
where $I$ is the action for the Euclidean (or complex, if rotating) black hole solution. 
Interpreting $I$ as a thermodynamic potential
$W$ divided by temperature $T$, $I=W/T$,
they calculated the entropy of these ensembles,
finding the Bekenstein-Hawking result.
They briefly treated de Sitter (dS) spacetime as a variant of this scheme, with $I$ equal to 
the action of Euclidean de Sitter space, 
and setting $W=-TS$, with no mass, charge, or  angular momentum terms, on the grounds that
the space has no boundary. It follows that 
$I = -S$, from which GH read off the entropy, which is equal to 
one fourth the de Sitter horizon area in Planck units.\footnote{There is a typo in the GH paper \cite{GH}: in their eq.\ (3.15) it is written $S = 12\pi \Lambda^{-1}$, while it should be 
$S = 3\pi \Lambda^{-1}$, where 
$\Lambda$ is the cosmological constant.}

But how can one interpret and justify this calculation? 
Because the spatial sections of de Sitter spacetime are closed, there is no outer boundary at infinity at which to impose the conditions defining the ensemble,
so it is not clear what ensemble has been defined. It cannot
be regarded as a canonical  ensemble, since there is no way to specify a temperature. 
The vanishing of the energy suggests that it should perhaps be 
interpreted as
a microcanonical ensemble.
In fact, if $H\rightarrow0$, the 
partition function $Z={\rm Tr}\,e^{-\beta H}$ becomes trace of the identity operator, 
which is
the total dimension of the Hilbert space at zero energy, so $\log Z$ becomes the microcanonical entropy. But why
can the GH path integral for $Z$ of dS be interpreted as computing
that dimension? 
In short, 
if the GH result is the answer, what exactly is the question? 
In section \ref{Gibbons-Hawking}
we offer a resolution to this puzzle, by
showing how the GH path integral can be seen as the limit of a well-defined partition function or density of states.

Another puzzling issue
in de Sitter thermodynamics is the sign of the temperature. Since addition of energy (in the form of matter or a black hole) to the static patch results in a \emph{decrease} of the generalized entropy, it has been argued \cite{KV,TJ&Manus} that 
the static patch must have 
{\it negative} temperature.
This appears to be incompatible with 
the positive Gibbons-Hawking temperature
of a quantum field in the static patch;
however, since the two temperatures refer to
different systems (the former being
the full gravitating system), perhaps
there is no contradiction \cite{TJ&Manus}.

In this work we attempt to 
clarify the statistical mechanics 
of de Sitter spacetime by 
requiring, first and foremost, a well defined ensemble.
We adopt the
viewpoint emphasized by York \cite{York86}, that the definition of a thermodynamic ensemble for a gravitational system necessitates the introduction of a boundary, on which ``quasilocal" 
thermodynamic data are fixed. 
This necessity is tied to the 
deep consequences of diffeomorphism invariance: the absence of local observables and the holographic nature of quantum gravity.\footnote{The
question of whether the phase space and Hamiltonian are well defined 
for a gravitational system within
a finite boundary is addressed in
section \ref{foundations}.}
We shall consider both canonical and microcanonical ensembles,
and allow for both black hole
and cosmological horizons. 
A configuration with cosmological horizon 
that dominates the canonical ensemble
will thus be identified as a thermodynamic {\it phase} of the system.

Thermodynamic ensembles 
defined by quasilocal boundary data have been
well studied in the black hole phase
(see e.g.\ \cite{York86,WY,York88,Martinez-York,Braden-etal,BY,Brown-Creighton-Mann, Andre-Lemos}).
One motivation for introduction of the quasilocal 
boundary in that case was to obtain a  
stable canonical ensemble. A stable black hole 
ensemble was first 
defined by Hawking and Page~\cite{HawkingPage}, 
for the case with 
a negative cosmological 
constant and
asymptotically 
anti-de Sitter boundary conditions.
At high enough temperature that 
system has a stable black hole phase, 
because of the radial growth of the redshift factor
which acts, in effect, like a confining box. 
As first noticed by 
York~\cite{York86}, 
the imposition of a quasilocal boundary condition 
produces a similar mechanism for stabilizing a black hole ensemble, without requiring a negative cosmological constant. 
The temperature of a black hole as measured at the boundary is governed 
by two factors: the size of the black hole determines its surface gravity
and therefore its Hawking temperature at infinity, 
while the location of the horizon relative to the boundary determines the blueshift 
of that Hawking temperature at the boundary.
These factors act in opposite directions, and the blueshift effect is unbounded,
hence at sufficiently high temperature there exists a stable ``large" black hole
configuration whose temperature at the boundary increases when it absorbs energy
and decreases when it emits energy. This large black hole phase thus has
positive heat capacity and is therefore stable. 
 
\paragraph{Summary of approach and main results} 
%Let us summarize our approach and main results here. 
We consider 4-dimensional Einstein gravity with a fixed positive cosmological constant, and with spatial boundary a round 2-sphere with a fixed area. 
We assume further that the thermal bath at the boundary is not rotating, i.e.\ that the Hamiltonian for the ensemble generates time flow orthogonal to the spherical slices.
For the canonical ensemble 
the temperature $\beta^{-1}$ 
is fixed at the boundary.
We employ the path integral representation of the partition function $Z(\beta)$, in which 
paths are periodic in time with period $\beta$ at the boundary. The integration contour is deformed such that it passes over \euc geometries including stationary points of the action, for which the integrand becomes $e^{-I}$, where $I$ is the \euc action. The canonical ensemble is thus
specified by the boundary area and time period.

To simplify the analysis, while still hopefully
capturing some essential physics, 
we restrict attention to
paths in the path integral that share the spherical symmetry of the boundary. 
The relevant stationary points of the action presumably fall within this class.
As in the symmetry reduced analysis of \cite{WY},
we impose the initial value constraint equations by hand at each time, so that the paths are trajectories in the physical phase space. 
This ensures that the path integral computes the physical Hilbert space trace in partition function. Apart from a special (Nariai) case 
in which all the 2-spheres have the same area,
the solutions to the constraints form
a one parameter family that corresponds 
in our adopted gauge
 to the initial data on static slices of the Schwarzschild-de Sitter (SdS) solutions with mass parameter $M$.\footnote{Solutions with a cosmological horizon and negative $M$ exist, but are excluded if one requires that a nonsingular continuation of the solution exists beyond the boundary, i.e.\ outside of the thermodynamic system.} These solutions include slices through
 any number of horizons in the maximal extension
 of Lorentzian SdS, however those that cross more than one horizon cannot be embedded in a smooth Euclidean 4-manifold. We adopt the hypothesis that the non-embeddable solutions to the constraints do not arise on paths produced by the Euclidean contour deformation, hence should be omitted. 
 The paths in the path integral have constant $M$,
 since the constraints impose that $M$ has vanishing time derivative on each slice. 
 For given boundary data, we identify the stationary points of the action $I$ with respect to variation of $M$, and study the behavior of the action as a function of $M$. We estimate the partition function by finding the path of minimum action.

For a given boundary size, at very low positive temperature there is
only one stationary point, whose spatial slice is a ball in empty dS surrounded by the boundary. As the temperature increases, another stationary point arises, for which the boundary is surrounded by a
portion of SdS ending at the cosmological horizon. 
(This includes pure dS space as a special case.)
At even higher temperatures, two further stationary points arise, which have black hole horizons surrounded by the boundary. The empty dS 
and black hole stationary points are essentially similar to those in the case with zero \cite{WY} 
or negative \cite{HawkingPage} cosmological constant.
Above some temperature, the cosmological horizon solution has the lowest action among the four stationary points, and one might therefore expect the partition function path integral is well-approximated by this solution. 
However, like for the smaller of the two black hole solutions, the cosmological horizon solution is a local {\it maximum} of the 
(spherical, constrained)
canonical \euc action as a function of $M$.  
This indicates instability of the cosmological horizon solution. This instability is a property
of the ensemble defined with an idealized reservoir of fixed temperature that is unaffected by the evolution of the system. We thus 
consider also a dynamical model for the reservoir that can stabilize the system. This issue does not
arise for the black hole phases since for them it makes physical sense to postulate an infinite reservoir whose temperature is fixed independently of the system dynamics.

In the microcanonical ensemble, 
instead of fixing the temperature,
the quasilocal energy density \cite{BY-quasi} is fixed on the system boundary. For the spherical boundaries with fixed metric that we consider, this is equivalent to fixing the total quasilocal energy $E$,
which together with the spherical boundary radius determines the microcanonical ensemble. (We assume vanishing rotation, so that
the quasilocal momentum density is set to zero.)
The path integral representation
for the density of states $\nu(E)$ is a sum over geometries satisfying these boundary conditions, 
weighting with $\exp(i\im)$, 
where $\im$ is the microcanonical action~\cite{BY}. 
Since the microcanonical path integral directly computes the density of states, it should be 
valid even for systems
whose canonical ensemble
is ill-defined due to instability and/or a divergent partition function. For each set of boundary data with nonzero energy,
it turns out that there is only one admissible solution to the constraint equations, which corresponds to a SdS solution with some (possibly negative) mass parameter. 
We identify this stationary point of the action and approximate the path integral using that solution, which gives the leading order expression for the 
entropy, $S \sim \log \nu(E)$. For spacetimes with a horizon
this yields 
the Bekenstein-Hawking entropy,
and for those without it yields zero entropy.

Motivated by previous suggestions \cite{KV,TJ&Manus,Jacobson:2019gco} that the de Sitter static patch should perhaps be understood as a {\it negative} temperature canonical ensemble, 
we explored this possibility for both ensembles. In a canonical ensemble at negative
temperature the free energy is {\it maximized}
rather than minimized. We therefore 
estimated the entropy using the configuration at which the (spherical) action has a local maximum. This yields a {\it negative}
entropy, indicating either an error in the approximation or in the implicit deformation of the integration contour, and we elaborate on these possibilities.
We also compute the microcanonical temperature, and find it to be positive. Therefore, neither the canonical nor the microcanonical ensembles  defined in this paper admit negative temperature phases. 
In the Discussion section we say a few words about how this result might be 
reconciled with the previous reasoning~\cite{KV,TJ&Manus,Jacobson:2019gco} in favor of a negative temperature interpretation. 

\paragraph{Relation to previous studies}
Thermodynamic ensembles with cosmological horizons have been studied previously
in a number of different settings. 
In \cite{Teitelboim,Gomberoff-Teitelboim} a boundary was placed at either the black hole or cosmological horizon. 
The canonical ensemble for de Sitter  space with a timelike boundary inside the cosmological horizon has been considered in~\cite{Hayward} (where a fluid matter distribution is also included), and in \cite{WangHuang,Saida} (where the cosmological constant is taken as variable). 
More recently, 
refs.\  \cite{Miyashita2,Svesko:2022txo,Draper-Szilard} appeared while we were preparing this paper.\footnote{A report on an earlier stage of our work is available in \cite{APStalk}.}
Where our work overlaps with aspects of these papers, our results are 
in agreement. The greatest overlap is with ref.~\cite{Draper-Szilard}
which, like our paper, 
considers both microcanonical and canonical ensembles in the spherically reduced theory, includes
both black hole and cosmological horizons,
identifies the same stationary points and stability properties,
and compares actions to determine the thermodynamic phases.
Our work differs in that 
we start from the reduced phase space formalism, 
in which the initial value constraints are imposed at the outset,
and off-shell configurations are included in the path integral.
Some of the points we focus on that do not figure in
or are not fully explored in previous work are (i) the underlying justification and limitations of the Euclidean path integral method, 
(ii) the complete classification of spatial topologies and solutions to the constraints, and the principles by which some of these are excluded from the ensemble, (iii) the possibility of transitions from black hole to cosmological horizon within the same ensemble, (iv) a thorough phase analysis revealing that sometimes endpoint configurations dominate the partition function and exploring their physical significance, 
(v) a physical explanation as to how a reservoir model can stabilize the cosmological horizon in the canonical ensemble, (vi) an exploration of the possibility of negative temperature ensembles, and (vii) a demonstration of how the Gibbons-Hawking partition function on the 4-sphere can be derived
as a limit of well-defined thermodynamic ensembles and computes the dimension of the Hilbert space of states within a cosmological horizon.

The remainder of the paper is structured as follows.
In section \ref{foundations} we review and discuss the fundamental underpinnings of the approach to gravitational thermodynamics undertaken in our study.
Section \ref{can} considers the canonical ensemble, and section \ref{micro} considers the
microcanonical ensemble. In section \ref{Gibbons-Hawking} we show how Gibbons and Hawking partition function for de Sitter space can be seen as a thermodynamic partition function in the limit when the system boundary vanishes, and section \ref{Discussion}
includes some remarks and further discussion. 
Finally, three appendices contain some computational details.
We use signature $({-}{+}{+}{+})$ for Lorentzian 
metrics, and work in Planck units, setting $c= \hbar = G = k_{\text{B}} = 1$.

\section{Foundational issues}
\label{foundations}

Before embarking on the 
semiclassical, symmetry reduced
analysis that forms the basis for this paper, we would like to briefly
address in this section a number of issues 
regarding the fundamental underpinning of this analysis.

\paragraph{GR as effective field theory}
Throughout the paper we consider pure gravity in four dimensions and employ the Einstein-Hilbert action (with appropriate boundary terms) in the path integral. Although general relativity is not UV-complete, 
it can be consistently quantized 
as a low-energy effective field theory (EFT)~\cite{Donoghue:2017}. 
The exact gravitational path integral  
is therefore presumably not well
defined, and in any case it could
not be evaluated 
exactly.
Nevertheless, beginning with the work of Gibbons and Hawking~\cite{GH},
there
is ample evidence that when explored in a semiclassical approximation around macroscopic stationary points it has an uncanny ability to make good sense of some aspects of quantum gravity. 
In this approach the main contribution to the path integral is presumed
to be captured by the stationary points of the action --- the classical solutions --- and neighboring configurations, which are geometries with  curvature low compared to the Planck scale.
For such purposes, the higher curvature terms expected in the true EFT action of quantum gravity would presumably be negligible. 
While this semiclassical computational scheme thus has a degree of self-consistency, 
we are not aware of
any analytical 
arguments that directly 
support the validity of ignoring contributions from highly curved configurations, which
would depend on the form of the 
higher derivative terms and, ultimately, on the UV completion of the theory. The only real evidence that it makes sense is just that it has given sensible looking results in the past. Aiming for more, perhaps  constraints on the effective action could be derived from the requirement that the semiclassical contributions indeed dominate the path integral.

Having neglected higher curvature terms in the action in the low energy EFT treatment, it would not be consistent to include configurations with diverging curvatures in the path integral. In particular, we do not consider contributions to the partition function from configurations with conical defects. As a conical singularity is approached the curvature may remain small, but at the vertex it diverges, producing a delta-function like contribution to the curvature (see e.g.\ \cite{Solodukhin} and the references therein for further detail). While the integral of the scalar curvature remains finite, the curvature squared terms in the effective action might contribute large amounts that could only be reliably evaluated in the UV completion of the theory. On this ground, we assume all the geometries contributing to the path integral are smooth manifolds.

\paragraph{Dirichlet boundary conditions}
The canonical and microcanonical thermodynamic ensembles 
we consider are specified by 
boundary conditions at a timelike boundary. The presumption is that this 
defines an autonomous classical system with a well-defined phase 
space that can be canonically quantized, and that is dynamically stable. This presumption underlies most prior work on the subject, but its validity has not been established, and in fact there are reasons to doubt it. The canonical ensemble is defined by 
Dirichlet boundary conditions on a timelike boundary, and it is not clear that 
the corresponing initial boundary value problem is well-posed~\cite{AnAnderson}. Moreover, even if well-posed, 
it is not necessarily stable. In fact, it was found numerically in \cite{Andrade-etal} that there exist unstable solutions to the 
Einstein equation, linearized about a dS background
inside a sufficiently large 
Dirichlet wall, or outside a Dirichlet wall when linearized
about an SdS background. 
If these instabilities exist also in the nonlinear theory,
they would appear to indicate that the system cannot 
achieve thermodynamic equilibrium, in which case canonical ensemble
we consider in this paper may not be well defined.  
\paragraph{Constrained phase space and covariant path integral}

When deriving the path integral representation 
of the evolution operator, the partition function, or the density of states using a phase space formulation, the paths
are initially trajectories through phase space,
and a configuration space path integral can then be obtained by integrating out the momenta.
In a field theory, 
this can lead directly to a path integral over 
covariant fields, but
for a theory with local gauge symmetry, such as electrodynamics or general relativity, the 
covariant fields must be restricted by
initial value constraints and gauge-fixed
in order to parametrize the physical phase space.
Depending on the particular theory, there may or may not be a way to express the resulting path integral
as an integral over covariant fields with a local action, 
and some specified contour of integration.
In the approach taken by Whiting and York~\cite{WY}, and in subsequent work along those lines, the action that appears in the exponential integrand of the path integral is 
the ``reduced action", in which the constraints
have been imposed. This is the approach we take in this paper.\footnote{Whiting and York \cite{WY} use the term ``reduced action" when referring to the action of constrained metrics. We call it a constrained action, and reserve the term ``reduced action'' for its spherical reduction.}

Once the constraint equations are implemented one way or another, the infamous conformal mode (which makes the \euc gravitational action unbounded below \cite{GHP}) is eliminated \cite{Schleich, Hartle-Schleich, Mazur-Mottola,Loll-Dasgupta}. 
The conformal fluctuations around a background metric are fixed by the constraint equations, and are not independent dynamical degrees of freedom (see e.g.\ \cite{Kuchar, Hajicek});
thus, upon solving the constraints there will be no kinetic term in the action involving the conformal mode to cause problems.
Therefore, when working with the constrained action, any kind of instability that may appear would be a non-conformal one, as we will see in the analysis of the canonical partition function. 

\paragraph{Gauge fixing}
While we impose the constraints ``by hand",
gauge fixing is still required to avoid summing over redundant, physically-equivalent paths.
To maintain gauge invariance
of the path integral, the 
measure must include a 
determinant that is usually
implemented by including a Gaussian Berezin integral over Faddeev-Popov ghost fields \cite{Faddeev-Popov} (which in general couple to the physical fields). We are able to ignore the ghost contributions, because 
we focus on the leading order
contribution, which comes from the action at the classical stationary point, divided by $\hbar$. The ghost contribution is $O(\hbar^0)$, so it could become comparable to that from the stationary point only if there is a parametrically large quantity that could overcome the Planck length suppression. It seems this cannot happen in a bounded region; thus, like our predecessors~\cite{York86,  WY,York88}, we simply ignore the ghost contribution.

\paragraph{Validity of semiclassical approximation}
If a stationary point of the action is truly to dominate the path integral evaluation of thermodynamic quantities, it must not only lie in the classical regime, but it must be {\it stable} against small fluctuations in its neighborhood.  
That criterion 
is plausibly required 
in order for the stationary point to characterize a stable or metastable phase within the ensemble. 
In particular, 
a stationary point (extremum) of the \euc action can provide 
a good approximation for 
the contribution of a phase to the 
partition function 
only if it is
a local minimum of the action 
along the integration contour.
Two approaches for examining whether a stationary point satisfies this criterion have been
considered in the literature.

One approach is to expand the \euc action to quadratic order around the stationary point
and check whether there are any fluctuations that decrease the action --- i.e.\ ``negative modes".\footnote{In this paper --- as in the works before us --- we are assuming that the positivity of the second variation of the action at a stationary point indicates a {\it local minimum}. However, rigorously speaking, the criteria for a stationary point to furnish a local minimum depends on the precise definition of {\it local} in the space of functions, i.e.\ how one defines the neighborhood of a function. There are simple examples where a stationary function at which the second variation of the functional integral is positive fails to be a local minimum, according to some definition of local~\cite{Calculus, calculus-Ball}.}
Gross, Perry and Yaffe \cite{GPY} applied this method to an asymptotically flat black hole. 
Among all
variations away from the Schwarzschild solution, they found only one (non-conformal) negative mode  which is a
 static, spherically symmetric,
 transverse traceless variation that lowers the  \euc  action
and therefore indicates that the asymptotically flat black hole is thermodynamically unstable. 
Allen \cite{Allen} placed the black hole in a spherical box, and found that when 
$R \lesssim 2.89 M$, where $R$ is the box radius and $M$ is the black hole mass,
the negative mode is eliminated. 
In ref.~\cite{Allen} 
a coordinate location of the boundary was fixed. 
Gregory and Ross \cite{Gregory-Ross}
fixed instead the box area, and obtained the condition $R<3M$ 
for eliminating the negative 
mode.\footnote{\label{foot1} York \cite{York86} had earlier found precisely the same stability condition simply from the 
requirement that the heat capacity
computed from spherical black hole solutions at fixed proper box temperature be positive. A similar result was previously obtained by Hawking and Page \cite{HawkingPage} for the case of a large, asymptotically anti-de Sitter black hole, and this was generalized 
in \cite{Brown-Creighton-Mann} to 
the case of a finite box.} 
The negative mode analysis 
was later carried out by Prestige \cite{Prestidge} for the case of 
Schwarzschild-anti-de Sitter black holes in a finite box. In particular, he
recovered the Hawking-Page result\footref{foot1} in the limit of an infinite box.

In another approach for analyzing 
stability, 
first taken by Whiting and York \cite{WY} with zero cosmological constant, one restricts to a highly symmetric,
e.g.\ spherical
class of configurations 
that are solutions to the constraints,
and examines how the action at the 
stationary points compares with the action 
for all other configurations
within that category. 
This is the approach we take in this paper. 
It is simpler to implement than
checking for negative modes, especially
since the latter generally requires numerical methods. It also has the advantage of allowing one to probe the configuration space more globally, rather than only in the infinitesimal neighborhood of the stationary points. For a complete stability analysis one should check that the \euc action has a local minimum everywhere in the  neighborhood of a stationary point;
however, in all cases that have been analyzed with both methods, no instabilities have been detected using the generic local method that were not also seen among the symmetric configurations.

For the microcanonical ensemble, one might expect that the accuracy of the semiclassical approximation can be examined following approaches similar to those used to check stability and dominance of a stationary point in the canonical ensemble.  The local 
analysis, checking the variation of the action due to quadratic fluctuations around each stationary point (perhaps numerically),
can indeed be applied.
While such an analysis is feasible, it lies beyond the scope of this paper.\footnote{Ref.~\cite{Marolf-Santos} recently examined numerically the
stability of the spherically reduced microcanonical ensemble with a black hole in a finite box
and $\Lambda\le0$.
An important difference from
our approach is that there the initial value constraints are not imposed.} 
It turns out that the 
symmetry reduced global method,
on the other hand, is not applicable.  Once the constraints are imposed, there is a unique solution to the constraints satisfying the boundary conditions,
hence
there is no free parameter in the action 
parameterizing different configurations.\footnote{In ref.~\cite{Miyashita} a
minisuperspace path integral method is employed (in the case of zero cosmological constant) to assess the behavior of the microcanonical path integral near the stationary points. A class of static, spherically symmetric metrics that coincide with the Schwarzschild metric except for an unspecified factor $N$ in $g_{rr}$ is considered. 
The integration contour in the complex plane for this $N$ can be chosen so that some stationary points give the largest contribution to the microcanonical path integral, and hence 
might be viewed as the dominant configurations. However, 
since in general these metrics 
do not satisfy the constraints, 
and the chosen class of metrics is somewhat arbitrary, the significance 
of this construction is not clear
to us.}
Here we therefore leave unassessed the accuracy of the approximation.

\section{Canonical ensemble} 
\label{can}

We begin with a review of the formal construction of a path integral representation of the partition function $Z=\text{Tr}\, e^{-\beta H}$ for gravitational systems. This construction is standard, however we shall draw attention to a subtlety concerning the integration over the lapse function and shift vector which has not been emphasized in the literature.
In particular, we will see that the standard ``Euclidean path integral" is not actually equivalent to the partition function.

We consider a family of surfaces $\Sigma$ parameterized by the time coordinate $\tau$, and a time flow vector $t^{\mu}$ satisfying $t^{\mu}\nabla_{\mu} \tau=1$. The vector $t^{\mu}$ can be decomposed into the lapse function and shift vector, $t^{\mu}=N u^{\mu}+\underline{N}^{\mu}$, where $u^{\mu}$ is the unit normal to the surface $\Sigma$, and $\underline{N}^{\mu}$ is the spatial part of $N^{\mu}=(N,N^i)$.
The classical Hamiltonian of general relativity is a linear combination of the constraints $\mc{H}_{\mu}$, plus a boundary term, 
\be
H=\int_{\Sigma}\! d^3x \, N^{\mu}\mc{H}_{\mu} + H_{\partial}\, .
\label{H}
\ee
The lapse function and shift vector appear as non-dynamical fields. 
The dynamical fields are the components of the 3-metric $h_{ij}$ on the constant-$\tau$ surface $\Sigma$ and their conjugate momenta $\pi^{ij}$ (see e.g.\ \cite{toolkit}).
The boundary term $H_\partial$ depends on the boundary conditions defining the system, which we will take to be 
fixed boundary metric (Dirichlet) conditions. The Hamiltonian \eqref{H}
generates evolution with respect to the time flow $t^\mu$, which is chosen to be a unit vector on the boundary, so that $H$ generates evolution with respect to the proper time of the ``boundary observers".

By breaking the time interval into infinitesimal steps and inserting a complete set of states at each instant, the partition function can be written as a phase space path integral over the canonical variables $h_{ij}$ and $\pi^{ij}$, 
\be
Z=\text{Tr} \, e^{ -\beta H }=\int\! \mc{D} h_{ij}\, \mc{D} \pi^{ij} \, \delta(\mc{H}_{\mu})\, \text{exp} \left [ i\int \! d\tau \! \int_{\Sigma} \! d^3x \, \pi^{ij}\dot{h}_{ij}\,  -\int\! d\tau\,  H_{\partial}\right ]\, .
\ee
The constraints are imposed by a delta function that restricts the paths to the physical phase space.
To implement the trace we take the $h_{ij}$ paths to be periodic in $\tau$,
and the period of $\tau$ is 
$\beta$, which corresponds to the proper time period at the boundary.
To eliminate redundant summation over gauge equivalent paths, one may fix a gauge and include a Fadeev-Popov determinant to maintain gauge invariance, or equivalently divide by the volume of the diffeomorphism group. As explained in section \ref{foundations}, the ghost contribution is subdominant in the stationary point approximation, and can thus be omitted here,
so in practice we will just fix a gauge. 
The delta function $\delta (\mc{H}_{\mu})$ 
can be rewritten as a functional Fourier integral over the auxiliary fields $\lambda^{\mu}=(\lambda,\lambda^i)$. Thus 
\be \label{ZNpih}
Z=\int \! \mc{D}\lambda^{\mu}\, \mc{D} h_{ij}\, \mc{D} \pi^{ij} \, \text{exp} \left [ i\int \! d\tau \! \int_{\Sigma} \! d^3x \, (\pi^{ij}\dot{h}_{ij}\, - \lambda^{\mu}\mc{H}_{\mu})\,- \, \int\! d\tau\, H_{\partial} \right ]\, .
\ee
Since the constraints are quadratic in $\pi^{ij}$ the integration over the momenta can be done explicitly, resulting in a functional integral over the configuration space. Performing the momentum integration is equivalent\footnote{More precisely, since the coefficient of $(\pi^{ij})^2$ involves the metric, this is true only up to a metric-dependent factor. Along with other metric dependence that may be in the integration measure, this can be neglected in the stationary point approximation, as it is subleading in $\hbar$.}
to evaluating the exponent at the stationary `point' when varying with respect to $\pi^{ij}$, i.e.\ at momenta that satisfy $\dot{h}_{ij}=\delta H/\delta \pi^{ij}$,
which determines the momentum in terms of the spacetime metric and its derivatives.\footnote{Explicitly  $\pi^{ij}=\sqrt{h}(K^{ij}-Kh^{ij})/16\pi$, where $K_{ij}=(\dot{h}_{ij}-\nabla_i \lambda_j-\nabla_j \lambda_i)/2\lambda$ is the extrinsic curvature of the constant-$\tau$ surfaces $\Sigma$ and $K$ is its trace.}
After imposing this relation, the path integral becomes (for details, see e.g.\  \cite{toolkit,HH})
\be \label{ZNh}
\begin{aligned}[b]
Z=\int \! \mc{D}\lambda^{\mu}\, \mc{D} h_{ij}\, \text{exp} \Big [ &\frac{i}{16\pi} \int_{{\cal M}}\!d^4x \, \lambda\sqrt{h}\, (R-6/\ell^2)\\  +&\frac{i}{8\pi} \int_{\partial{\cal M}} \! d^3x \, \lambda \sqrt{\sigma}(s\, u^{\mu}u^{\nu}\nabla_{\mu}n_{\nu})
+2i\int_{\partial{\cal M}} \! d^3x\sqrt{\sigma}\,h_{ij}\lambda^i\,n_k\pi^{jk}/\sqrt{h}\\
 +& \frac{1}{8\pi}\int_{\partial{\cal M}} \! d^3x \, N_{\partial}\sqrt{\sigma}\, k -2\int_{\partial{\cal M}} \! d^3x\sqrt{\sigma}\, h_{ij}N^i_{\partial}\,n_k\pi^{jk}/\sqrt{h}\Big ]\, ,
\end{aligned}
\ee
where $\ell = \sqrt{3/\Lambda}$ is the dS length scale,  
and the last two terms are equal to $-\int\!d\tau\,H_{\partial}$,
expressed in an arbitrary gauge.\footnote{Later we are going to choose the boundary conditions where the shift vector vanishes on the boundary, $N^i_{\partial}=0$; but here we keep the expressions in the general form.} 
In the above equation $u^{\mu}$ is the unit normal to the slices $\Sigma$ with $u^\mu u_\mu = s =\pm1$, $n^{\mu}$ is the outward unit normal to the manifold 3-boundary, $\sigma$ is the determinant of the induced metric on boundary 2-surface $\partial \Sigma$, and $k$ is the trace of its extrinsic curvature as embedded in $\Sigma$, which we define with respect to the outward normal and is equal to $k=(g^{\mu \nu}-s\,  u^{\mu}u^{\nu})\nabla_{\mu}n_{\nu}$. We chose the slices $\Sigma$ to intersect the manifold 3-boundary orthogonally, thus the normal $n^{\mu}$ is also the normal to the 2-boundary surface in $\Sigma$. The path integral is taken over geometries that are periodic in the time coordinate,
with proper time period $\beta$ at the boundary.

Note that the functions $\lambda$ and $\lambda^i$ in the second line of (\ref{ZNh}) are the boundary values of the auxiliary field $\lambda^{\mu}$ that was 
used in the Fourier representation of the constraint delta function, whose original contour of (functional) integration was along the real axis. Hence the first two lines in the exponent of 
\eqref{ZNh} are pure imaginary, whereas the third line is real. To arrive at the 
usual ``Euclidean partition function" requires some hocus pocus. Imagine, first, that 
we could rotate the $\lambda^{\mu}$ contour 90 degrees in the complex plane, 
i.e.\ let $\lambda^{\mu}=-iN^{\mu}$ and integrate over real $N^\mu$. 
The original integral over the real bulk values of $\lambda^{\mu}$ imposes
the constraints everywhere apart from the boundary, so by
continuity it also imposes the constraints at the boundary. The
boundary values of $\lambda^{\mu}$ therefore need not be integrated
over. In particular, if we could impose 
$\lambda^{\mu}_{\text{boundary}}=-iN^{\mu}_{\partial}$, 
where $N^{\mu}_{\partial}$ is the real lapse and shift fixed by the boundary data, 
then the boundary integrals in the exponent in (\ref{ZNh})
would combine to $-\frac{1}{8\pi }\int_{\partial \mc{M}}\!d^3x\sqrt{\gamma}K$,
where $\gamma$ is the determinant of the induced metric on the boundary and
$K$ is the trace of the extrinsic curvature of the boundary surface, 
defined with respect to the outward normal vector.
Having rotated the $\lambda^\mu$ contour of integration as just described,
and denoting $N^{\mu}$ and $h_{ij}$ jointly as $g_{\mu \nu}$, the partition function
\eqref{ZNh}
would thus become 
\be 
Z=\int \! \mc{D} g_{\mu \nu } \,  e^{-I[g]}\, ,
\label{Zc}
\ee
where the \euc canonical action $I$ consists of the Einstein-Hilbert term on the manifold $\mc{M}$
together with a boundary term which, for the canonical ensemble, is the Gibbons-Hawking-York (GHY) boundary term on $\partial \mc{M}$:
\be
I[g]=-\frac{1}{16\pi }\int_{\mc{M}}\!\!d^4 x \sqrt{g}\,(R-6/\ell^2)-\frac{1}{8\pi }\int_{\partial \mc{M}}\!d^3x\sqrt{\gamma}K\, .
\label{Icov}
\ee
Thus, had this contour rotation been valid, we would have arrived at the standard ``Euclidean path integral" representation for the partition
function.\footnote{Unlike the asymptotically flat \cite{GH, BY-quasi} or anti-de Sitter space \cite{Bala-Kraus} cases, $\partial \mc{M}$ is a finite boundary, hence there is no need to add a boundary counterterm to the action in order to render it finite.} 

Past experience shows that {\it something} is correct about this Euclidean path integral,
but it is certainly not entirely valid. The $\lambda^\mu$ contour rotation is not consistent with 
the Fourier representation of the constraint delta functions, and in fact it is well known that
the constraints are not imposed by the Euclidean path integral. In particular, this is why
the Euclidean action is unbounded below because of the conformal mode, which should be eliminated by the 
constraints. But how then can we understand the fact that the stationary points of the Euclidean
path integral can provide physically sensible estimates for thermodynamic quantities? 
The answer, we suppose, is that although in fact the integration contour for $\lambda^\mu$ 
can {\it not} be rotated 90 degrees, it can be {\it deformed} in such a way that it does pass through the 
Euclidean (or complex) stationary point.  However, 
non-Euclidean contributions along the deformed contour 
are  essential to maintaining equivalence with the partition function that we started from.
In what follows we shall adopt this Euclidean path
integral, imposing the constraints by hand, as discussed in section \ref{foundations}.

For a nongravitational system, the path integral for the partition function would be over field configurations on a fixed Euclidean manifold $\Sigma\times S^1$, where $\Sigma$ is the spatial manifold at one time and $S^1$ is the Euclidean time circle with period $\beta$  equal to the inverse temperature.
To specify a gravitational canonical ensemble instead one fixes a {\it boundary} of topology ${\cal B}\times S^1$, 
where ${\cal B}$ is the ``box" whose time-independent geometry must be specified, and the proper period $\beta$ of the Euclidean time circle $S^1$ is equal to the inverse  temperature at the boundary. Conceptually, the box is in contact with a reservoir that fixes the proper temperature at the box.\footnote{In an equilibrium state, the local proper temperature away from the boundary is related to the boundary temperature by the inverse of the local norm of the time translation Killing vector (the ``Tolman redshift" factor).}

In the nongravitational case, the phase space is independent of the thermodynamic parameters, and in particular the spatial topology and geometry are fixed {\it ab initio} by the domain of the system. In the gravitational case,
having fixed the boundary topology
${\cal B}\times S^1$, 
the (Euclidean) spacetime topology is not yet determined. 
If topology is a dynamical degree of freedom in quantum gravity, then one should presumably sum over topologies, including ones that change as a function of the Euclidean time. That said, it is not clear whether and how the notion of a self-adjoint Hamiltonian for the system would carry over to a system with dynamical topology, so the statistical significance
of the canonical partition function 
becomes rather murky in that case.\footnote{\label{JT}Some progress in understanding the sum over topologies for a two dimensional gravity model has recently been made in ref.~\cite{Saad:2019lba}.} 
In the present study we  
assume that it is consistent with quantum gravity,
at least in a suitable 
approximation,
to restrict attention to time-independent spatial topologies, and we 
shall make that restriction.

The seminal work of Gibbons and Hawking \cite{GH} demonstrated that 
even when the spatial topology is constant in time,
the Euclidean
{\it spacetime} topology
should not be restricted to a product form $\Sigma\times S^1$ when computing 
the partition function.
In particular,
the Bekenstein-Hawking entropy 
for an ensemble dominated by black hole configurations
is captured by the Euclidean 
Schwarzschild stationary point,
in which the Killing time 
circle shrinks to a point at the horizon. 
The topology of 
this Euclidean black hole is 
not of the form $\Sigma\times S^1$, but is rather $S^2\times D^2$, where $D^2$ is a disc whose boundary $S^1$ is the boundary time circle, and whose center lies at the Euclidean horizon.
Why this topological 
transmutation 
should give what is believed to be the right result for the entropy is not entirely clear.
It seems that the macroscopic manifestation of the gravitational Hilbert space relevant to the static black hole entropy is correctly captured by the geometry {\it outside} the horizon. It is as if the path integral ``knows enough" to set up the {\it relevant} Hilbert space, even if that may not be the {\it whole} Hilbert space, and even if the path integral is not 
well defined outside of certain approximate forms.\footnote{Recent progress on the black hole information paradox has been made involving another topological transmutation mechanism~\cite{Penington:2019kki,
    Almheiri:2019qdq,Almheiri:2020cfm}. } 
We  thus 
assume  that the topology of the 
4-manifold $\mc{M}$, whose boundary is 
$\partial\mc{M}=\mc{B} \times S^1$,
is either 
$\Sigma \times S^1$ 
with $\partial \Sigma=\mc{B}$,
 or is such that
the time circle $S^1$ shrinks to zero size in the interior
at a set of fixed points of the time flow where the lapse function vanishes. 
This fixed point set is called a (Euclidean) {\it horizon}, or {\it bolt}.\footnote{While for \euc metrics such a codimension 2 fixed point set is usually called a ``bolt"~\cite{Gibbons:1979xm}, we will use the term ``horizon" also for the Euclidean case,
in view of its role in the context of gravitational thermodynamics.}

Although we have derived the covariant form of the
action \eqref{Icov}, as explained in 
section \ref{foundations} we shall 
restrict
the path integral
to paths that satisfy the constraints,
so as to obtain an integral over paths in the physical phase space. Our next step is to evaluate the action subject to the constraints. 
The Euclidean time function $\tau$ defines the foliation with respect to which the constraints are imposed. 
When a horizon is present, however, the constant-$\tau$ surfaces all meet at the horizon, so they do not provide a proper foliation of the manifold there. 
To finesse this complication, we follow the common procedure \cite{BTZ} of cutting the manifold in the vicinity of the horizon, and evaluating the action of the interior and exterior regions separately in the limit as the cut surface approaches the horizon.
The action in the interior part of the manifold near the horizon consists only of the Einstein-Hilbert bulk integral in (\ref{Icov}), which vanishes in the desired limit due to the zero volume of that region and regularity of the scalar curvature.
The remaining integration domain has the topology $\Sigma \times S^1$, with a horizon boundary component, $\partial \Sigma_h \times S^1$, due to excision of the horizon, at which a boundary term in the action arises.
As explained below, this horizon boundary term is proportional to the area of the horizon, and becomes the entropy at the stationary points.\footnote{\label{conical}It is worth emphasizing that, although we step slightly outside the strictly Hamiltonian framework to assign an action to paths with horizons, that assignment does not entail any dependence on the lapse function (except at the system boundary) provided the 4d manifold remains smooth. 
In fact,  one obtains the same 
value of the action even for a lapse that produces a conical singularity at the horizon. (For an explicit demonstration see appendix A of ref.~\cite{Gregory-Moss}; 
presumably, it may also be seen by 
considering a 1-parameter family of lapse choices with smooth geometries whose limit has a conical singularity). We note, however, that for geometries with large or singular curvature, the neglect of higher curvature corrections to the Einstein-Hilbert action does not appear to be justified.}

In \euc signature we have $R=-2(G_{\mu \nu}-R_{\mu\nu})u^{\mu}u^{\nu}$,
where $u^\mu$ is the unit normal to the hypersurfaces of constant $\tau$, 
and is related to the lapse $N$ by $u_\mu = N\nabla_\mu \tau$. 
The first term together with the cosmological constant term gives the Hamiltonian constraint, and the second term is \cite{HH}
\be
R_{\mu \nu}u^{\mu} u^{\nu} = K^2-K^{ij}K_{ij}-\nabla_{\mu}(u^{\mu}\nabla_{\nu}u^{\nu})+\nabla_{\nu}(u^{\mu}\nabla_{\mu}u^{\nu})\, .
\label{Ruu}
\ee 
The first total derivative in (\ref{Ruu}) contributes nothing to \eqref{Icov} since $u^{\mu}$ is tangent to the boundary, but the last term gives rise to a boundary integral.  
Thus we have
\be
\begin{aligned}[b]
-\frac{1}{16\pi }\int_{\mc{M}}\!d^4 x \sqrt{g}\,(R-6/\ell^2)= &\,\frac{1}{8\pi}\int_{\mc{M}}\!d^4 x \sqrt{g} \Big [(G_{\mu \nu}+(3/\ell^2) g_{\mu\nu}) u^{\mu} u^{\nu} + K_{ij}^2-K^2\Big ]\\
&+\frac{1}{8\pi}\int d\tau  \int_{\partial \Sigma}\! d^2x  \sqrt{\sigma}\, N \,u^{\mu}u^{\nu}  \nabla_{\mu}n_{\nu}
\, ,
\end{aligned}
\label{action}
\ee 
where $n^{\mu}$ is the outward unit normal to the boundary, and the boundary integral has come from the last term in (\ref{Ruu}) after integrating by parts and using $u^{\mu}n_{\mu}=0$. 
The boundary $\partial \Sigma$ in (\ref{action}) includes the intersection of the slices $\Sigma$ with the system boundary and, if there is a horizon,
also the stretched horizon cross-section $\partial\Sigma_h$.
As mentioned above, we impose directly
the Hamiltonian constraint
$(G_{\mu \nu}+(3/\ell^2) g_{\mu\nu}) u^{\mu} u^{\nu}=0$.
The terms quadratic in the extrinsic curvature are present in general, and we evaluate them later for the spherically symmetric geometries. Next we evaluate the boundary terms in the action.

In addition to the boundary term in \eqref{action}, there
is also the GHY term in (\ref{Icov}), which is only present at the system boundary. Thus at the system boundary the combined boundary integral becomes 
$-(1/8\pi)\int \! d^3x N\sqrt{\sigma}  (g^{\mu \nu}-u^{\mu} u^{\nu})\nabla_{\mu}n_{\nu} = -(1/8\pi)\int \! d^3x N\sqrt{\sigma}\, k$, where $k$ is the trace of the extrinsic curvature of the 2-boundary $\mc{B}$ as embedded in $\Sigma$.
To evaluate the integrand in the last term in (\ref{action})
at the horizon boundary we need the relation
\be
Nu^{\mu}u^{\nu}\nabla_{\mu}n_{\nu}=n^{\mu}\nabla_{\mu} N\, ,
\label{nN}
\ee
which follows from $u_\mu = N\nabla_\mu \tau$, $u^2=1$,  
and $n^\mu \nabla_{\mu}\tau=0$.\footnote{
%A derivation goes as follows: 
$Nu^{\mu}u^{\nu}\nabla_{\mu}n_{\nu}= -Nu^{\mu}n^{\nu}\nabla_{\mu} u_{\nu} = -Nu^{\mu}n^{\nu}\nabla_{\mu} (N\nabla_{\nu} \tau)= -N^2u^{\mu}n^{\nu}\nabla_{\mu} \nabla_{\nu} \tau
= -N^2u^{\mu}n^{\nu}\nabla_{\nu} \nabla_{\mu} \tau= -N^2u^{\mu}n^{\nu}\nabla_{\nu} (N^{-1}u_{\mu})= -N^2 n^{\nu}\nabla_{\nu} N^{-1}= n^{\mu}\nabla_{\mu} N$.
} 
(Note that the result (\ref{nN}) is independent of the choice of gauge; in particular, it does not depend on the choice of the shift vector on the horizon.)
If the manifold is smooth where time circle closes off at the horizon,
the derivative of the proper circumference of the time circle with respect to the proper distance away from the horizon must be $2\pi$, so we have
$\oint (-n\cdot\nabla N) d\tau = 2\pi$.\footnote{The
minus sign appears because the unit normal
$n^{\mu}$ points outward from the integration volume, which on $\partial\Sigma_h$ is {\it toward} the horizon.}
The boundary integral in the last term of \eqref{action}, evaluated at the inner boundary as it limits to the horizon, thus yields  $-1/4$ times the horizon area $A_h$ in Planck units.
Putting all the terms together, the canonical \euc action for constrained geometries becomes
\be
I=\frac{1}{8\pi}\int_{\mc{M}}\!d^4 x \sqrt{g}( K^{ij}K_{ij}-K^2)-\frac{1}{8\pi}\int_{\partial \mc{M}} \! d^3x N\sqrt{\sigma}\, k-A_h/4\, ,
\label{constrained}
\ee
where the area term is present only if there exists a horizon, and the boundary integral is evaluated at the system boundary. 
The term involving the integral of 
$k$ is equal to $\beta$ times 
the Brown-York quasilocal energy (cf.\ 
\eqref{E} in section \ref{micro}).

\subsection{Spherical reduction} 
\label{reduction}

The functional integral (\ref{Zc}) with the action (\ref{constrained}) defines the partition function as a sum over constrained 4-geometries
satisfying the boundary conditions on
the boundary with topology $\partial \mc{M}\! = \! \mc{B} \times S^1$. 
For definiteness and convenience we choose the boundary ${\cal B}$ to be a round 2-sphere, with area $4\pi R^2$. We further choose the shift vector to vanish on the system boundary.
Thus the partition function is a function of the boundary radius $R$ and the proper circumference of the boundary time circle $\beta$.  
Given a spherically symmetric boundary condition, there exist solutions to the Einstein equation --- the stationary points of the action (\ref{Icov}) --- that are locally spherically symmetric. Those solutions are fully classified by a generalization of the Birkhoff theorem, which  states that the spherically symmetric solutions of the Einstein equation 
with a positive cosmological constant are locally the class of \sds spaces or the Nariai solution \cite{Birkhoff,Birkhoff2,Birkhoff1}.\footnote{\label{-it}
In \cite{Birkhoff,Birkhoff1} this theorem is proved in Lorentzian signature, but since the Einstein equations are analytic in the time coordinate, if $g_{\mu\nu}(t,x)$ is a solution, so is $g_{\mu\nu}(-it,x)$. This allows us to safely view the \euc version of the mentioned geometries as all the possible Euclidean stationary points, because if there were additional \euc solutions, after continuing analytically back to the Lorentzian signature they would correspond to some Lorentzian solutions that have not been already considered, which contradicts the Birkhoff theorem.
(In \euc signature the Nariai solution is 
a special case of the SdS solution.)
Ref.~\cite{Birkhoff2} contains a much more general theorem, which applies to multiple dimensions, signatures and geometries, and includes matter sources.}
In constructing the reduced phase space, we shall restrict attention to solutions to the constraints that are {\it globally} spherically symmetric.\footnote{A remarkable infinite class of locally but not globally spherically symmetric solutions to the vacuum constraints with a positive cosmological constant was constructed and studied in refs.~\cite{Morrow-Jones-Witt,Schleich-Witt}.}\footnote{We remind the reader that, as discussed in section \ref{foundations}, 
linearized nonspherical solutions were found in \cite{Andrade-etal},
some of which are unstable. Whether corresponding nonlinear solutions exist is not yet known; for this paper we provisionally assume that no such unstable solutions exist.} 
This amounts to 
a restriction on the possible topologies, so it can be regarded as a
specification of the nature of the
system (at least in the effective 
field theory with which we are
approaching quantum gravity).\footnote{Whether other topologies would 
be consistent with a semiclassical 
partition function is an interesting
question that we will not address here.}
In restricting to spherical configurations, we are neglecting the nonspherical graviton fluctuations (and those of other fields if present). Within the effective 
field theory framework for quantum gravity,
the contribution of those modes to thermodynamic quantities is negligible in the presence of a horizon. For the purpose of classifying thermodynamic phases, they may therefore be neglected.

A stationary point of the action can correspond to a stable or metastable equilibrium state only if it is at least {\it locally} a minimum in the (Euclidean) configuration space. As discussed in section \ref{foundations}, the 
thorough method to check this
local minimum criterion is to carry out a perturbative analysis around the relevant stationary point, to determine whether
there are any ``negative modes" --- regardless 
of symmetry --- that decrease the action \cite{GPY,Allen,Gregory-Ross, Prestidge}. 
Previous work using this method found that when the boundary conditions are spherical the only negative modes
were also spherical.
The approach we take in this paper, following the
original approach of ref.~\cite{WY}, restricts
attention to the spherical configurations.
%which
This restriction, while
strong,
nevertheless
allows us not only to detect any spherical negative mode
in the infinitesimal neighborhood of a stationary point, but 
also 
to compare the action globally for all spherical configurations 
(including ones
at which the action is nonstationary)
with prescribed topologies.

Having assumed that 
the spatial topology is constant,
and that the spatial slices are globally
spherically symmetric,
we may choose a coordinate 
gauge such that
the line element 
takes the form 
\be
ds^2=U(\tau, y) d\tau ^2 +V(\tau, y)^{-1} dy^2 +r(y)^2 d\Omega^2\, ,
\label{metric}
\ee
where $d\Omega^2$ is the metric on a unit 2-sphere.
The system boundary value of the radial coordinate is $y_{\partial}=0$,
and the remaining 
$y$ reparametrization freedom is not yet fixed.
$U$ and $V$ are periodic functions in the time coordinate $\tau$, which is now chosen to have period $2\pi$. 
On the system boundary we restrict to static geometries corresponding to equilibrium states, so that the metric components at the boundary can be chosen to be $\tau$-independent functions that determine the boundary data $R\equiv r(0)$ and $\beta\equiv 2\pi \sqrt{U(0)}$, specifying the canonical ensemble. 
For each value of $\tau$ the ``spatial" submanifold is foliated, 
apart from an inner end where the topology may change, by 2-spheres, one of which is the system boundary. 
There are three possibilities for the inner terminus:
the sphere can collapse to a point, 
it can be antipodally identified to become the projective plane
${\rm RP}^2$, 
or it can remain a finite area sphere.
In the first two cases 
the topology of the 4-manifold $\mc{M}$ is 
$\Sigma^3\times S^1$, where $\Sigma^3$ is a 3-ball
$B^3$, or projective space minus a point ${\rm RP}^3\backslash\{\rm pt\}$, respectively.
In the third case the absence of an inner boundary
requires that the time circle closes off at a horizon,
so the topology of the 4-manifold $\mc{M}$ is $S^2\times D^2$. 
Thus there are three possible manifold topologies for $\mc{M}$,  
\be
B^3\times S^1,
\qquad {\rm RP}^3\backslash \{{\rm pt}\}\times S^1,
\qquad S^2 \times D^2 \, .
\label{topology}
\ee 
We shall sometimes refer to these topologies  as the ``ball'', the ``RP$^3$ geon'',\footnote{In Lorentzian signature this refers to a quotient of the maximally extended Schwarzschild spacetime \cite{Friedman:1993ty}.
The thermodynamics of and quantum field theory on that quotient was studied in detail in ref.~\cite{Louko:1998dj}. This however is quite different from the present setting, since
here the geon is a horizonless configuration,
and the boundary is $S^2\times S^1$, not ${\rm RP}^2\times S^1$.} 
and the ``horizon topology''.

In the gauge (\ref{metric}) we have 
$K_{ij}\propto \partial_\tau {h}_{ij}\propto \delta_i^y\delta_j^y$, 
hence $K^{ij}K_{ij}-K^2=0$, so  the 
bulk term in (\ref{constrained}) vanishes.
Therefore, 
for spherical metrics the only surviving terms in the constrained action (\ref{constrained}) are the Brown-York quasilocal energy term and the horizon area term. 
With our choice of radial parameterization the outward unit normal to the boundary is $n=-\sqrt{V}\partial_y$. Thus we have\footnote{Our expression for the action matches the results found by Hayward \cite{Hayward}, who considered spherical geometries with positive cosmological constant. We have relaxed the stationarity condition present in \cite{Hayward}, but the general form of the action remains the same. This general form was already manifest in the case with zero cosmological constant as shown by Whiting and York \cite{WY}.}
\be
I = -\frac{1}{8\pi}\int \! d\tau \sqrt{U_{\partial}} \int \! d^2x \sqrt{\sigma} \, k-A_h/4=\beta R\, (r'\sqrt{V})_{\partial}-\pi r_h^2 \, ,
\label{Ih}
\ee 
where prime denotes the derivative with respect to $y$, and $r_h$ is the horizon radius. The last term appears only if there exists a horizon, which here is a spherical set of fixed points of the time flow 
at which the function $U$ vanishes. 
Note that \eqref{Ih} has the form of 
$\beta$ times a Helmholtz free energy, 
$I=\beta F= \beta E-S$, where 
$S$ is the Bekenstein-Hawking horizon entropy term and
$E$ is the Brown-York energy. Indeed, this will be its interpretation in the stationary point approximation of the partition function.
The action \eqref{Ih} is the spherical reduction of the constrained action
\eqref{constrained}, and we shall refer to it as the {\it reduced action}.

\subsection{Reduced action and  stationary points} 
\label{spt}

In this section we find the solutions to the constraints, argue that some of those solutions should not be included on paths in the path integral, evaluate the action and find the contributing paths for a given boundary radius, and 
identify the nature of the paths for which the action is stationary.

\subsubsection{Constraints}
Recall that the constraints were already imposed in going from the action \eqref{action} to the form \eqref{constrained}, whose spherical reduction is \eqref{Ih}. However, the full impact of the constraints is not yet manifest, since we have not determined how the constraint equations restrict the 
 metric components
in the line element \eqref{metric}.
The constraint equations 
 on constant $\tau$ slices 
are part of the full set of (Euclidean) Einstein equations, namely %they are 
the 
four equations $G^{\tau}_{\,\,\mu}+\Lambda g^{\tau}_{\,\,\mu}\!=\!0$. The $\tau \tau -$equation is the Hamiltonian constraint, and the remaining three are the momentum constraint. The angular part of the momentum constraint is automatically satisfied for the metric (\ref{metric}) due to spherical symmetry, so there are just two constraints,
\begin{align}
    G^{\tau}_{\,\,\tau}+ \Lambda g^{\tau}_{\,\,\tau }&= \frac{1}{r^2}\left ( V(r'^2+2 r r'')+ V' r r' -1\right )+ \frac{3}{\ell ^2} =0\label{Gtt}\\
    G^{\tau}_{ \,\,y }+ \Lambda g^{\tau}_{ \,\,y}&=-\frac{r' \dot{V}}{rU V}=0 \, ,\label{Gty}
\end{align}
 where prime and overdot denote partial derivative with respect to $y$ and $\tau$, respectively.
We see that 
there are two distinct cases for satisfying the conditions on $V$ and $r$: 
either $r'=0$, or $r'\neq 0$.

For the $r'=0$ solutions, 
all 2-spheres have the same area, and the Hamiltonian constraint (\ref{Gtt}) implies
that their radius is $r=\ell/\sqrt{3}$. 
This solution can meet the boundary condition only if the radius of the boundary $S^2$ has the same special value,  
$R=\ell/\sqrt{3}$.
Since the radius of the 2-spheres is constant,
and we assume the spatial boundary has only one component, 
the topology (\ref{topology}) in this case has to be either the geon
${\rm RP}^3\backslash \{{\rm pt}\}\times S^1$
or the horizon $S^2 \times D^2$.
When $r'=0$ the constraints 
do not impose any 
conditions on $U$ and $V$.
The proper distance
at constant $\tau$ from the boundary to the inner terminus (the RP$^2$ or the horizon) is determined by $V$, 
and is therefore not fixed. 
In the horizon topology case
$U$ vanishes where 
the time circle contracts to zero size in the interior.
The action (\ref{Ih}) for 
the RP$^3$ geon is $I=0$, and that for the 
horizon topology is $I=-\pi \ell^2/3$, 
independent of the temperature. 
We will see that the Nariai solution \cite{Nariai} is in this class of configurations.

For the $r'\! \neq \!0$ solutions,  the radial momentum constraint \eqref{Gty}  implies that $\dot V=0$. 
In this case we can further fix the gauge by setting $y$ equal to the radial proper distance, which amounts to setting $V=1$ in (\ref{metric}). The line element is then
\be
ds^2=U(\tau,y)d\tau^2+dy^2+r(y)^2d\Omega^2\, ,
\label{sp2}
\ee
and the Hamiltonian constraint \eqref{Gtt} becomes
\be
r'^2+2rr''-1+3r^2/\ell^2=0\, .
\label{r-eqn}
\ee
After multiplication  by $r'$, (\ref{r-eqn}) can be expressed as $(rr'^2 -r + r^3/\ell^2)'=0$, which implies 
\be
r'^{\,2}=1-2M/r-r^2/\ell^2 \equiv f(r) \, ,
\label{rprime}
\ee
where $M$ is an integration constant with dimension of length.
Although
(\ref{rprime}) cannot be solved analytically for $r(y)$,
it determines the 
boundary value 
$r'(0)
=\pm \sqrt{f(R)}$,
which suffices
for evaluating the action (\ref{Ih}) 
(in the gauge $V=1$). 
The plus (minus) sign is associated to the case where the 2-sphere area initially grows (shrinks) away from the boundary, corresponding to negative (positive) extrinsic curvature of the boundary.\footnote{Recall that our convention is that the extrinsic curvature is defined with respect to the normal direction {\it out} of the system.} 
Note that this sign is {\it not} determined by the boundary data in the canonical ensemble. 
Regularity at the horizon (where $U=0$) requires that $r'$ also goes to zero there.\footnote{\label{Vzero}As discussed in section \ref{foundations},
the action for configurations with large curvature 
would depend on higher curvature corrections in the UV completion of the theory, so presumably cannot be accurately treated using the Einstein-Hilbert action. 
To see that regularity at $U=0$ requires $r'=0$, 
recall that the curves of constant $y$ are loops around
the $U=0$ ``origin" point in the $\tau$-$y$ subspace. 
If $r'$ is not zero at the origin, 
then it flips sign discontinuously 
on a line that passes through the origin.
This shows that the curvature 
would diverge strictly at the origin,
but not necessarily as the origin is approached. 
However, the sectional curvature 
in a surface generated by the $\partial_\tau$
direction and a 
tangent to the sphere is  $-2(U'/U)(r'/r)$, which indeed diverges as $U\rightarrow 0$ unless also $r'\rightarrow0$.
This requirement was also argued for in \cite{WY} using the Euler character.}
This requirement relates the integration constant $M$ to the horizon radius $r_h$
in each configuration via\footnote{Though perhaps not usefully, this can also be solved explicitly for the three roots, two of which are positive,
\[
r_k = \frac{2}{\sqrt{3}}\cos[\frac13\cos^{-1}(-3\sqrt{3}M)-\frac{2\pi k}{3}], \qquad k=0,1,2.
\]
}
\be
2M=r_h(1-r_h^2/\ell^2).
\label{M-rh}
\ee
Our next task is to classify the solutions
of \eqref{r-eqn}.

\begin{figure}[t!]
\centering
\includegraphics[scale=1]{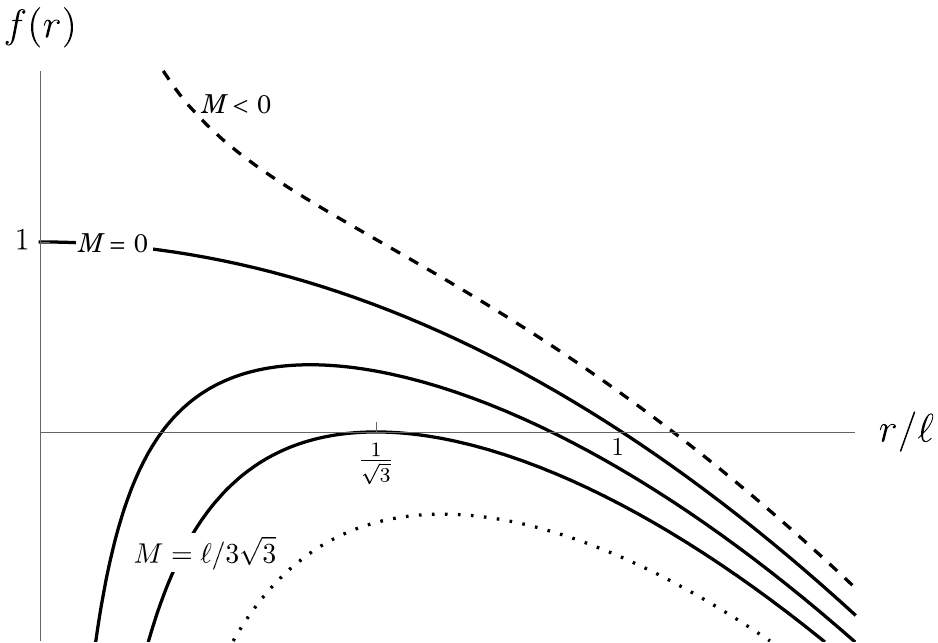}
\caption{The function $f(r)$ for different values of the mass parameter. 
Real solutions for $r'$ (and therefore for the Riemannian spatial metric) exist only where $f(r)\ge0$, and a horizon can only exist where
$f(r)=0$.
}
\label{f(r)}
\end{figure}

In order for a real solution for $r'$ --- and thus for a Riemannian spatial metric --- to exist, the function $f(r)$ in (\ref{rprime}) must be positive.
Figure \ref{f(r)} shows $f(r)$ for different values of the mass parameter $M$.
To picture the solutions $r(y)$ of  
\eqref{r-eqn}  it is useful to think of $y$ as a pseudo time coordinate, and interpret 
(\ref{rprime}) as the energy conservation equation for a particle with coordinate $r$ in one dimension, with kinetic energy $r'^{\,2}$, potential energy $-f(r)$, and total energy equal to zero. Then the roots where $f(r)=0$ correspond to turning points in the motion.
The character of the solutions depends on the value of $M$, and there are 
five qualitatively different
cases, which we now consider in turn, in decreasing order of $M$.

There is no solution with $M>\ell/3\sqrt{3}$ (cf.\ the dotted curve in figure \ref{f(r)}), as $f(r)$ is then negative for all (positive) values of $r$. At the critical value  $M=\ell/3\sqrt{3}$, the function $f(r)$ is negative except at $r=\ell/\sqrt{3}$, where $r'^{\,2}=f=0$. This is the case discussed previously, for which the radius of the 2-spheres is the same for all $y$,  and the manifold closes off at some value of $y$ with either a horizon or an RP$^2$.

\begin{figure}[t!]
\centering
\includegraphics[scale=0.7]{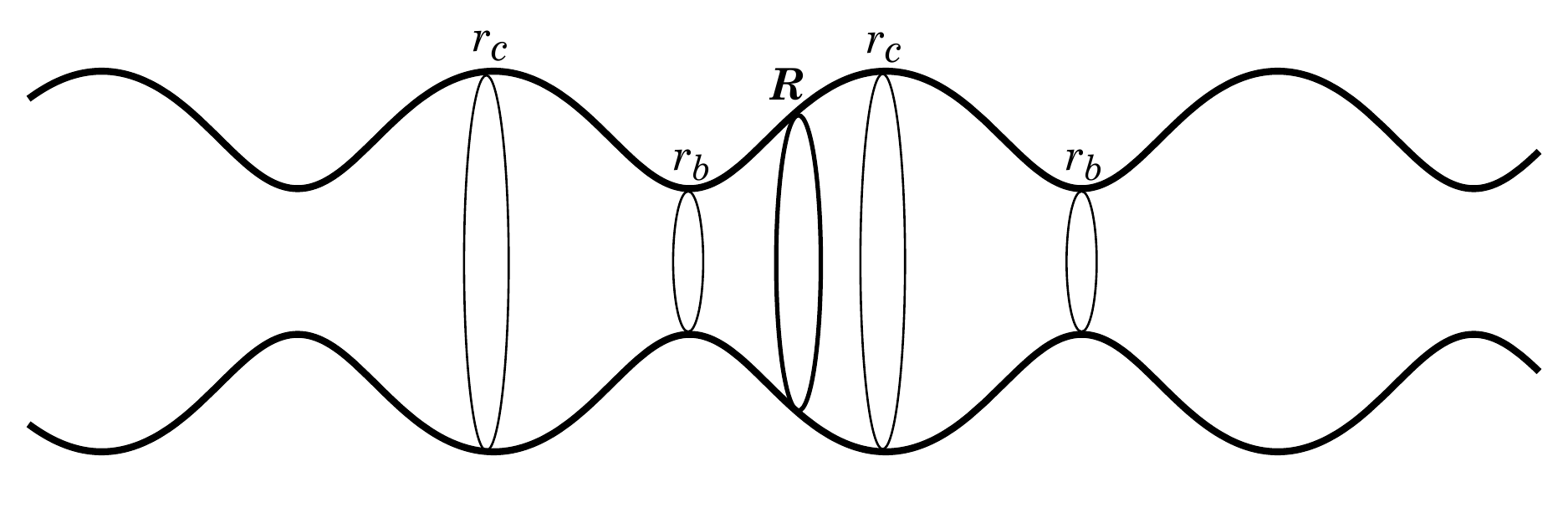}
\caption{The oscillation of radius $r$ in the constant-$\tau$ slice. The 2-spheres are shown by circles. Starting from the boundary at $R$, the radius may initially increase (decrease), reach $r_c$ ($r_b$) and continue past that. A horizon or an RP$^2$ may lie on any neck or waist with radius $r_b$ or $r_c$, respectively. Four possible cases involving the least number of oscillations are indicated.}
\label{embed}
\end{figure}

For $0<M<\ell/3\sqrt{3}$, the function $f(r)$ 
is positive between two zeros which we call $r_b$ and $r_c$,
with $r_b$ the smaller of the two, i.e.\ $r_b<r_c$.
Integrating the constraint equation \eqref{r-eqn} starting from 
the system boundary radius $R$ at $y=0$, 
the integration constant
$M$ fixes the magnitude but not the sign of $r'(0)$.
Thus, given $M$,
$r$ can initially increase or decrease, with rate
 $r'(0)=\pm \sqrt{f(R)}$. 
In either case, the radius reaches a root of $f(r)$, 
where $r'=0$. At that point 
a smooth 4-geometry can 
be obtained by closing off the time circle with $U=0$,
i.e.\ by a horizon, or by closing off the spatial 
manifold with an RP$^2$.
Alternatively, the root can be a turning point for the $y$
evolution. Any number of oscillations of $r(y)$ between the turning 
points is possible, 
with the manifold finally closed off at either an $r_b$ or $r_c$ end
by a horizon  or an RP$^2$. 
Figure \ref{embed} depicts  a Euclidean
embedding of a constant-$\tau$ slice 
(with one dimension suppressed)
for an oscillating $r(y)$ solution to the
constraints. 
That 3-geometry coincides with a static slice in the maximally extended Lorentzian SdS spacetime.

For $M=0$, if $r'(0)<0$, then the spatial manifold
closes off at an origin at $r=0$. If instead $r'(0)>0$, then there are again three possibilities: either the manifold closes off at a horizon with de Sitter radius $r_c = \ell$, it closes off with an RP$^2$, or there is
a turning point there after which the manifold contracts to $r=0$, 
giving the spatial manifold a $B^3$ topology.
Finally, for $M<0$, (cf.\ the dashed curve in figure \ref{f(r)}), 
$r'$ diverges at $r=0$, so 
a nonsingular manifold is obtained
only if $r=0$ is avoided. 
This can happen only if $r'(0)>0$ and the manifold closes off
at a horizon or an ${\rm RP}^2$ when $r$ 
reaches the root of $f$, at $r_c>\ell$.

\subsubsection{Requirement of embeddability in a solution}
As discussed above, 
the solutions to the constraints
include cases where $r'(y)=0$ 
at one or more values of $y$,
and at each such a point 
there are three possibilities 
for a solution to the
constraints: the manifold can close off at a horizon, it can close off at
an RP$^2$, or it can continue past that value of $y$. However, we now argue that only the
horizon possibility should be retained
for the paths in the path integral. 
If the turning point solutions
were to be included, they would introduce 
an infinite degeneracy of paths with identical action, whose spatial geometry  oscillates an arbitrary number of times between two radii. 
While that certainly sounds problematic, to exclude them there should be a first principles reason. That reason, we suggest, is that --- as will be shown below --- such spatial geometries are not embeddable 
in any four dimensional Euclidean solution.
Our rationale for the requirement of embeddability is the following. 

The constraints are being imposed because 
in the phase space path integral representation of the partition function 
one should sum over paths in the physical phase space. A point in the physical phase space can be thought of as initial data for a real time solution to the equations of motion, so it goes without saying that such a solution to the constraints is embeddable in a Lorentzian solution.   
The metric components are composed of
the lapse and shift and the spatial metric $h_{ij}$. 
In passing to a Euclidean path integral,
the lapse and shift contours have been 
deformed, in effect replacing 
the lapse and shift by  $-iN^{\mu}$, 
with real $N^{\mu}$.
Motivated by the problematic aspect described above, and by the notion that the analytic continuation should somehow respect the space of solutions, 
we conjecture that  
solutions to the constraints 
that are not embeddable in  solutions to the Euclidean field equations, are 
not reached by such contour deformation.
On this basis we shall 
adopt the working hypothesis
that such solutions to the constraints should be excluded from the configurations in
the Euclidean path integral.

One way to establish the required condition for embeddability in a solution to the field equations is to invoke the Euclidean version of the Birkhoff theorem: the only four dimensional spherical solutions are a portion of Euclidean \sds (excluding a 
conical singularity at one of the horizons which is absent in the Nariai limit). None of these solutions contains a slice in which $r'=0$ at a location that is not a horizon. 
Alternatively, without invoking Birkhoff theorem, we can establish this condition directly using the Einstein equations as follows.

For a diagonal metric
the field equation, 
$G_{\mu \nu}=-\Lambda g_{\mu \nu}$, implies the relation 
$g^{\tau \tau}G_{\tau\tau}=g^{yy}G_{yy}$. 
For the spherically symmetric line element (\ref{metric}) these components of the Einstein tensor are
\be
\begin{aligned}[b]
G_{\tau \tau}&=\frac{U}{r^2}\Big ( rr'V'+V(r'^2+2rr'')-1 \Big)\, ,\\
G_{yy}&=\frac{1}{Vr^2}\Big (Vr'(r'+rU'/U)-1 \Big )\, .
\end{aligned}
\ee 
For the case where $r(y)$ is not a constant function, the above relation yields 
\be\label{Usol0}
\frac{U'}{U}=\frac{V'}{V}+2\frac{r''}{r'} \quad \implies \quad U(\tau,y)=u(\tau)V(\tau, y) r'^{\,2}(y) \, \, , 
\ee 
where $u(\tau)$ is an arbitrary function.
It therefore follows 
that, 
if $r(y)$ is not constant,
$U$ must vanish at a location where
$r'=0$. Solutions to the constraints that close with an RP$^2$ or continue past such a location
are therefore not embeddable in a smooth 
four dimensional solution, since to be smooth they would
need to have $U\ne0$ there.
In the case where $r(y)$ is constant, as seen above 
the Hamiltonian constraint requires $r=\ell/\sqrt{3}$, and the $\tau \tau$, $\tau y$, and $yy$ components of the field equation are satisfied without specifying $U$ and $V$. In this case, to determine the solution one needs to impose also the angular part of the Einstein equation. 
This yields uniquely the Euclidean Nariai geometry, $S^2\times S^2$, where both 2-spheres have the same, constant curvature.\footnote{In an orthonormal basis 
the only nonzero components of the Riemann tensor 
for metric that is the product of two, 2-dimensional Riemannian manifolds
are $R_{1212}$ and $R_{3434}$ (and index permutations of 12 or 34), 
where 1 and 2 span the $\tau$-$y$ subspace and 3 and 4 span the transverse spherical directions. 
The Ricci tensor is thus diagonal, with components
$R_{11}=R_{22}=R_{1212}$ and $R_{33}=R_{44}=R_{3434}$.
The Einstein equation implies $R_{\mu\nu} = \Lambda g_{\mu\nu}$, so we have $R_{11}=R_{22}=R_{33}=R_{44}=\Lambda$, and hence
$R_{1212}=R_{3434}=\Lambda$.}
Note that this uniqueness rules out an RP${}^2$ terminus in this case as well.

\subsubsection{Action and contributing paths}

Having identified all the spherical paths that contribute to the constrained path integral for the partition function,
we next write the action for these paths.
The action (\ref{Ih}) subject to the constraints takes the more explicit form
\be
I[M;\beta,R]=\pm \beta R \left [ 1-\frac{2M}{R}-\frac{R^2}{\ell^2} \right ] ^{1/2}\! -\pi r_h^2\, ,
\label{Ibc}
\ee
where the $\pm$ sign corresponds to the sign
of $r'(0)$, and in the case with no horizon 
$r_h$ and $M$ should be set to zero. 
The mass parameter $M$, or equivalently the horizon radius $r_h$ \eqref{M-rh}, is not determined 
by the canonical boundary parameters $R$ and $\beta$ until a stationary point is selected. Thus the horizon radius $r_h$ is the single degree of freedom that appears in the reduced action.
We next assess which configurations contribute
to the path integral for different cases of system ``size".

For boundary radius $R>\ell$, a real metric exists only if $M<0$. Excluding a singularity at
$r=0$, the system can only consist of the region
with $r>R$, and must be closed off by a cosmological horizon at $r_c > \ell$. The action for such configurations is unbounded below, since 
for large $|M|$ we have $r_h\sim (-M)^{1/3}$.
It is questionable whether negative $M$ should be included, and we shall discuss that question 
below.

For boundary radius $R=\ell$,  
$M$ can be negative as in the previous case, 
or it can be zero, corresponding to 
the value for de Sitter space. Since the system boundary in this case sits right on top of the dS horizon, there is no c-configuration, and thus the only possibility is that the configuration corresponds to a spatial slice of the static patch of de Sitter space.

For boundary radius $R <\ell$, both the $M<0$ and $M=0$ cases discussed previously are possible. In addition, since the boundary is
now smaller than the de Sitter horizon, there is also the possibility for the system to be the region between $R$ and the de Sitter horizon.
Moreover, in this case $M>0$ is
possible, bounded above by $2M=R - R^3/\ell^2$. Excluding a singularity at $r=0$,
there are two possible configurations for positive $M$: the system can be the region between $R$ and the b-horizon, $r_b<r<R$, or it can be the
region between $R$ and the c-horizon, $R<r<r_c<\ell$. 
Note that in the limit $M\rightarrow 0$ of the
b-horizon case the action becomes identical to that of the empty de Sitter case, although there
is a discontinuous topological transition between the two cases.

\subsubsection{Stationary points and solutions}
\label{sp-sol}

The reduced action as a function of $M$ is stationary at 
paths corresponding to 
spherical solutions to the full set of Einstein equations that meet the boundary
conditions and are smooth at a horizon if one exists. 
These solutions will be candidate equilibria, 
which are studied as a function of temperature in the following subsection.
According to the Birkhoff theorem, 
they correspond to portions of the
Euclidean Schwarzschild-de Sitter.
We emphasize however that the constraint equations alone 
do not specify the function $U$ in (\ref{sp2}) 
(which is the square of the lapse function for the $\tau$ foliation),
and the constrained action (\ref{constrained}) is independent of $U$ 
except for its boundary value that is fixed by the temperature.
For any stationary path of the constrained action, $U$ can be found which,
together with that path, makes the metric (\ref{sp2}) a 4-dimensional solution to the 
Einstein equation in the chosen gauge,
but that 4-dimensional interpretation is not inherent 
in the constrained phase space 
description of the thermodynamic ensemble.\footnote{As expressed in \cite{WY}: ``The ordinary thermal equilibrium is thought of as occurring {\it in} spacetime. In the present case, the concept of spacetime is merely heuristic."\label{heuristic}}

The stationary points can also be identified by just
imposing sufficient components of the Einstein equation, beyond the constraints.  
In fact, 
the result \eqref{Usol0}, derived from the 
$\tau\tau$ and $yy$ components of the Einstein equation, can be used 
to determine completely the form of $U$ in a solution to the Einstein equation.
We found when imposing the constraints \eqref{Gtt} and \eqref{Gty}
that $V$ is independent of $\tau$
(apart from the exceptional case $r'(y)=0$),
which allows the gauge choice $V=1$
and yields the relation 
\eqref{rprime}. With this gauge choice,
\eqref{Usol0} becomes
$U=u(\tau)r'^{\,2}$.
Moreover, $u(\tau)$ can be set to a constant by reparametrization of the
$\tau$ coordinate, and
the canonical boundary data imposes the relation  $\sqrt{U(0)}=\beta/2\pi$, 
so we find that 
\begin{equation}
U(y(r))=(\beta/2\pi)^2\, f(r)/f(R),
\label{Usol}
\end{equation}
where $f(r)$ is the function defined in \eqref{rprime},  $f(r)=1 -2M/r -r^2/\ell^2$. 
If $M\ne0$ then there is either a b- or c-horizon in the system configuration, and
the condition of smoothness at the horizon fixes the value of $M$ in terms of $R$ and $\beta$. Alternatively, this relation is determined by requiring that the action \eqref{Ibc} be stationary with respect to variation of $M$.\footnote{For 
configurations with other values of $M$,  the action is {\it not} stationary with respect to variation of $M$. One could still locally construct the corresponding solution to the Einstein equation using the lapse function \eqref{Usol}, but it would have a conical singularity at the horizon where the lapse vanishes. 
As we are working within a reduced phase space formulation of the theory, however, such constructed spacetime solutions play no role (cf.\ also footnote \ref{heuristic}).} 
The $M=0$ case corresponds locally
to de Sitter space. It  admits  a 
c-configuration, for which the system 
boundary is surrounded by a horizon (which
is smooth for the correct relation between $R$ and $\beta$). It also admits a
$B^3$ configuration, for which the system 
boundary encloses a ball-region that is locally 
de Sitter space, as explained below. 
The $B^3$ configuration is an isolated point rather than a stationary point of the reduced action,
yet it is indeed a solution to the Einstein equation.

To help visualize the character of the stationary points
and solutions we now provide
depictions of their spatial slices, 
as well as Penrose diagrams for the corresponding
Lorentzian versions of these solutions,
classified by topology.

\begin{figure}[t!]
\centering
\includegraphics[scale=0.35]{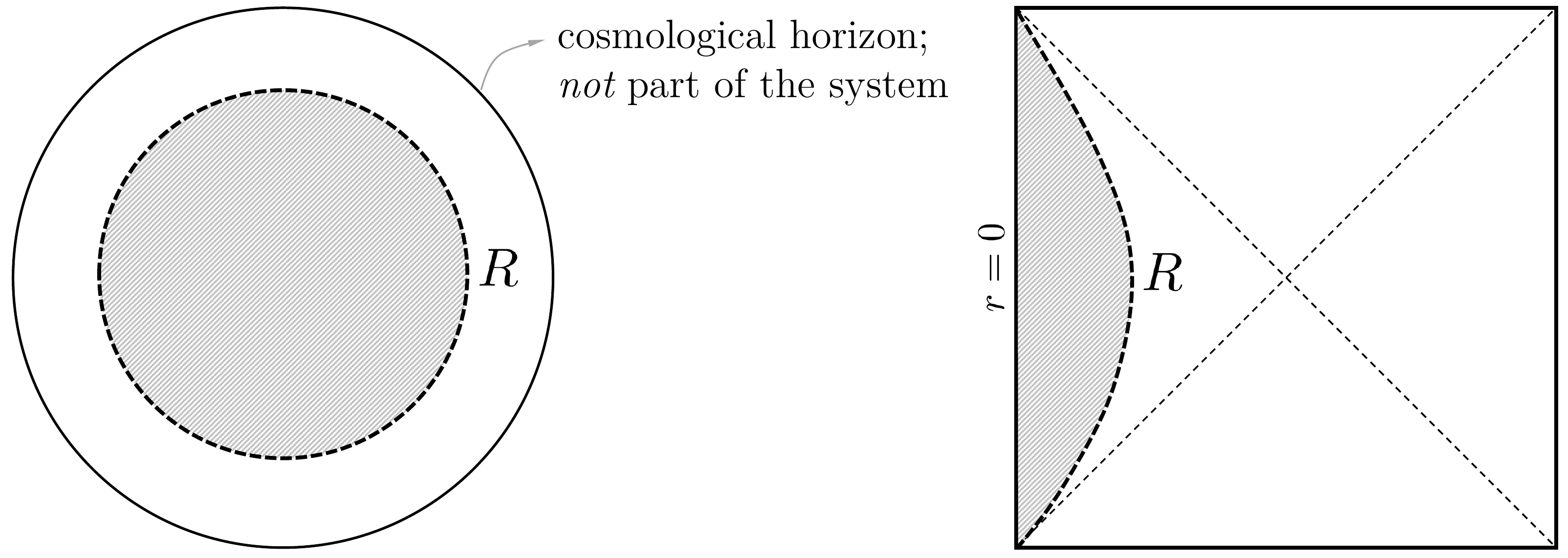}
\caption{The stationary point ``thermal de Sitter" as part of the static patch of de Sitter space, shaded on a space diagram (left) and on the Penrose diagram for the corresponding Lorentzian solution (right). The cosmological horizon lies outside the system.}
\label{thermal-dS}
\end{figure}

The configuration with topology $B^3\times S^1$ 
has $M=0$ and
exists for $R\le \ell$ and any $\beta$.
The 4-geometry with this spatial slice
that satisfies the full set of Einstein equations
is --- locally in $\tau$ --- a portion of de Sitter space,
but embeds globally into Euclidean de Sitter,
i.e.\ the round 4-sphere of radius $\ell$,  only if
$\beta^{-1}$ matches the redshifted Gibbons-Hawking 
temperature at the boundary radius $R$. Otherwise, 
it is ``cold de Sitter space'' or ``hot de Sitter space''. 
As such, we refer to the general case as 
``thermal de Sitter space".
The Lorentzian counterpart of this geometry 
is part of the static patch of de Sitter space.
The spatial slice and the Lorentzian counterpart
are illustrated in figure \ref{thermal-dS}.

For the topology $S^2 \times D^2$,
the time circle closes off at
the center of the disc, where the $S^2$ factor has a finite area and corresponds to the horizon bifurcation surface in 
the associated Lorentzian geometry. 
Depending on whether the 
horizon radius $r_h$ is smaller or larger than the boundary 
radius $R$, we call the central $S^2$ 
a ``black hole horizon" with radius $r_b$, or a ``cosmological horizon" 
with radius $r_c$, respectively. 

\begin{figure}[t!]
\centering
\includegraphics[scale=0.3]{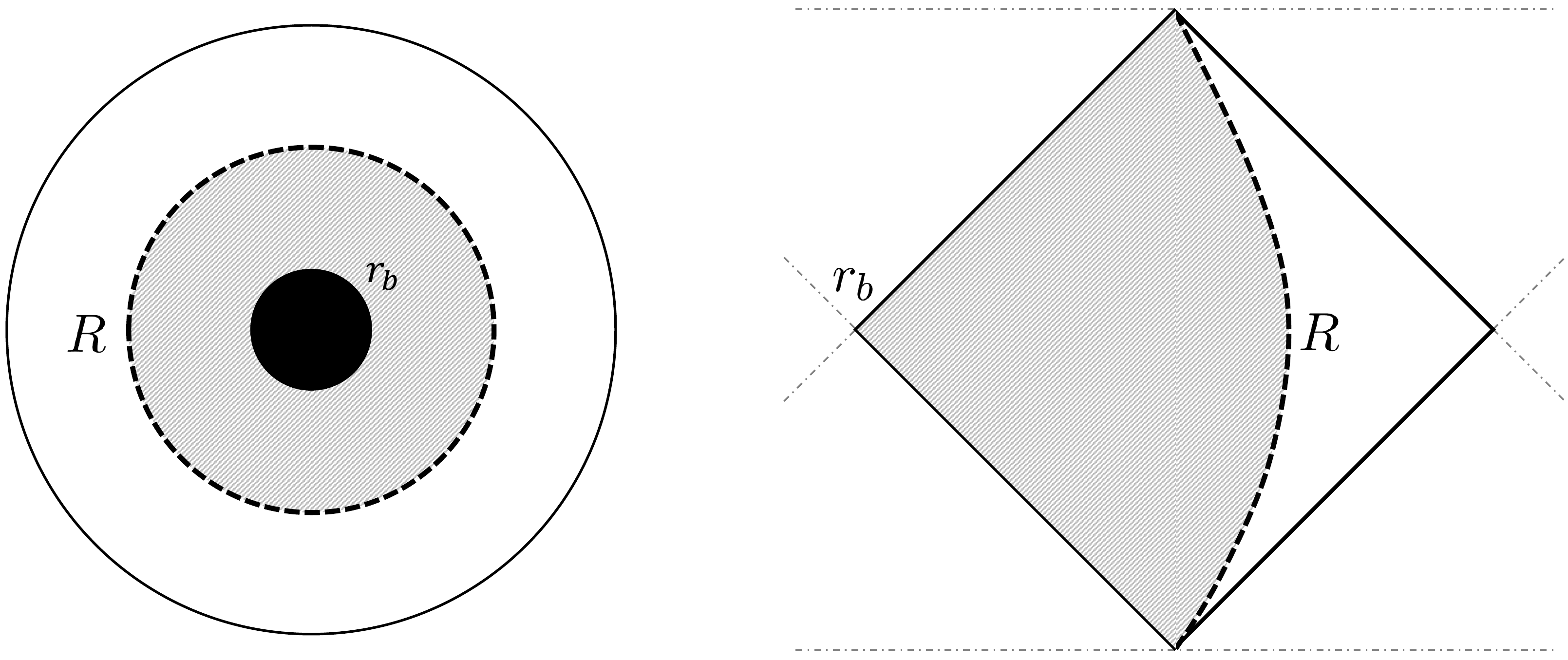}
\caption{The black hole configuration. The system is the shaded region that extends from the black hole horizon $r_b$ to the system boundary $R\, \,  (\neq \ell/\sqrt{3})$,
on a space diagram (left) and a Penrose diagram for the corresponding Lorentzian solution (right).
The cosmological horizon lies outside the system.}
\label{b-system}
\end{figure}

The geometry of a
black hole configuration is shown in figure \ref{b-system}. 
It corresponds
to a bounded region in the static patch of the \sds space that 
includes the black hole horizon but not the cosmological horizon. 
If the spatial slice is analytically 
extended ``outward'' beyond the system boundary to the
cosmological horizon, a conical singularity would be present in the Euclidean solution, as the two horizons have different temperatures. 
However, it makes no sense physically to analytically extend the solution beyond the system boundary. Either that region of space is not in our
model system or, if the system is realized physically in ambient 
spacetime, beyond the boundary lies the reservoir that imposes the canonical boundary conditions.
The $r_b\rightarrow 0$ limit of the black hole configuration 
is a topological transition to
the thermal de Sitter space. 

\begin{figure}[t!]
    \centering
    \begin{subfigure}{0.49\textwidth}
        \centering
        \includegraphics[scale=0.45]{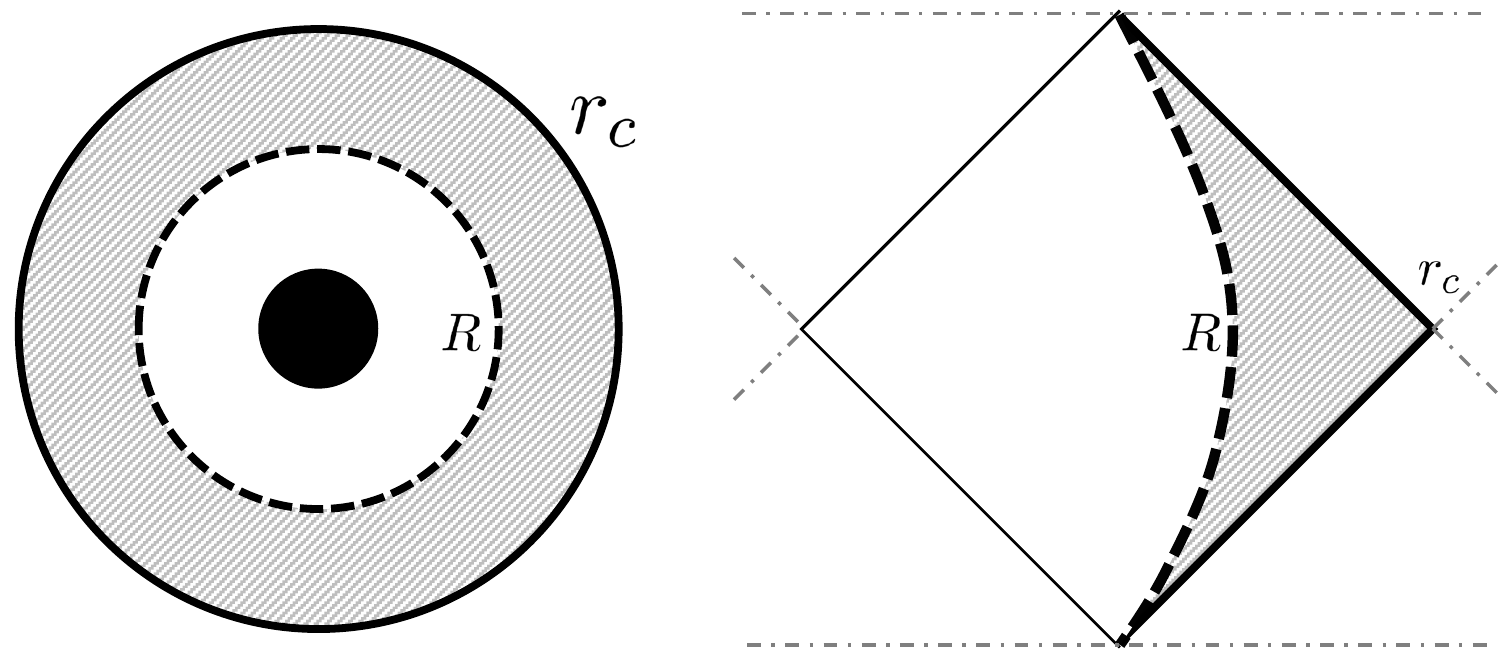}
        \caption{}
        \label{c-system1}
    \end{subfigure}%
  ~   
    \begin{subfigure}{0.49\textwidth}
        \centering
        \includegraphics[scale=0.45]{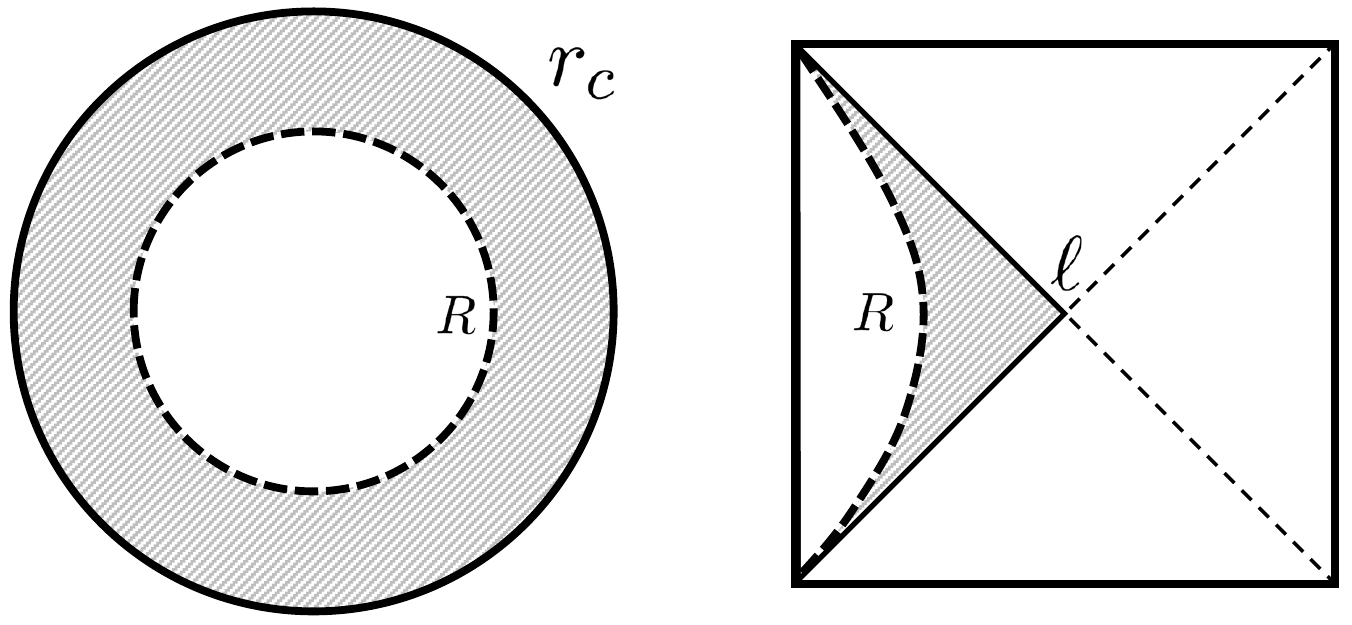}
        \caption{}
        \label{c-system2}
    \end{subfigure}
    \caption{The cosmological configurations, when the mass parameter is positive (a) or zero (b). The system is the shaded region, in either the space diagram (left) or the corresponding Lorentzian Penrose diagram (right). For a given temperature, the mass parameter of the stationary point in (a) is different from the one in figure \ref{b-system}, so a given SdS space is not both a b-horizon and a c-horizon stationary point. }
    \label{c-system}
\end{figure}

The cosmological horizon solution is the region with greater spherical radius than the system boundary ($r>R$), and includes the cosmological horizon. 
The cosmological configurations are well-defined regardless of what configuration occupies the region $r<R$,
which is not part of the system configuration
(but we will discuss later the question of its physical significance).
There are three possible behaviors if the spatial slice is analytically extended ``inward'' beyond the system boundary: if $M>0$  there is a black hole inside, if $M=0$ it is empty de Sitter space,
and if $M<0$ there is a naked singularity.
The embedding of such configurations in SdS, for $M>0$ and $M=0$, are shown in figure \ref{c-system}.

There is a maximum for the mass parameter in the \sds geometry, $M=\ell/3\sqrt{3}$, at which the black hole and the cosmological horizon have the same area, with $r_b\!=\!r_c\!=\! \ell/\sqrt{3}$. This is the Nariai limit \cite{Nariai}, with \euc topology $S^2 \!\times \! S^2$, and in suitable coordinates the \euc line element can be written as \cite{Ginsparg-Perry}
\be
ds^2=\frac{\ell^2}{3}\left( \sin^2 \chi d\tau^2+ d\chi^2 + d\Omega^2
\right)\, .
\label{Nariai}
\ee
All of the transverse 2-spheres have the same radius $\ell/\sqrt{3}$, and the proper circumference of the \euc time circle on the $\tau$-$\chi$ sphere is $(2\pi \ell/\sqrt{3}) \sin \chi$, which varies between zero at the poles  and the maximum value on the equator. For a system boundary located at $\chi=\chi_0$, this circumference is identified with the local inverse temperature,  $\beta= (2\pi \ell/\sqrt{3}) \sin \chi_0$. The boundary radius $R=\ell/\sqrt{3}$ and temperature (which ranges from
$\sqrt{3}/2\pi \ell$ to $\infty$)
define a canonical ensemble that includes configurations corresponding to both sides of the boundary. These configurations have horizons (at $\chi=0$ or $\chi=\pi$) with the same area but different proper distances from the boundary.\footnote{\label{failed}It is interesting to note that a particular proper distance from the boundary is selected if one starts with an SdS configuration with boundary radius equal to the Nariai value, $R=\ell/\sqrt{3}$, and takes the limit as $M$ approaches the maximal value $M=\ell/3\sqrt{3}$. One then arrives at the particular
Nariai case in which the boundary is located at $\chi_0=\pi/2$, i.e., at the equator of the $\tau$-$\chi$ sphere. See appendix \ref{NariaiApp} for an explicit calculation showing this. (We tried and failed to find a way to anticipate this simple result.)
The corresponding temperature is the minimal possible one for the
system boundary in the Nariai case, $\sqrt{3}/2\pi\ell$.}

In this section we have identified the spherically symmetric configurations that satisfy the constraints, obtained the ``reduced action" for such paths in the path integral for the partition function, and identified and characterized the stationary points of the action.
For given boundary radius $R$ and inverse temperature $\beta$ some of these configurations 
are admitted, and depending on those values stationary points of the action --- i.e.\ solutions to the field equations --- may be admitted. 
In the following we discuss which configurations dominate the path integral, and investigate the corresponding metastable or stable equilibrium phases.

\subsection{Thermodynamic phases}
\label{stab}

The action $I$ in the integrand of 
the partition function integral (\ref{Zc}) 
is a function of the boundary data $(R,\beta)$ 
and the cosmological length scale $\ell$, as well as 
possibly other variables determining the path. 
In the spherical reduction (\ref{Ibc}), 
the only other continuous variable
is the mass parameter $M$, or equivalently the horizon radius $r_h$ (related to $M$ by \eqref{M-rh}),
which at a stationary point is
determined by the previous parameters. 
Planck's constant enters only in the denominator of the dimensionless
exponent $I/\hbar$, which is thus $l_{\text{P}}^{-2}$ times
a function of $R, \beta, \ell, r_h$
with dimension (length)$^2$ that
takes the explicit form
\be\label{Igen}
\frac{I}{\hbar}=\frac{1}{l_{\text{P}}^{2}} \left\{\mp\beta R \,\left [ 1-\frac{r_h}{R}\left (1-\frac{r_h^{2}}{\ell^2}\right )-\frac{R^2}{\ell^2}\right ]^{1/2}- \pi r_h^{2}\right \}.
\ee
The sign in front of the first term is that of the Brown-York energy, which is negative for thermal de Sitter ($r_h=0$) or 
the black hole ($r_h=r_b$), 
and positive for the cosmological horizon ($r_h=r_c$).
For on-shell actions, i.e.\ at stationary points,
$r_h$ is a function of the other parameters, including $\beta$.
Note that the cosmological length $\ell$ must appear with negative powers, as it should drop out 
in the $\ell\to\infty$  limit.

We restrict attention to the ``semiclassical'' setting, where  $I/\hbar$ is very 
large, which should generically be the case
if the lengths $R$, $\beta$, $\ell$ and $r_h$ are all 
very large compared to the Planck length,
and proceed with the usual assumption that 
the partition function can then 
be approximated at leading order by the contribution of the geometry that minimizes the \euc action (\ref{constrained}), or rather its spherical reduction (\ref{Ih}) or \eqref{Igen}. 
That is, $Z\sim e^{-I^*/\hbar}$, where $I^*$ is the action of the corresponding geometry. This can be called the ``zero-loop" approximation \cite{WY}.
A metric that locally minimizes the action (compared to neighboring metrics) might be a stationary point, i.e., a solution to the field equations, or it might sit at an ``endpoint" of the configuration space.
These minima correspond to candidate equilibria. 
Endpoint minima of the free energy of an ensemble
can occur, 
for instance,  when the configuration space
is subject to a constraint. In that case, the constraint
would presumably be physically realized by some external potential. If that potential were included in the action
governing the dynamics, the ``endpoint" would actually
become a stationary point.
In our case, in contrast, the endpoint
configurations correspond to configurations in which the system under consideration cannot be physically
realized. In particular, as discussed below in section \ref{c-instability},
they correspond to cases in which the ``reservoir" mass 
becomes negative, or the reservoir becomes a black hole that swallows the system boundary, or the cosmological horizon coincides with the system boundary, squeezing the system to zero volume.
When such an endpoint dominates the ensemble, that
indicates an instability whose physical nature can be understood
only when including the nonspherical, radiative modes in the dynamics.
We shall consider such endpoints as candidate
unstable ``phases".
The accuracy of the ``zero-loop" approximation at an endpoint configuration is discussed in appendix \ref{endpoints} where we find that, provided the ensemble parameters are much larger than the Planck length and the system volume remains finite, the leading order approximation is valid at any candidate dominant configuration.

The global minimum characterizes the stable phase, and local minima that are not global minima correspond to metastable phases.
In the system under consideration, however, the relevant meaning of ``global" is not entirely clear, since thermal dS, b-horizon, and c-horizon configurations are not continuously connected. As an initial data set (phase space point) with our $S^2$ boundary condition, thermal dS has topology $B^3$ while the horizon configurations both have topology $S^2\times [0,1]$ 
since they have a boundary also at the horizon 
(on one time slice, as opposed to in the Euclidean 4-geometry). 
And although the b- and c-horizon configurations
have the same {\it manifold} topology,
within the spherical reduction of the phase space 
with boundary radius $R$ they are topologically disconnected in configuration space, except in the special case when $R=\ell/\sqrt{3}$.

For example, if $R<\ell/\sqrt{3}$, then the largest possible b-horizon radius is $r_b=R$, whereas the smallest possible c-horizon radius is $\widehat R>R$ ($\widehat R$ is defined in eq.~\eqref{Rhat}).
There is no continuous path through the (one-dimensional) space of spherical initial data, labeled by the horizon radius or mass parameter, that connects the b- and c-horizon configurations.
We take this to suggest that no transition is possible, even allowing for quantum tunneling. (We have not determined whether the configurations could be connected by a path including nonspherical data, but we suspect they cannot be.)
On the other hand, if --- for some reason --- one should integrate over
paths through the space of complex 3-metrics, the b- and c-horizon configurations can be connected.
For example, for a given $R<\ell/\sqrt{3}$, starting from zero mass parameter and increasing $M$, the black hole horizon grows until its radius meets $R$.
Increasing $M$ further produces complex geometries, since $r'^2=f(r)$ (\ref{rprime}) is negative for some values of $r$. 
If $M$ is increased to the Nariai point, where two roots of $f(r)$ coincide
(cf.\ figure \ref{f(r)}), it can then be decreased
while the range of $r$ is extended to the larger root, 
corresponding to the cosmological horizon. Eventually $M$ decreases
sufficiently for the c-horizon configuration to become Riemannian. This
defines a path through complexified configuration space joining the b- and c-configurations.
A similar story holds for $R>\ell/\sqrt{3}$.
In the special case when the
boundary radius is precisely $R=\ell/\sqrt{3}$,
however, b-c transitions are possible via the intermediate constant radius configuration described in footnote \ref{failed}.

At issue
is whether transitions 
between topologically disconnected configurations
occur in quantum gravity. If not, then 
the disconnected sectors
are distinct thermodynamic systems that should be specified in the definition of the ensemble, and whose phase structures should be separately analyzed.
The canonical quantization of a phase space with disconnected components would presumably produce a direct sum of quantum theories, each a superselection sector, and no transitions connecting them would take place.
Since, however, the spatial manifolds of the horizon systems are truncated
at the horizon, we may not be dealing with a complete phase space. 
Also, it is not known to what extent quantum gravity is correctly captured by canonically quantized general relativity. 
In fact it is traditionally presumed that topology change
occurs in quantum gravity, but that question remains open. 

It seems plausible that thermal de Sitter and the black hole configurations are dynamically connected despite their different classical topology, because as the black hole mass parameter tends to zero, the metric and extrinsic curvature of a spatial slice approaches that of de Sitter outside a sphere of ever shrinking size. The curvature blows up at a point in the limit, but the region where the two solutions differ has then
shrunk to zero size, so we are inclined to suppose that such a transition is indeed allowed in the full quantum theory. The situation is different,
however, for b-c transitions, since
b- and c-horizon configurations with the same boundary radius $R$ are macroscopically separated (at least in the spherical reduction). 
In view of our uncertainty regarding the existence of b-c transitions in quantum gravity, we shall consider in the following both possibilities.

\begin{figure}[t!]
    \centering
    \includegraphics[scale=0.63]{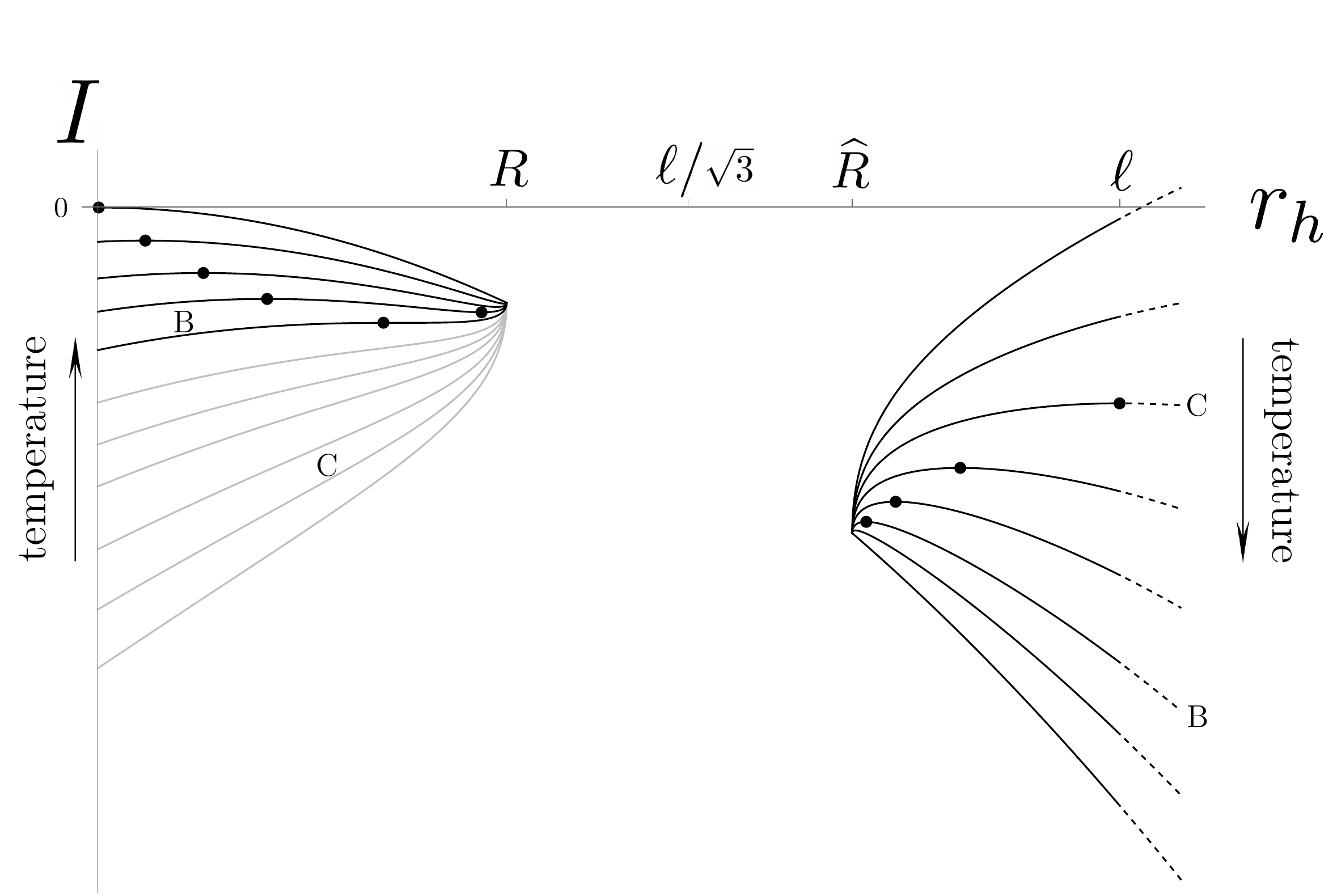}
    \caption{
    The canonical action as a function of horizon sizes,  for b-horizons (the curves on left) and c-horizons (the curves on right), for a case with $R<\ell/\sqrt{3}$ (numerically, $R=0.4\,\ell$). 
    The action of the endpoints at $R$ and $\widehat R$ is minus the horizon entropy, $-\pi R^2$ and $-\pi \widehat R^2$, respectively.
    The gap between the b-curves and the c-curves decreases as $R$ tends to $\ell/\sqrt{3}$.
 When $R=\ell/\sqrt{3}$, the two branches join at a point with $I=-\pi \ell^2/3$.
    For b-horizon configurations, at temperatures above that of curve B, 
    there are two stationary points, indicated with solid dots, corresponding to ``small'' and ``large'' black holes. At lower temperatures there are no stationary points. For c-horizon configurations there is one stationary point at any temperature.  For temperatures below that of curve C, the stationary point lies on the dashed part of the graph ($r_c\!>\!\ell$), corresponding to a negative mass parameter in the metric.}
    \label{actionsbc}
\end{figure}

Our next task is thus to determine, for each temperature, how the actions of the different paths compare.
Figure \ref{actionsbc} shows plots of the action as a function of $r_h$,
for a given system boundary size $R$ and
at different temperatures,
with b-horizon configurations on the left and 
c-horizon configurations on the right.
For the b-horizon configurations the plot has the same features as the case with zero cosmological constant, discussed by Whiting and York \cite{WY}.
The action of thermal de Sitter space 
corresponds to the
$r_b\rightarrow 0$ limit of $I_b$.\footnote{Thermal dS is topologically distinct from the black hole configurations, so the reduced action
near $r_b=0$ in figure \ref{actionsbc} does not depict variations of
thermal dS.}

The plots look qualitatively the same at different values of $R$, except for the ranges of $r_b$ and $r_c$, which are determined by requiring that the metric be Riemannian, which corresponds to the condition that the argument of the square root in \eqref{Igen} be positive.
In particular, for boundary radius $R<\ell/\sqrt{3}$ (which is the case in figure \ref{actionsbc}), the horizon ranges are $0<r_b<R$ and $r_c> \widehat{R}$, where 
\begin{equation}\label{Rhat}
    \widehat{R} :=\sqrt{\ell^2-3R^2/4}-R/2
\end{equation} 
is the larger root of $f(r)$ \eqref{rprime} when the smaller root is $R$. This lower bound on $r_c$ corresponds to an upper bound on $M$, and can be seen as the condition that the boundary of the c-region not be swallowed by the black hole horizon that would lie outside the system if the solution to the constraints were extended as SdS beyond the system boundary.
For boundary radius  $R>\ell/\sqrt{3}$, the horizon ranges are $0<r_b< \widehat{R}$ and $r_c>R$, where now  $\widehat{R}$ is the smaller root of $f(r)$ when the larger one is $R$. This upper bound on $r_b$ corresponds again to an upper bound on $M$, and in this case can be seen as the condition that the boundary of the b-region not be swallowed by the would-be cosmological horizon.

As $R$ approaches $\ell/\sqrt{3}$, 
the right endpoint of the b-horizon curves
in figure \ref{actionsbc} approaches the left endpoint 
of the c-horizon curves, and both the black hole and cosmological horizons
approach the system 
boundary. At precisely $R=\ell/\sqrt{3}$, there is a qualitative
change in the nature of the endpoint configurations in two ways. 
First, the distance from the boundary to both horizons takes the 
nonzero value $\pi \ell/2\sqrt{3}$ (see appendix \ref{NariaiApp}).
Second, in addition to the SdS b-horizon and c-horizon configurations
there is another configuration, the Nariai
solution (\ref{Nariai}), that also meets the boundary conditions
for a certain range of temperatures.
This configuration
has the same action as the b- and c-horizon configurations. 
Its horizon lies at
a temperature-dependent proper distance  from the system boundary,
$L=(\ell/\sqrt{3})\sin^{-1}(\sqrt{3}\beta/2\pi\ell)$
($L$ is a two-valued function of $\beta$). 
At precisely the minimal temperature, $\beta=2\pi \ell/\sqrt{3}$,
the distance becomes $L=\pi \ell/2\sqrt{3}$, which agrees with the distance
for the b- and c-horizon configurations. At this temperature, 
the slopes of the b- and c- action graphs coincide where the endpoints meet
at the Nariai solution. The action has a local maximum at this configuration,
hence the system in this configuration is unstable; see the end of section \ref{phasdiag} for further discussion.

Configurations with negative mass parameters $M<0$ correspond to nonsingular Riemannian metrics within the system domain only in the case of c-horizon configurations. 
Their horizon radii are larger than the de Sitter radius $\ell$,
and their action is indicated by the dashed segments of figure \ref{actionsbc}
(see also the dashed curve in figure \ref{f(r)}).
Moreover, if  $R>\ell$
these are the {\it only} configurations.
These configurations are everywhere regular in the system domain, 
between the boundary and the horizon. Their action is unbounded below as 
$r_c\to\infty$, so if physically admissible they would always
dominate the partition function, indicating an instability.
In the following we shall exclude them from consideration, on the grounds that 
a physical realization of the reservoir that 
selects the canonical ensemble presumably could not source a configuration with negative $M$.

\begin{figure}[t!]
    \centering
    \includegraphics[scale=0.4]{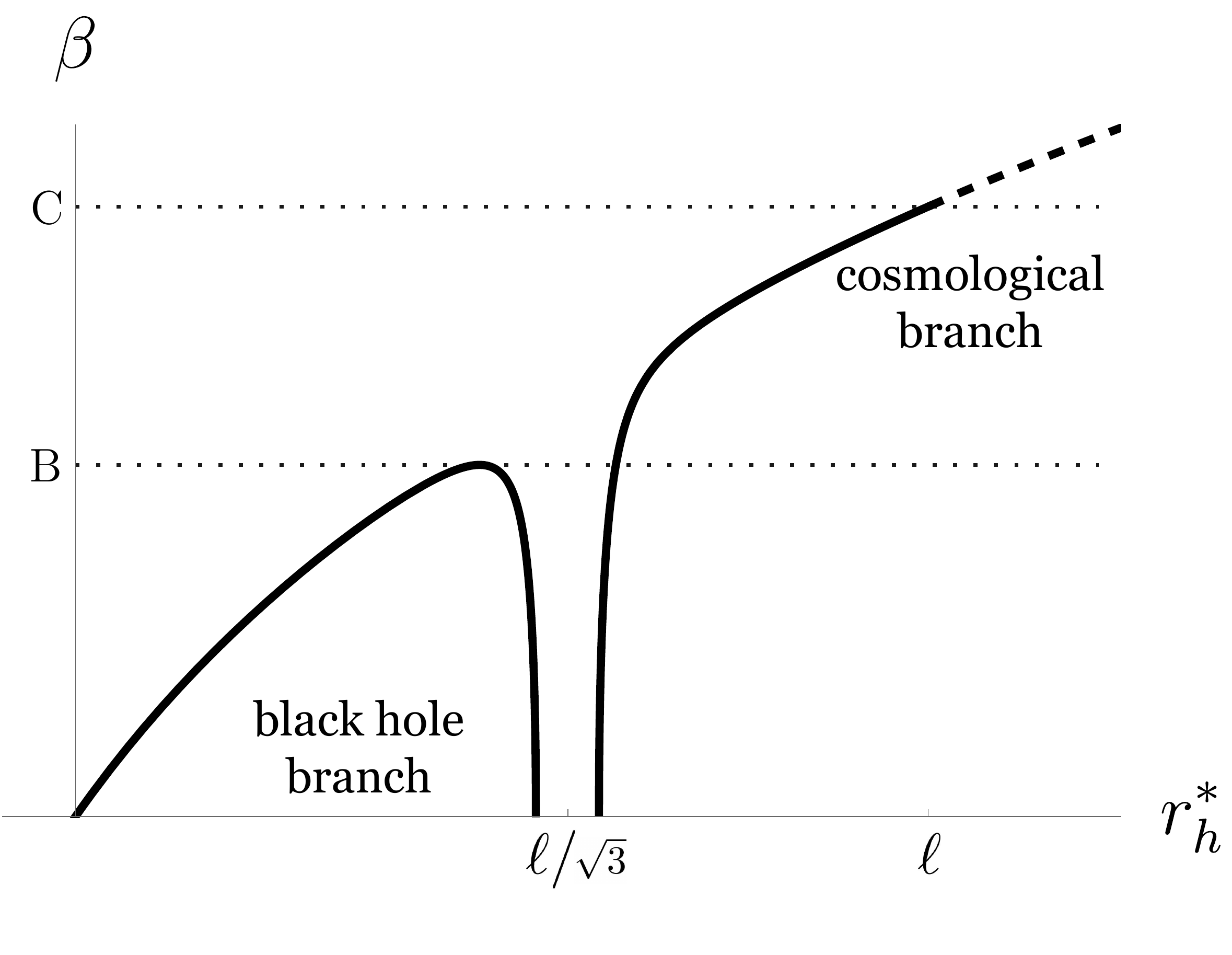}
    \caption{The inverse temperature 
    $\beta$ as a function of the horizon radius for the black hole ($r_b^*$) and cosmological ($r_c^*$) solutions, as given by (\ref{temp}), for any boundary sizes $0\!<R\!<\ell$
    (here $R=0.54\ell$)
    except the special value $R\!=\!\ell/\sqrt{3}$ (the $\beta$ and $r_h^*$ axis scales
    differ).
    When $R\!=\!\ell/\sqrt{3}$ the two branches merge in the middle on a vertical line. For
    $\beta \!<\! \beta_{\text{B}}$ there are three stationary points. The dashed part of the cosmological branch ($r_c^*\!>\!\ell
    \, , \,\beta\!>\! \beta_{\text{C}} \, =2\pi \ell \sqrt{1-R^2/\ell^2}$) corresponds to negative mass parameters, ascending monotonically as $M^*$ becomes more negative. 
    }
\label{beta-rh}
\end{figure}

Since the horizon size is the only remaining variable in the reduced action, the stationary points are readily obtained from the condition $\partial I / \partial r_h\!=\!0$. 
This gives the relation between the horizon radius $r^*_h$ that extremizes the action and the boundary data $R$ and $\beta$,  
\be
\beta=\frac{4\pi r^*_h}{|1-3r_h^{*2}/\ell^2|}\left [ 1-\frac{r^*_h}{R}\left (1-\frac{r_h^{*2}}{\ell^2}\right )-\frac{R^2}{\ell^2}\right ]^{1/2} .
\label{temp}
\ee
 To characterize the stationary points it is helpful to plot  $\beta$ (\ref{temp}) versus the horizon radii $r_h^*$ for the cosmological horizon and black hole branches, as shown in figure \ref{beta-rh} 
for a value of the system boundary radius $R\ne \ell/\sqrt{3}$. 
The critical
temperatures B and C specified in figure \ref{beta-rh} are defined in the same way as those indicated in figure \ref{actionsbc}. The dashed segment on the cosmological branch corresponds to negative mass parameter.
As $R$ approaches the value $\ell/\sqrt{3}$ from both below and above, the two branches in figure \ref{beta-rh} approach each other until they meet at $r_h^*=\ell/\sqrt{3}$ when $R=\ell/\sqrt{3}$.

Next we determine the equilibria and dominant configurations as a function of temperature, starting with low temperature. 
As can be seen from figure \ref{beta-rh},
for temperatures below $T_{\rm C}
=(2\pi \ell \sqrt{1-R^2/\ell^2})^{-1}$,
which is the blueshifted dS temperature at the boundary, 
the only solution with a horizon is the cosmological one with $r_c>\ell$, corresponding to a negative mass parameter. As can be seen from figure \ref{actionsbc}, 
all of the cosmological horizon solutions are local maxima of the action, although the plot there does not extend far enough to display the maxima for $\beta>\beta_{\rm C}$. 
Beyond that maximum the action decreases without bound
if negative mass configurations are admitted.
As previously mentioned, such configurations would entail an instability, and we shall characterize the phases excluding them. 
Apart from negative mass configurations,
the configuration with the lowest action
at such low temperatures is thermal dS, whose action corresponds to the $r_b=0$ endpoint of the $I_b$ plot in figure \ref{actionsbc}.

Between inverse temperatures $\beta_{\rm C}$ and $\beta_{\rm B}$, again the only solution with a horizon is the cosmological one, now with $r_c<\ell$. For yet higher temperatures, $\beta<\beta_{\rm B}$,
there are three stationary points, corresponding to three different mass parameters: one cosmological horizon solution, and two black hole solutions --- the ``small" and ``large" black holes.
It is clear from figure \ref{actionsbc} that,
between the two black hole solutions, the large black hole dominates, as the small black hole is a local maximum of the reduced action, while the large black hole is the adjacent local minimum. (This can also be shown analytically in the same way as in the $\Lambda \!=\! 0$ case \cite{York86}.)
For any boundary size $R$, at sufficiently high
temperature the large black hole has a smaller action than thermal de Sitter. Regarding the cosmological horizon solution, it can be proved  analytically (see appendix \ref{c-dominance}) that this stationary point has a lower action than that of the large black hole at 
all temperatures where the black hole solutions exist ($\beta \! < \! \beta_{\text{B}}$ in figure \ref{beta-rh}), 
and hence always dominates over the black hole solutions. 
However, since that stationary point is a  maximum of $I_c$, the relevant comparison is
actually with the endpoints, at the smallest and
largest possible values of $r_c$, which have yet lower action. It turns out 
that in fact the largest horizon, $r_c=\ell$,
always dominates  over $r_c=R$ or $r_c=\widehat R$ whenever 
the large black hole
dominates over thermal de Sitter.\footnote{Although the spatial geometry of the $r_c=\ell$ configuration coincides with a portion of a spatial slice of de Sitter space, we do not refer to this as a ``thermal de Sitter" phase because, unless $\beta$ equals the inverse dS temperature (corresponding to curve C in figure \ref{actionsbc}), the 4-dimensional \euc manifold has a conical singularity at the horizon for any choice of lapse function that vanishes at the horizon. 
}

\subsubsection{Phase diagrams}
\label{phasdiag}

We next present phase diagrams constructed by a quantitative analysis
of the relative size of the action for the various spherical configurations. 
As explained above, since we do not know if b-c transitions are allowed, we shall consider 
both possibilities, which calls for three separate phase diagrams.
The first includes only thermal dS and b-horizon configurations, the second includes only c-horizon configurations, and the third includes all configurations.
The diagrams were constructed using Mathematica, since the comparisons can generally not be done analytically.
Each phase diagram indicates which configuration dominates, i.e.\ has the lowest action, for each boundary size $R$ and inverse temperature $\beta$, within the given class of configurations. 
The dominant configuration gives the leading order approximation to the free energy $F = I/\beta$, so minimizing the action corresponds to minimizing the free energy. From the action \eqref{Igen} we see that the free energy 
has the form $E-TS$, where $E$ is the Brown-York energy and $S=\pi r_h^2$ is the Bekenstein-Hawking entropy of the horizon (in Planck units).\footnote{Alternatively, in terms of the partition function, the entropy is given by
\be
S = (1-\beta\partial_\beta)\log Z \approx (\beta\partial_\beta - 1)I= \pi r_h^2,
\ee
where $I$ is the action of the dominating configuration. Although the value of $r^*_h$ at a stationary point depends on $\beta$, that dependence does not contribute to the entropy precisely because the action is stationary with respect to variation of $r_h$. And when a c-configuration endpoint dominates, $r_c$ does not vary with 
$\beta$, since $r_c$ is either $R$, $\widehat R$, or $\ell$.} 
For thermal dS or b-horizon
configurations, the energy is negative so it lowers the free energy, whereas for c-horizon configurations it is positive and thus raises the free energy. The entropy always lowers the free energy.

\begin{figure}[t!]
    \centering
    \includegraphics[scale=0.4]{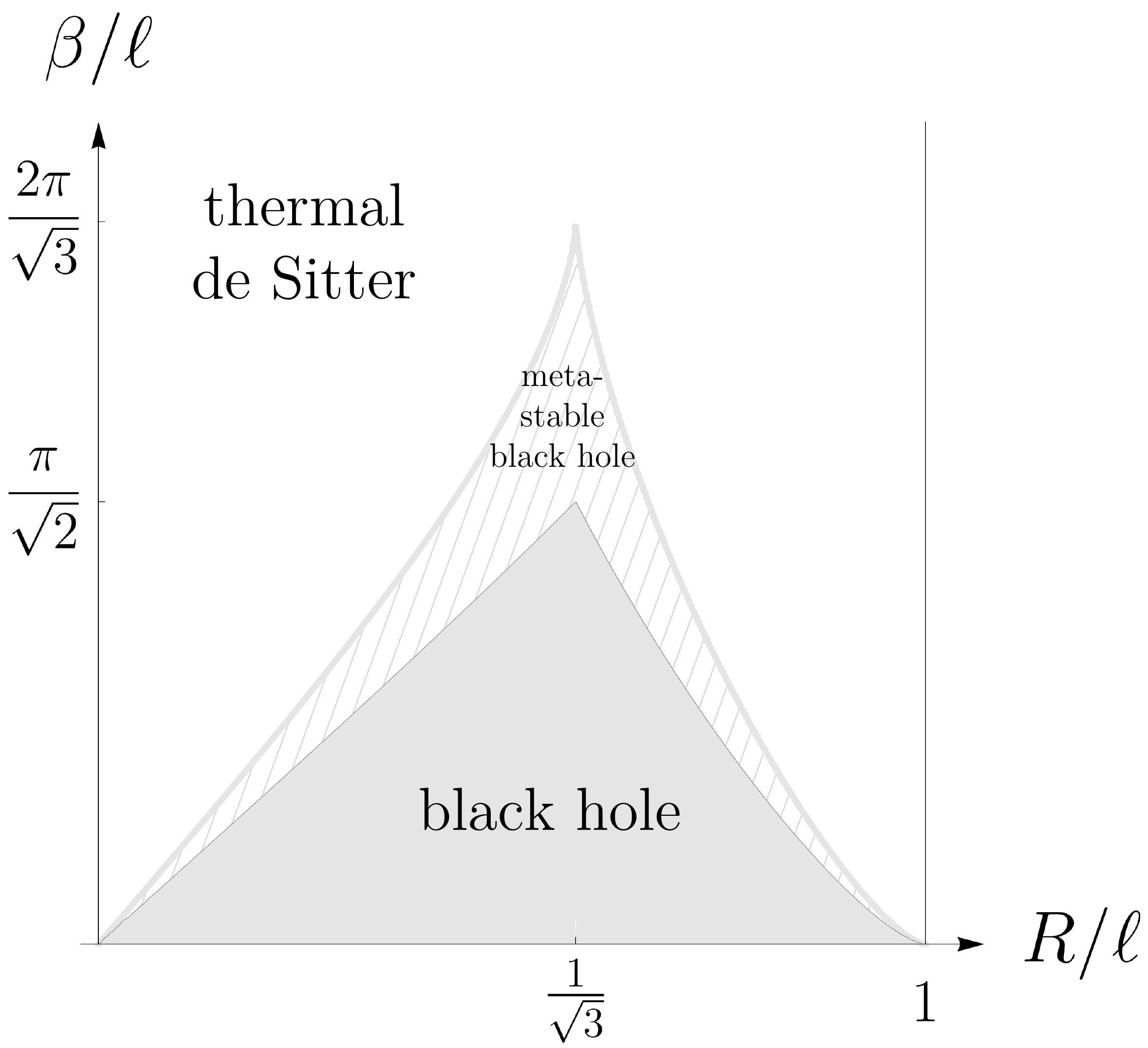}
    \caption{Phase diagram for thermal dS and b-horizon configurations, with
    boundary size $R$ and inverse temperature 
$\beta$.}
    \label{b-phase}
\end{figure}

\begin{figure}[t!]
    \centering
    \includegraphics[scale=0.4]{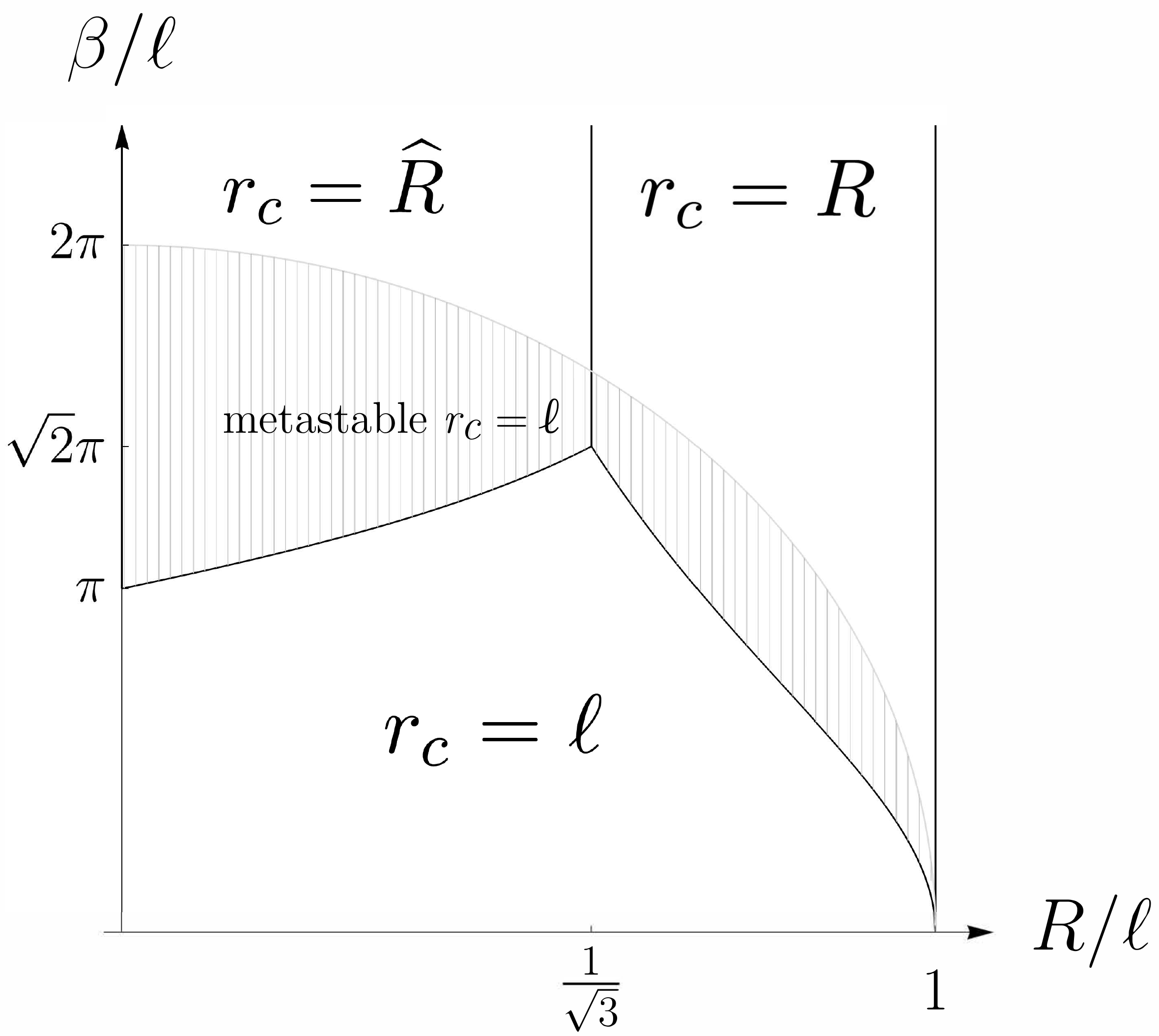}
    \caption{Phase diagram for c-horizon configurations, with
    boundary size $R$ and inverse temperature 
$\beta$.  
The upper boundary of the metastable 
$r_c=\ell$ region corresponds to the blueshifted dS temperature. As shown in figure \ref{actionsbc}, below that temperature --- marked by C --- the $r_c=\ell$ endpoint is no longer a local 
minimum of the action (even excluding negative mass parameters).
    }
    \label{c-phase}
\end{figure}

\begin{figure}[t!]
\centering
\includegraphics[scale=0.4]{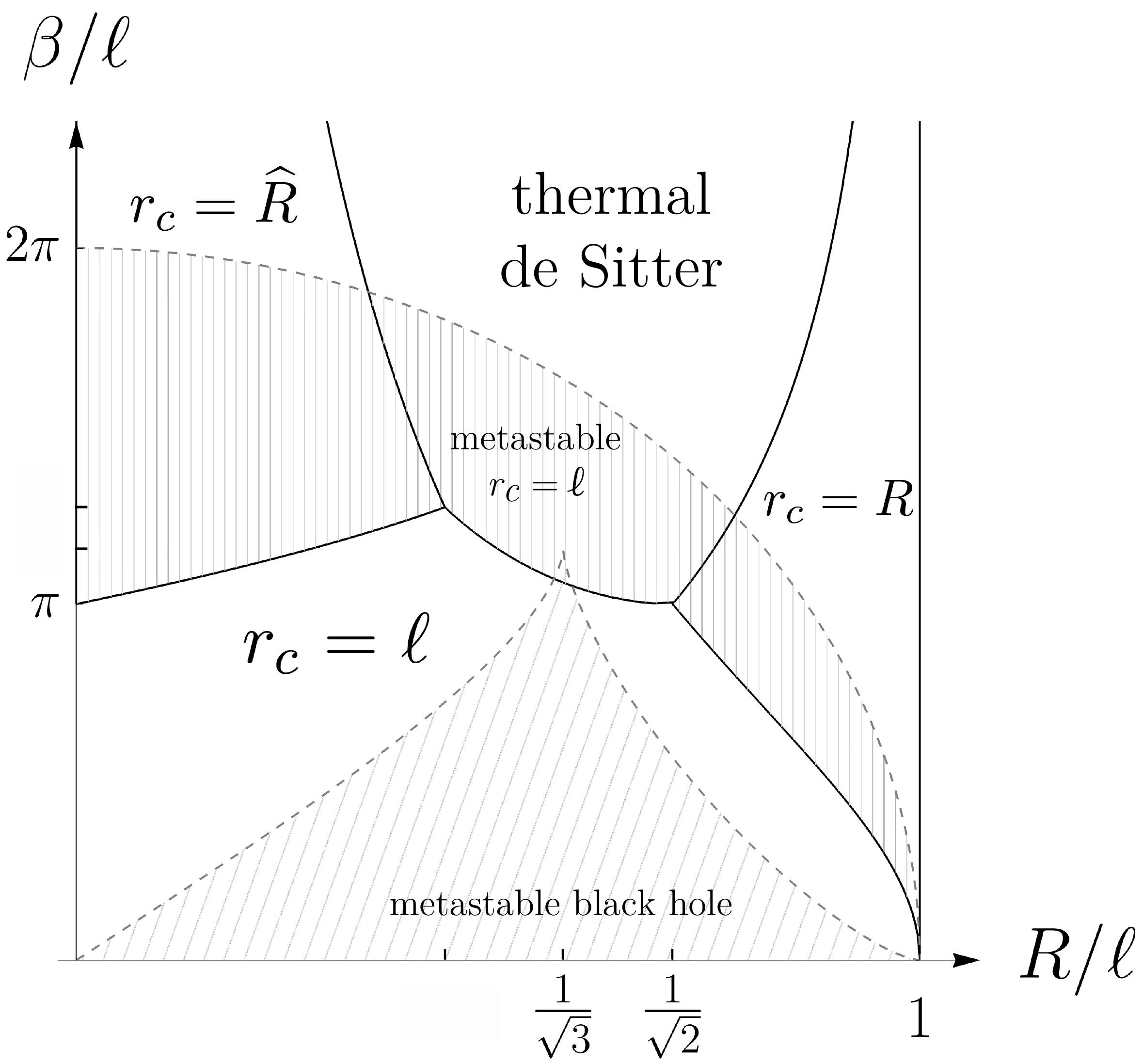}
\caption{Phase diagram for combined thermal dS, b- and c-horizon configurations,
allowing for b $\! \leftrightarrow \!$ c transitions,
with boundary size $R$ and inverse temperature $\beta$. In the region with diagonal lines the small and large black hole solutions exist. (The coordinates $(R/\ell,\beta/\ell)$ of the two triple points are 
$(\sqrt{3-\sqrt{5}}/{2},\pi\sqrt{(\sqrt{5}+1)/2})$
and $(1/\sqrt{2},\pi)$, and the tip of the metastable black hole region is at $(1/\sqrt{3},2\pi /\sqrt{3})$, where the minimal temperature Nariai sits.) 
}
\label{beta-R}
\end{figure}

The phase diagram for thermal dS and b-horizon configurations is shown in figure \ref{b-phase}. 
For any system boundary size $R<\ell$, at low enough temperature there is no black hole equilibrium, so thermal de Sitter is the only candidate phase. 
The thermal dS action coincides with that of the left endpoint of the b-horizon  graphs \eqref{actionsbc}.
It is the absolute minimum of the action at low enough temperatures, hence there is no other phase in the spherical sector that thermal dS can decay to.\footnote{As for 
nonspherical fluctuations, thermal dS is also stable against gravitational clumping of thermally excited (massless) modes (Jeans instability \cite{Jeans}), since 
at temperatures for which thermal dS dominates,
the finite size of the cavity is smaller than the Jeans wavelength. (This is analogous to the stability of ``hot flat space" when confined to a box (cf., e.g., \cite{GPY}).)
In fact, even without a box Ginsparg and Perry \cite{Ginsparg-Perry} showed that for classical perturbations around the de Sitter metric there is no analog of the Jeans instability in de Sitter space.} 
As the temperature is raised, two black hole 
equilibria appear, the larger of which is
locally stable, but its free energy is initially higher than that of thermal dS so it is a metastable phase. This region is indicated by the diagonal fill lines. 
At yet higher temperature 
the contribution of the entropy of the black hole overcomes the fact that thermal dS has a lower energy, so the black hole becomes the stable phase.  
(This is similar to 
what happens in anti-de Sitter space \cite{HawkingPage} where, as the temperature is raised, the locally stable black hole solution appears before the Hawking-Page transition, i.e., before it has a lower free energy than thermal AdS.)

The phase diagram for c-horizon configurations is shown in figure \ref{c-phase}, which displays a single triple point.
At low enough temperature the
dS horizon radius configuration
$r_c=\ell$ is not a local minimum of the action, so it is not a candidate equilibrium.
For system boundary size $R<\ell/\sqrt{3}$, at low enough temperature the only phase is the $r_c=\widehat R$ configuration,
which corresponds to left endpoint on the c-branch
 in figure \ref{actionsbc}. While a local minimum, it
is not a stationary point of the action, so
its horizon radius is 
{\it not} related to $\beta$ by the formula (\ref{temp}).
At higher temperature, the action of the 
$r_c=\ell$ configuration becomes
a local minimum,
corresponding to right endpoint on the c-branch
in figure \ref{actionsbc}.
As the temperature rises, its action is initially
greater than that of the $r_c=\widehat R$ configuration
because its greater energy outweighs its greater entropy.
At yet higher temperature, however, it becomes the stable phase. 
Once the temperature goes above $\ell/\pi$, which is twice the Gibbons-Hawking temperature of de Sitter space,  $r_c=\ell$ is 
always the stable phase 
for any $R<\ell/\sqrt{3}$.
A similar pattern occurs for the larger system boundary radius $R>\ell/\sqrt{3}$, but now with 
$r_c=R$. However, 
note that a configuration with boundary radius $R$ and 
$r_c=R$ has zero spatial volume,
which calls into question the validity of our semiclassical evaluation of the partition function for this phase.
Another qualitative difference from the $R<\ell/\sqrt{3}$ case is that, as $R$ approaches $\ell$, the  transition temperature to the stable $r_c=\ell$ phase goes to infinity.

Finally, the phase diagram for the combined b-c system is shown in figure \ref{beta-R}. For any
boundary radius $R$, at low enough temperature thermal dS dominates, and at high enough temperature, the $r_c=\ell$ horizon configuration dominates.
The full phase structure contains two triple points, and is complicated to describe, so we refer the reader to figure \ref{beta-R} for the specifics.  
Note that while there is a region
where metastable black hole configurations exist, they are never the dominant configuration,
if transitions to c-horizon configurations are possible.

The 
$R=\ell/\sqrt{3}$ case is exceptional, since in addition to b- and c-horizon configurations there exists the Nariai solution if $0\!\leq\!\beta/\ell\!\leq\!2\pi/\sqrt{3}$. 
The Nariai phase
is never dominant at any temperature. To see this,
note first that 
its action $I\!=\!-\pi \ell^2/3$ is equal to that of the b- and c- configurations with $r_b=r_c=R$,
so to determine its potential dominance we can use the corresponding black hole action.
The Nariai ensemble
parameters $(R,\beta)$ correspond to the vertical line from the tip
to the base
of the metastable black hole
region with diagonal lines in figure \ref{beta-R}. As seen in 
the figure,
at low temperature thermal dS dominates (with temperature greater than the GH temperature
at the boundary)
while
at high temperature the $r_c=\ell$ configuration dominates.\footnote{The transition happens at $\beta/\ell=
({2\pi}/{\sqrt{3}})\sqrt{27/32}$.}
This is 
consistent with the
instability of the Nariai phase discussed in \cite{Bousso-Hawking-anti,Bousso-adventures}, where it is shown that perturbing the Nariai solution (by including scalar fields) makes the black hole horizon evaporate completely, leaving empty de Sitter space.

\subsubsection{Cosmological horizon instability and reservoir dynamics}
\label{c-instability}

We have identified the spherical, purely gravitational configurations 
and compared their free energies towards determining
the stable, metastable, and unstable phases corresponding to each spherical 
boundary size $R$ and inverse temperature $\beta$. 
When these parameters are restricted 
to be much larger than the Planck length, 
one expects that the  spherical
gravitational contribution dominates over nonspherical and nongravitational contributions.  Nevertheless, the latter
contributions can provide a channel through which the 
system evolves from an unstable configuration to one
with smaller free energy. 
In this subsection we consider the dynamics of this channel,
mediated by the Gibbons-Hawking radiation from the cosmological
horizon. We attempt here only a qualitative account of that radiative process. 
A more precise account of the quantum field dynamics with back-reaction on the Schwarzschild-de Sitter background 
might be pursued along the lines of ref.~\cite{Zhao:1994hm}.

For a consistent physical model of the system, 
one should also take into account the evolution of the reservoir, 
because the reservoir is enclosed by the horizon and therefore 
sources the $M$ dependence of the spacetime. 
This is an essential difference from the b-system case,
where the boundary encloses the system.
In that case, the reservoir has no effect on the gravitational field (spacetime geometry) inside the boundary, and can be taken arbitrarily large, 
so its only role is to fix the temperature of the ensemble. 
For the c-system case, since we shall interpret $M$ as sourced by 
the reservoir, we exclude negative $M$ values as unphysical. 
We imagine that the reservoir has a thermostat and internal degrees
of freedom enabling it to maintain itself at a fixed temperature as long as $M$ is nonzero.\footnote{One could instead define the system abstractly
without a physical model of the reservoir, with 
a temperature fixed by definition regardless of the value of $M$. 
In that case there would be no reason to exclude negative $M$
from the phase space of exterior configurations, 
and the system would be unstable to unbounded growth of the horizon.}

As illustrated in figure \ref{actionsbc}, at every temperature there is a cosmological horizon radius at which the action is a local maximum.
However, below some temperature that local maximum occurs with a negative mass parameter $M$, which we exclude. 
For $M=0$ the local maximum sits at the de Sitter radius
$r_c=\ell$, and the temperature of the ensemble matches the blueshifted Gibbons-Hawking temperature of de Sitter space 
at the location of the boundary $R$.\footnote{The blueshifted
Gibbons-Hawking temperature is 
$\hbar\kappa/2\pi$, 
where $\kappa$ is the horizon surface gravity defined with respect to the horizon Killing vector with unit norm at the boundary.}
This corresponds to the black dot at the right endpoint of the solid black portion of curve C in figure \ref{actionsbc}.
In the phase diagram figure \ref{c-phase} this occurs on the circular arc. 
At higher temperatures
the local maximum, illustrated 
by the black dots on the curves below C
in figure \ref{actionsbc},
corresponds to a SdS solution 
with $M>0$, and $r_c<\ell$.
Let us consider an initial configuration of this type, 
and ask how it evolves if perturbed.
The endpoint of the instability 
depends upon $R$ and $\beta$,
and corresponds to 
the phase shown in figure \ref{c-phase}.

The instability is similar to the familiar one for the small black hole
configurations, which correspond to the
local maxima of the black hole branch of the action plot figure \ref{actionsbc}, so let us begin by recalling how that works. The small black hole starts out in equilibrium with the radiation bath.  If 
it then emits a little more energy than it absorbs  
it shrinks, which raises its temperature above that of the bath and leads to a runaway evaporation. Conversely, if it absorbs a little more than it emits it grows, which lowers its temperature below that of the bath and leads to a runaway growth until it reaches the size of the stable, large black hole.

Now let us follow the instability
of a cosmological horizon with $r_c<\ell$ that starts out in equilibrium. If it emits a little less energy than it absorbs from the bath, then there is a net flux of energy outward across the horizon, which causes the horizon to grow. This growth can be inferred using the Raychaudhuri equation on the horizon, or by reasoning that a loss of energy from the reservoir must entail a decrease of the mass parameter $M$ in the solution outside the reservoir. (One can presumably define $M$ as a dynamical quantity that satisfies a conservation law together with the flux of the energy momentum tensor, but we will not try to 
formalize that here.) As the horizon grows its Gibbons-Hawking 
temperature 
measured at the boundary
drops below that of the reservoir, 
so it absorbs yet more and continues to grow in a runaway process. Eventually, however, 
the reservoir mass drops to zero,
at which point there is presumably no way for it to maintain the fixed temperature of the canonical ensemble, and the system settles to empty dS with 
$r_c=\ell$ and a boundary temperature that 
coincides with the ambient blueshifted 
Gibbons-Hawking temperature of dS.
This final state corresponds to the right endpoint of the solid black part of curve C on the cosmological branch in figure \ref{actionsbc}. 
The reservoir thermostat is no longer controlling the temperature, since there is
no reservoir mass available. If the reservoir absorbs energy in a fluctuation, then its thermostat is reactivated, and it immediately rejects the extra energy and the system settles back to dS, at least with this interpretation of the reservoir dynamics.\footnote{It would appear from curve C in figure \ref{actionsbc} that the right endpoint is a marginally stable state that should decay to smaller $r_c$;
however, curve C corresponds to an ensemble
with boundary temperature fixed at the blueshifted Gibbons-Hawking temperature,
which as just explained is not the setting on the reservoir thermostat.}

Instead of positing that the reservoir holds a fixed temperature as the mass parameter decreases all the way to $M=0$, one could imagine a more physically
plausible reservoir whose temperature drops as $M$ drops below some threshold, and goes to zero as $T_{\rm res}\propto M^\alpha$ for some exponent $\alpha$. 
(For example, a cavity filled with radiation in thermal equilibrium and energy $E$ would have $T\propto E^{1/4}$.) Then $dT_{\rm res}/dM\propto M^{\alpha-1}$ would diverge as $M\rightarrow 0$ provided $\alpha < 1$. The blueshifted horizon temperature and its derivative with respect to M on the other hand approach finite values in this limit. In this case, we expect that the system would stabilize at some small value of $M$ and a horizon size smaller than the dS horizon. This would presumably happen at the value of $M$ that would make $T_{\rm res}$ equal to the blueshifted horizon temperature at the reservoir.

If instead the cosmological horizon emits a little more
energy than it absorbs from the bath, then the mass of the reservoir grows and the horizon shrinks, which raises its temperature and leads to a runaway shrinking. The endpoint of that shrinking corresponds to the left endpoint of the curves on the cosmological branch in figure \ref{actionsbc}, where the Brown-York energy vanishes and the free energy is just the negative of the horizon entropy times the temperature. For the case shown in that figure, where $R$ is smaller than the Nariai radius 
$\ell/\sqrt{3}$, that endpoint sits at $\widehat R$, 
so the system volume between the boundary and the horizon remains nonzero. A physical interpretation
of this endpoint configuration
is that the reservoir has absorbed too much energy to be confined within a
boundary of radius $R$, so has become a black hole that then swallows the boundary, destroying the ensemble.
For the case where $R$ is greater than the Nariai radius, the left endpoint corresponds instead to the system boundary $R$ itself. 
In that case, the horizon has shrunk down to coincide with the system boundary so the system volume vanishes. 
We are doubtful that the semiclassical treatment we are employing accurately describes this peculiar 
configuration. A physical interpretation in this case could be that the reservoir
boundary falls across the cosmological 
horizon, again destroying the ensemble.

\subsection{Negative temperature?}

As discussed in the Introduction, the fact that the de Sitter static
patch entropy {\it decreases} when energy is added has led to the
suggestion that this thermodynamic system has negative temperature.
This motivates us to ask whether in the presence of a system boundary
a negative temperature canonical 
ensemble can be defined and, if so, whether it
can be stable. 
Mathematically, the 
question is whether the partition function 
is finite. A prerequisite for this in a normal
system is that the 
Hamiltonian should be bounded above. 
Figure \ref{ER} shows that the Brown-York energy of spherical configurations with different boundary radius is indeed bounded above, provided negative $M$ configurations are excluded (as we have argued they should be). So that criterion seems to be met.
A way to ensure convergence of the negative temperature partition function is to have a finite dimensional Hilbert space.
The covariant entropy bound \cite{Bousso99} suggests that the entropy is indeed bounded, and 
hence that the Hilbert space of the quantum states of the system is finite dimensional. However, 
as we now explain, the negative temperature ensemble
with a boundary seems not to exist, or at least not to admit a semiclassical path integral description.

The partition function for negative temperature ensembles is $Z=\text{Tr}\, e^{-\beta H} = \text{Tr}\, e^{+|\beta| H}$.
To express this as a path integral one follows the same steps as the positive temperature case 
and obtains \eqref{ZNh}, but with the signs of the last two integrals flipped. Thus in order to get the ``Euclidean path integral" one must
deform the integration contour of $\lambda^{\mu}$ such that it becomes $\lambda^{\mu}=+iN^{\mu}$ with real $N^\mu$, at least at the relevant stationary points.
For the analytically continued configurations 
the exponent in (\ref{Zc}) thus becomes $+I$. This implies that in a negative temperature canonical path integral the configuration with the largest \euc action dominates, which translates to the fact that the free energy of a system with negative temperature is in fact maximum at equilibrium. That free energy is maximized rather than minimized is a standard result for negative temperature ensembles~\cite{Wisniak}. 

 From figure \ref{actionsbc} one might think that the cosmological horizon solution at a local maximum of the action might be locally stable at sufficiently negative temperature.
 (Similarly when it exists the small black hole solution is a local maximum, hence might be 
 a candidate for a negative temperature equilibrium.) 
However, if the path integral is approximated with either of those solutions then, since $I$ is given by eq.~\eqref{Igen} with $\beta\rightarrow |\beta|$,
one obtains for the entropy a {\it negative} result:
\be
S = (1-\beta\partial_\beta)\log Z \approx (1-|\beta|\partial_{|\beta|})I= -\pi r_h^2.
\ee
Since the entropy of a quantum system is by definition positive, this
indicates that some error has been made.
One possible source of error is disregarding
the nonspherical modes, which contribute 
positively to the action, at least perturbatively~\cite{GPY, Gregory-Ross}.\footnote{While these cited references do not discuss geometries with cosmological horizon, we expect that a similar result holds for perturbations of the cosmological horizon solution.} 
This suggests that the spherical local maximum may not provide a good approximation to the
partition function, although to verify that
one would need to confirm that there exist
such nonspherical modes that satisfy the constraints as well as the system boundary conditions. 
Another possible source of error lies in the hocus pocus of the contour deformation. 
In fact, one could consider the spherical
reduction as a theory in itself, in which case
the nonspherical modes are irrelevant, and in that case we still seem to arrive at an inconsistent, negative entropy. Perhaps while it is possible to deform the contour so that it passes through a Euclidean local minimum of the action, the same cannot be done for a local maximum. It may be possible to
check this explicitly in spherically reduced GR or other 2d gravity theories.

\section{Microcanonical ensemble}
\label{micro}
 
For a quantum system with Hamiltonian $H$ the density of states $\nu(E)$ at energy $E$ is
\be \label{nu(E)}
\nu(E)=\text{Tr} \, \delta (E-H). 
\ee
In nongravitational 
statistical mechanics $\nu(E)$ 
is related to 
the partition function $Z(\beta)$ by an inverse Laplace transform (ILT). 
Computation of the density of states via the ILT in the context of black holes has been discussed
by various authors~\cite{Hawking-PI,HawkingPage,York86,
BWY,Louko-Whiting,Melmed-Whiting,Miyashita}. 
Since the gravitational partition function is 
only computable approximately, its analytic structure is not known, and the known approximations tend to be restricted to certain temperature ranges.
While for some energies the ILT method converges and yields 
the expected density of states using the available 
approximations to $Z(\beta)$, a naive application 
of the method gives a divergent integral in cases
where a canonically unstable black hole dominates the microcanonical ensemble~\cite{Hawking-PI,HawkingPage,York86}.
In asymptotically flat space, and for energies
corresponding to a small black hole in asymptotically
anti-de Sitter space, it was argued \cite{Hawking-PI,HawkingPage} that one can obtain
the expected black hole density of states via a procedure
involving rotation of the temperature contour in the ILT, 
as well as rotation of the integration contour for the unstable mode in the partition function. 
No justification for this procedure has been spelled out,
as far as we know, although it does appear to yield 
sensible results. However, Brown and York (BY) \cite{BY}
introduced a different approach to computing $\nu(E)$ 
that starts from a path integral representation
for $\nu(E)$ itself, rather than using the ILT of the partition function. This method avoids the
difficulties that arise when attempting to 
compute the ILT of the partition function.
Here we adopt this BY approach to directly estimate $\nu(E)$ for the bounded system with a positive
cosmological constant.

The first step in constructing the 
path integral representation for 
$\nu(E)$ is to express the delta function 
in \eqref{nu(E)} using the Fourier transform, 
\be \nu(E)= \frac{1}{2\pi} \int_{-\infty}^{+\infty} \!dT \, e^{iET} \left( \text{Tr} \, e^{-iHT}\right)\, . 
\label{nu} \ee
The Hamiltonian in 
(\ref{nu}) is the gravitational
Hamiltonian \eqref{H} 
with Dirichlet boundary conditions.
The trace can thus be expressed as a functional integral over configurations that are periodic in time,
weighted by $\exp(i I)$, where $I$ here denotes the Lorentzian action 
suitable for Dirichlet boundary conditions, whose \euc version for 
gravitational systems was discussed in section \ref{can} when constructing 
the partition function for the canonical ensemble.
When followed by the integral over $T$, weighted by $\exp(iET)$, this  becomes a path integral over 4-metrics,
\be
\nu(E) = 2 \Re \int  \mc{D} g\, e^{i\im}\, ,
\label{dos}
\ee
where $I_{\rm mic}=I + ET$ is the microcanonical action.
Twice the real part appears, because the integral in (\ref{nu}) is over both positive and negative $T$, 
and the factor $1/2\pi$ is absorbed into the measure.\footnote{See also \cite{BY-dos} for derivation in the case of nonrelativistic quantum systems,  and the example of a simple harmonic oscillator, where the microcanonical path integral can be evaluated exactly.} 
For gravitational systems the time interval $T$ is the proper time of the boundary observers, $T=\int \! d\tau N_\partial$, the integral of which is a sum over the boundary value of the lapse function, written in (\ref{dos}) as part of the metric integral. The boundary term in $I$  that involves the lapse function (cf.\ (\ref{constrained})), together with $ET$ produces
\be
i\int \! d\tau\,  N_{\partial} \left ( E+\frac{1}{8\pi}\int \! d^2x  \sqrt{\sigma} k \right )
\ee
in the exponent.
The integral over $N_\partial$ 
therefore sets  $-(1/8\pi)\int d^2x \sqrt{\sigma} k$ equal to $E$,
hence in effect changes the boundary conditions from fixing the lapse to 
fixing the BY quasilocal energy \cite{BY-quasi} (which is reviewed below) to be equal to $E$.
The path integral (\ref{dos}) is thus over periodic configurations with arbitrary period $T$ 
and fixed 
BY energy, hence may naturally
be referred to as the ``microcanonical path integral".

The BY quasilocal stress tensor for a gravitational system 
is defined by the functional derivatives of the Dirichlet action with respect to the boundary metric $\gamma_{ij}$ \cite{BY-quasi}:
\be
T^{ij}=\frac{2}{\sqrt{-\gamma}}\frac{\delta I}{\delta \gamma_{ij}}\, .
\ee
The energy surface density $\mce$ is obtained from the normal projection of the stress tensor on the 2-surface $\partial \Sigma$, 
\be
\mce\equiv u_i u_j T^{ij}=-  k/8\pi \, ,
\label{eps}
\ee
where again $u$ is the future-pointing timelike unit normal to $\Sigma$ and $k = \sigma^{ab}k_{ab}$ is the mean extrinsic curvature of $\partial \Sigma$ as embedded in $\Sigma$, which we define with respect to the outward normal.\footnote{Here there is an apparent sign discrepancy with the BY notation, since they define $k$ with respect to the boundary normal that points inward toward the system whereas our definition uses the outward normal.}
The energy surface density $\mce$ is the value (per unit boundary area) of the Hamiltonian that generates
\textit{proper} time translations on $\partial \mc{M}$ in the spacetime direction orthogonal to $\Sigma$; i.e., it is the energy surface density measured by a congruence of boundary observers on the timelike boundary whose four-velocities are hypersurface orthogonal \cite{BY-quasi}. 
The proper momentum surface density on the boundary is also defined from the stress tensor, and is part of the microcanonical boundary condition in general \cite{BY}. The spherical symmetry in the present study requires the momentum density to vanish, hence it is absent in our analysis.
The total BY quasilocal energy\footnote{Note that had we included a subtraction term in the Dirichlet action $I$ the boundary stress tensor and the energy density would be modified as well. Specifically, the energy density would have become $\mce=-(k-k_0)/8\pi $, where $k_0$ is the mean curvature of the boundary embedded in a reference geometry. The microcanonical action is independent of the choice of the reference space, because the boundary integral in (\ref{Imic}) cancels the same term that is already present in the (Lorentzian) Dirichlet action (whose \euc version is given in (\ref{constrained})); however as already discussed in the canonical section below (\ref{Icov}) we set the subtraction term to zero.}
is the integral of the density (\ref{eps}) over the 2-boundary $\partial \Sigma$,
\be 
E_{\rm BY}\equiv \int _{\partial \Sigma}\! d^2 x \, \sqrt{\sigma}\, \mce = -\frac{1}{8 \pi }\int _{\partial \Sigma}\! d^2 x \, \sqrt{\sigma}\,  k \, ,
\label{E}
\ee
which is indeed the quantity that is set equal to $E$ as a result of the boundary lapse integral.

After performing the boundary lapse integral, what remains in the exponent 
is the Dirichlet action with a boundary term removed, i.e.\  the microcanonical action, which can be written as
\be
\im=I+\int_{\partial \mc{M}}\!d^3x \sqrt{\sigma}N\mce\, ,
\label{Imic}
\ee
where the last term is present to cancel the boundary integral in $I$.
As shown explicitly in \cite{BY},
the solutions to the Einstein equation extremize the microcanonical action (\ref{Imic}) when the geometry of $\partial \Sigma$ as well as the energy density $\varepsilon$ (and momentum density, in general) are held fixed on the timelike boundary. 
In the case of our spherically symmetric boundary where the area of the boundary 2-sphere is fixed to $4\pi R^2$,
the microcanonical boundary condition reduces to specifying $R$ together with a uniform energy density $\mce$ on the boundary 2-sphere, or equivalently the total BY energy (\ref{E}), $E=4\pi R^2 \mce$.\footnote{In the nonspherical case there is no unique {\it energy} to fix. Rather, Brown and York 
proposed defining a quasilocal gravitational ``microcanonical" ensemble \cite{BY} with fixed quasilocal energy-momentum density on the boundary.
In general this is infinitely more data than simply specifying a total energy. In the present analysis, however, where we take the boundary manifold to be $S^2 \times S^1$ and restrict to spherical symmetry, it amounts to just one parameter.}

The entropy in a microcanonical
ensemble is the logarithm of the number of orthogonal states with energy in a band of width $\D E$ about $E$, where the choice of
$\D E$ is irrelevant as long as it does not grow exponentially with the size of the system.\footnote{Typically the number of states is exponential in the size of the system,
so as long as $\D E$ does not have a comparable
exponential dependence (which would in any case make no sense for a microcanonical ensemble) the dependence on $\D E$ is negligible.}
The microcanonical entropy for a gravitational system in a bounded region with density of states (\ref{dos}) is thus given by
$S = \log[\nu(E)\D E]$. 
Suppressing the irrelevant $\D E$ dependence
(with abuse of notation), we write this 
as $S = \log[\nu(E)]$.
The path integral is estimated semiclassically by the value of the integrand at a stationary point of the action,
i.e.\ at a configuration that satisfies the specified microcanonical boundary conditions and extremizes $\im$. 
With the spherically symmetric boundary conditions characterized above,  we expect the stationary points of the action have the same symmetry.\footnote{See, however, the discussion of this point in section \ref{reduction}.} Therefore, we look for spherically symmetric solutions to the Einstein equation that can be put on a manifold with a single boundary
with topology $S^2 \times S^1$.

For spherically symmetric solutions with positive cosmological constant, we again refer to the Birkhoff theorem \cite{Birkhoff, Birkhoff1} which states that the only solutions are locally isometric to the Schwarzschild-de Sitter family and the Nariai space. However, these Lorentzian geometries cannot be placed on a manifold with a {\it single} boundary, because across their bifurcation surface there is an asymptotic region with another boundary.\footnote{As an alternative 
approach, one could introduce an inner boundary at the bifurcation surface on the spatial slices, and require that the lapse function vanish there. After identification of two such spatial slices, the resulting Lorentzian spacetime would possess no inner boundary, but would have a conical singularity. As discussed in \cite{Marolf:2022ybi}, in the context of a path integral for the partition function, the action acquires an imaginary contribution from such a singularity, which results in the same semiclassical approximation to the partition function as is obtained from the Euclidean stationary point.}
Lorentzian stationary points are therefore not in the original domain of integration. However, as discussed in section \ref{can} when obtaining the ``Euclidean path integral", eq.~\eqref{Zc}, the contour of integration may be deformed into complex values, in particular for the lapse, in such a way that the contour may pass through a Euclidean stationary point with the required topology, i.e.\ a metric that satisfies the Einstein equation everywhere on a manifold with the desired boundary $\partial \mc{M}=S^2\times S^1$.\footnote{If $\partial \mc{M}$ consists of two disconnected timelike boundaries, then the path integral can be approximated by Lorentzian stationary points as discussed in \cite{Martinez}, where the microcanonical entropy of a two-sided black hole is evaluated. The microcanonical action vanishes in that case, hence the entropy is zero to leading order, which is the correct result for a pure state.}

Since 
the microcanonical boundary conditions on $\mce$ and $\sigma_{ab}$ remain the same upon the contour deformation of the lapse function, the \euc solutions must satisfy the same microcanonical boundary conditions as the Lorentzian ones mentioned above. 
 For  given boundary data $(E, R)$, 
to estimate the density of states
we must therefore identify the \euc stationary point obtained by 
analytic continuation of the lapse, $N\rightarrow\pm iN$.
The sign of the imaginary lapse should presumably 
be chosen so that the exponent of the integrand in \eqref{dos}
is positive, since otherwise one would not expect the stationary
point to yield a good approximation to the integral.
This sign turns out to be $N\rightarrow-iN$.\footnote{This sign happens to agree with the one required for the Euclidean path 
integral representation of the partition function, as explained
in detail at the beginning of section \ref{can}, but the rationale for that sign choice does not appear to be relevant in the present, 
microcanonical context.} 
Note that if this  \euc solution is to actually yield a good approximation
to the path integral, the exponent of the integrand 
should be a local maximum on the integration contour.
(Some issues that arise in trying to check whether this 
condition is satisfied are addressed
at the end of section \ref{foundations}.)
The microcanonical action $\im^{\text{E}}$ 
at this Euclidean stationary point is obtained from
(\ref{Imic}) by $N\rightarrow-iN$,
and the density of states (\ref{dos}) 
is approximated
by $\nu\approx \exp(-\im^{\text{E}})$.
The Euclidean microcanonical action is then the leading order approximation to the entropy, $S \approx -\im^{\text{E}}$.

The \euc microcanonical action $\im^{\rm E}$ differs from the \euc action $I$ (appearing in the exponent of the partition function path integral \eqref{Zc}) only by the absence of the BY energy term.
The \euc action $I$ was derived in (\ref{constrained}) for configurations that satisfy the constraints, which becomes (\ref{Ih}) after spherical reduction. The  microcanonical entropy is thus given by
\be\label{Smic}
S\approx -\im^{\rm E}=A_h/4=\pi r_h^2\, .
\ee
In the following, we obtain the entropy as a function 
of the boundary data $E$ and $R$, by identifying the relevant
stationary point and evaluating its action.

\subsection{Microcanonical phases}

The sign of the 
energy density \eqref{eps} is 
determined by that of the extrinsic
curvature which can have either sign. Hence, when specifying
a microcanonical ensemble the 
{\it sign} of the energy
must be specified. For a metric of the form \eqref{sp2}, the extrinsic curvature of the boundary is  $k=-2r'/r$ (recall our convention that $y=0$ at the boundary and increases into the system).
The sign of the energy is therefore positive (negative) if 
the system lies at greater (smaller) radii than the boundary.
For the \sds family,  $r'^2=f(r)$ (\ref{rprime}), the total energy \eqref{E} is therefore $-{\rm sgn}(k)R\sqrt{f(R)}$, i.e.\
\be
 E_{b,c}=\mp R \sqrt{1-2M/R-R^2/\ell^2} \, ,
 \label{Ebc}
\ee
where $b\, (c)$ and minus (plus) refer to the case where the system contains the black hole (cosmological) horizon.\footnote{As mentioned in section \ref{can} we do not include any counterterm in the action, since we are considering only bounded manifolds with finite actions, and also because there is no preferred reference state.
Were a flat space subtraction term  included, the quasilocal energy in the black hole case would have become $E_b=R(1-\sqrt{1-2M/R-R^2/\ell^2})$, which produces the familiar ADM mass \cite{ADM} when the cosmological constant is zero and the boundary is sent to infinity.} Therefore 
the boundary values $R$ and $E$
uniquely determine the mass parameter $M$,
unlike in the case of the canonical ensemble
where $M$ remains a free parameter that is 
fixed only when restricting to a stationary
point of the action.\footnote{For this 
reason we cannot probe the dominance of a
microcanonical stationary point by comparing
its action with that of other configurations
in the ensemble. In the spherical reduction, there is only one configuration. For further discussion of this point see the end of section \ref{foundations}.}
For the Nariai solution (\ref{Nariai})
all the 2-spheres have the same area, so the mean extrinsic curvature of the boundary vanishes and the total energy is zero.

\begin{figure}[t!]
\includegraphics[scale=0.54]{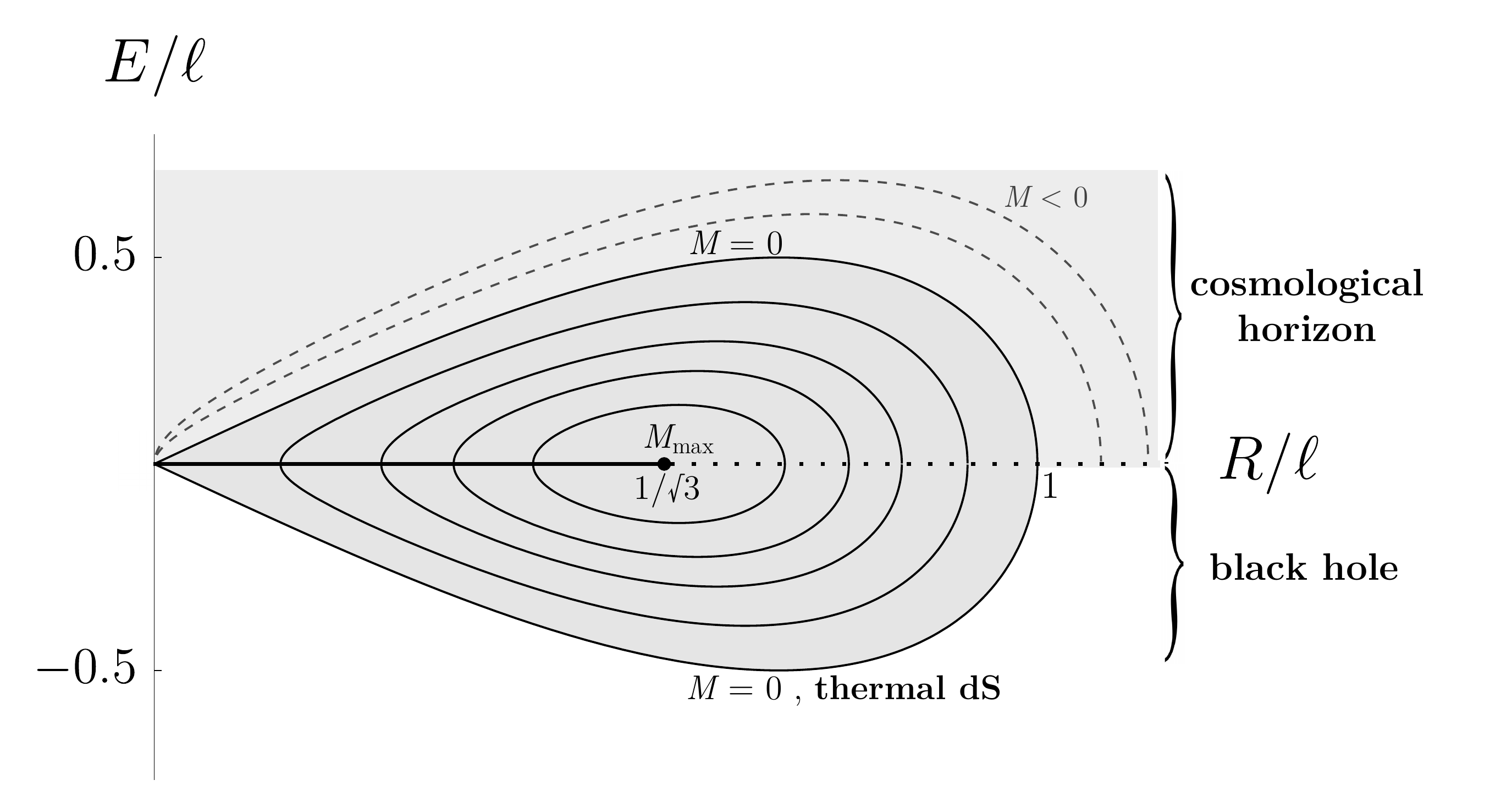}
\centering
\caption{The space of boundary data $R$ and $E$. The gray region corresponds to regular solutions (that avoid naked singularities). The contours indicate constant-mass curves, which are also constant entropy curves in the upper and lower half-planes, since a given $M$ specifies $r_b$ or $r_c$ uniquely. The mass contours shrink to the Nariai point with $R=\ell/\sqrt{3}$ and $E\!=\!0$, corresponding to the maximum mass, $M=\ell/3\sqrt{3}$. The dashed curves indicate negative mass parameter, which do not have counterparts in the black hole section. 
The solid segment on the horizontal axis from the origin to the Nariai point is the case where the system boundary and the black hole horizon coincide. On the remaining (dotted) part of the axis, the system boundary lies on the cosmological horizon.}
\label{ER}
\end{figure}

Figure \ref{ER} shows the constant mass contours
for the solutions determined by the boundary data $(R \,, E)$, and indicates the nature of the corresponding configuration.
For $R>\ell$, the mass parameter is necessarily negative in order to have a positive definite metric in the entire manifold. This yields a negative mass naked singularity on the black hole side and hence is presumably excluded by a stable UV completion of the theory.
On the cosmological side, however, the naked singularity is not in the part of the spacetime describing the bounded gravitational system. If that is considered as a stand-alone dynamical system --- without requiring that it be realizable as a subsystem with a physical boundary embedded in a complete spacetime --- then the solutions with negative mass parameter may take on a physical significance.

For boundary radius $R\leq \ell$, we distinguish different cases depending on 
the value of the energy: 
\ben[I.]
\item $|E|=R\sqrt{1-R^2/\ell^2}\equiv E_0(R)\,$.

This special case corresponds to the pure dS solution with zero mass parameter. 
For negative energy, $E=-E_0$ (the lower $M\!=\!0$ curve in figure \ref{ER}), the boundary encloses ``thermal de Sitter space",
i.e.\ 3-ball spatial patch of dS rotated
with arbitrary period in the Euclidean time direction (see section \ref{sp-sol}
and figure \ref{thermal-dS}). 
In this case the system contains no horizon, hence the entropy is zero for any 
such period. (A nonzero thermal 
entropy arising from the 1-loop contribution
would depend upon the period, as is the
case for ``hot flat space".)
For positive energy, $E=E_0$ (the upper $M\!=\!0$ curve in figure \ref{ER}), the system includes the cosmological horizon at $r\!=\!\ell$,
shown in figure \ref{c-system2},
and the entropy is given by the area of the cosmological horizon, $S=\pi\ell^2$.

\item $|E|>E_0\,$.

This case corresponds to a solution with a negative mass parameter (dashed curves in figure \ref{ER}), which yields a negative mass naked singularity on the black hole side ($E<-E_0$) and a regular
geometry on the cosmological side ($E>E_0$),
with entropy $S=A_c/4$.

\item $|E|<E_0\, , \, E \neq 0\,$.

This boundary condition gives the \sds solution, with the mass parameter $M\!=\!(E_0^2-E^2)/2R$.  For $-E_0<E<0$ (the inner curves in the lower half-plane in figure \ref{ER}) the system contains the black hole horizon at $r\!=\!r_b$ (the smaller positive root of $f(r)$), and the entropy is $S=A_b/4$. For  $0<E<E_0$ (the inner curves in the upper half-plane in figure \ref{ER}) the boundary is enclosed by the cosmological horizon at $r\!=\!r_c$ (the larger positive root of $f(r)$), and the entropy is $S=A_c/4$. %The \euc topology of both systems is $S^2 \times D^2$, and 
The corresponding Lorentzian geometries are depicted in figures \ref{b-system} and \ref{c-system1}, respectively.

\item $E= 0\,$.

The nature of the zero energy configurations depends on the boundary radius $R$, as indicated in the three distinct parts of the $E\!=\!0$ line in figure \ref{ER}. Since $E=0$ both 
black hole and cosmological configurations exist. 
The geometry in either case  
corresponds to the SdS space, with one of the
horizons located at the boundary, except
for $R=\ell/\sqrt{3}$ which is the Nariai geometry.
The configurations associated with the different ranges of boundary radius are as follows:
\ben[i.]
\item $R<\ell/\sqrt{3}\,$. In this case it is the black hole horizon that is located at the boundary, $r_b \rightarrow R$.
The black hole configuration is the region between the boundary and the black hole horizon, which has zero volume. 
Nevertheless, the entropy is $S=A_b/4=\pi R^2$.
The cosmological configuration is the region between the boundary
and the cosmological horizon at $r_c=\widehat R$ (\ref{Rhat}), 
with entropy $S=A_c/4=\pi \widehat R^2$.
The cosmological horizon has greater entropy,
so that configuration dominates in these $E=0$
ensembles.

\item $\ell/\sqrt{3} < R < \ell\, $.
In this case it is the cosmological horizon that is located at the boundary, $r_c \rightarrow R$.
The cosmological configuration is the region between the boundary and the cosmological horizon, which has zero volume, with entropy $S=A_c/4=\pi R^2$.
The black hole configuration is the region between the boundary and the black hole horizon at $r_b=\widehat R$, with entropy $S=A_b/4=\pi \widehat R^2$. Again, the cosmological horizon has greater entropy, so that configuration dominates in these $E=0$ ensembles. (Since the system volume vanishes,
however, the semiclassical treatment may not be valid.)
 
\item $R=\ell/\sqrt{3}\,$. In this case the geometry 
corresponds to the Nariai solution, with the maximum mass parameter $M_{\text{max}}=\ell/3\sqrt{3}$. There is an infinite degeneracy
however, since the proper distance from the boundary to the horizon
can be anything between 0 and $\pi\ell/\sqrt{3}$. All of those
configurations have the same entropy, $S=\pi\ell^2/3$.

\een

\een

\subsubsection{Microcanonical temperature}

\begin{figure}[t!]
\includegraphics[scale=0.54]{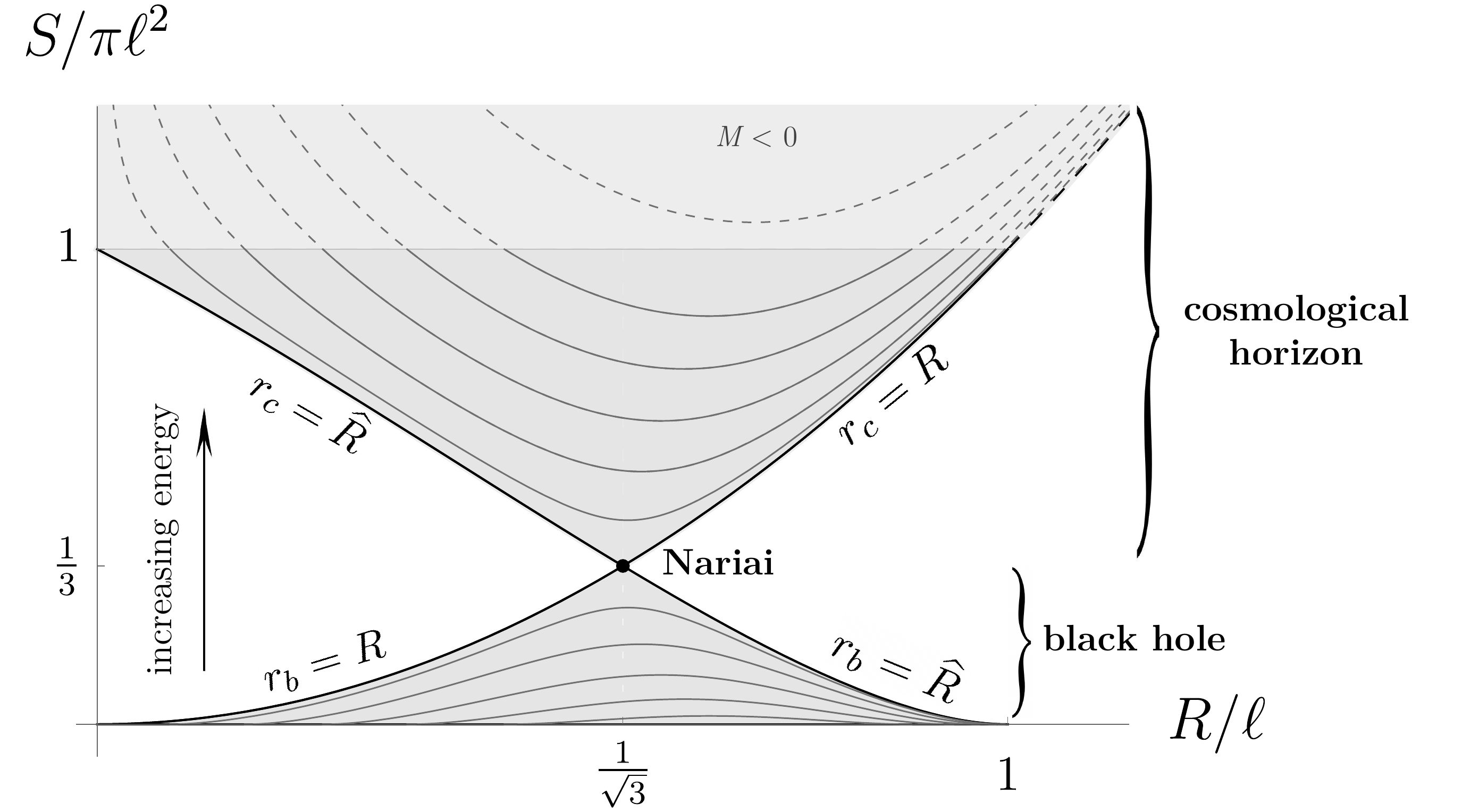}
\centering
\caption{Entropy vs.\ the system boundary radius. Constant-$E$ contours are shown, and the dashed curves correspond to negative mass parameter. The thick curves --- intersecting at the Nariai point --- indicate $E=0$, where either the black hole or the cosmological horizon radius coincides with the system boundary. The entropy increases as energy becomes larger, thus the temperature is positive for both black hole and cosmological horizon cases.}
\label{SR}
\end{figure}

Given the microcanonical entropy as a function of the energy, we can calculate the temperature from the thermodynamic definition $\beta=\partial S/\partial E$. The result is identical to the relation found between the inverse temperature and the horizon radius for stationary points in the canonical ensemble
(\ref{temp}), namely
\be
\beta_{b,c}=\frac{4\pi r_{b,c}}{|1-3r_{b,c}^2/\ell^2|}\left [1-\frac{r_{b,c}}{R}\left (1-\frac{r_{b,c}^2}{\ell^2}\right )-\frac{R^2}{\ell^2}\right ]^{1/2}\, ,
\label{Tmic}
\ee
where $b$ and $c$ correspond to the black hole and cosmological horizon solutions, respectively.\footnote{The agreement of microcanonical
temperature at energy $E$ with canonical temperature $T_c(E)$ for which the mean energy is $E$ 
holds in general if the microcanonical entropy at energy $E$ 
is equal to the canonical entropy at temperature $T_c(E)$.}
We see that the temperature is positive for both black hole and cosmological horizon cases.
Figure \ref{SR} depicts
how the black hole and cosmological horizon entropies compare in radius-entropy space, and
manifestly shows that the temperature is positive on both sides, since entropy increases with energy. 

\section{Relation to Gibbons-Hawking de Sitter partition function}
\label{Gibbons-Hawking}

In the Introduction we raised questions
concerning the interpretation and justification
of the GH approach \cite{GH} to de Sitter thermodynamics, specifically, their computation of the entropy.  The motivation of our work was, in part, to attempt to illuminate these questions. 
In this section we 
show how our results allow 
for the GH analysis to be 
realized as a special case of the thermodynamics
of well-defined ensembles, 
by considering the c-horizon configurations discussed in section \ref{can} and taking the limit where the system boundary shrinks to zero size. 
In order for the limit $R\rightarrow 0$ to yield 
a regular geometry, the mass parameter $M$ in (\ref{rprime}) must go to zero.
This implies that in this limit
the cosmological horizon radius \eqref{M-rh}
takes only the de Sitter value $r_c=\ell$ in spherical solutions to the constraints.
After examining this limit in the spherically reduced setting, we explain how the limit of the exact partition function results in the GH path integral over geometries on the 
4-sphere.

We first consider the microcanonical ensemble,
which is specified by $R$ and the BY energy 
$E$. According to \eqref{Ebc}, 
$E\rightarrow0$ in the limit $R\rightarrow0$.
(Although the extrinsic curvature diverges as $1/R$, the boundary area vanishes as $R^2$, so the total energy vanishes as $R$.)
The limiting microcanonical ensemble is thus the one
with parameters $(R,E)=(0,0)$. 
As mentioned above,
regularity of the stationary point
geometry as 
$R\rightarrow0$ requires 
$r_c\rightarrow\ell$, hence
the entropy \eqref{Smic} in this limit is just $A_c/4=\pi\ell^2$, which agrees with the 
GH result.
The microcanonical temperature
can also be computed in this limit, by taking the 
limit of the expression \eqref{Tmic}, which yields 
$T=1/2\pi\ell$,\footnote{\label{M/R}According to
\eqref{M-rh} the potentially indeterminate
term in \eqref{Tmic} (or \eqref{temp}) is $\sim M/R$,
which must go to zero
if the curvature  $\sim M/R^3$ is to remain finite.} 
the GH temperature
evaluated at the center of the static patch. 

The canonical ensemble is specified by $R$ and 
the temperature $T$ on the boundary. At a stationary
point of the action the relation between these parameters and the horizon radius is given in \eqref{temp}. 
Since in the limit $R\rightarrow0$
the horizon radius must approach $\ell$, the only temperature for which a stationary
point exists is the de Sitter temperature, $T=1/2\pi\ell$.\footref{M/R}  Also the entropy in the canonical ensemble is always the horizon area, hence it approaches the GH result. Moreover, since the BY energy term vanishes, the
canonical Euclidean action \eqref{Ih}
becomes $I=-\pi \ell^2$,
in agreement with the action of Euclidean de Sitter space 
found by Gibbons and Hawking. 

We have thus recovered the GH temperature and entropy 
from the vanishing boundary limit of the spherically reduced  
microcanonical and canonical ensembles. This symmetry reduced,
stationary point analysis does not, however, provide an understanding of 
what the ``exact" path integral signifies, if anything. To
arrive at such an understanding, we can begin with the full \euc
path integral representation of the partition function in the 
presence of a boundary, and follow the limit as the boundary 
shrinks to zero size.

\begin{figure}[t!]
\includegraphics[scale=0.54]{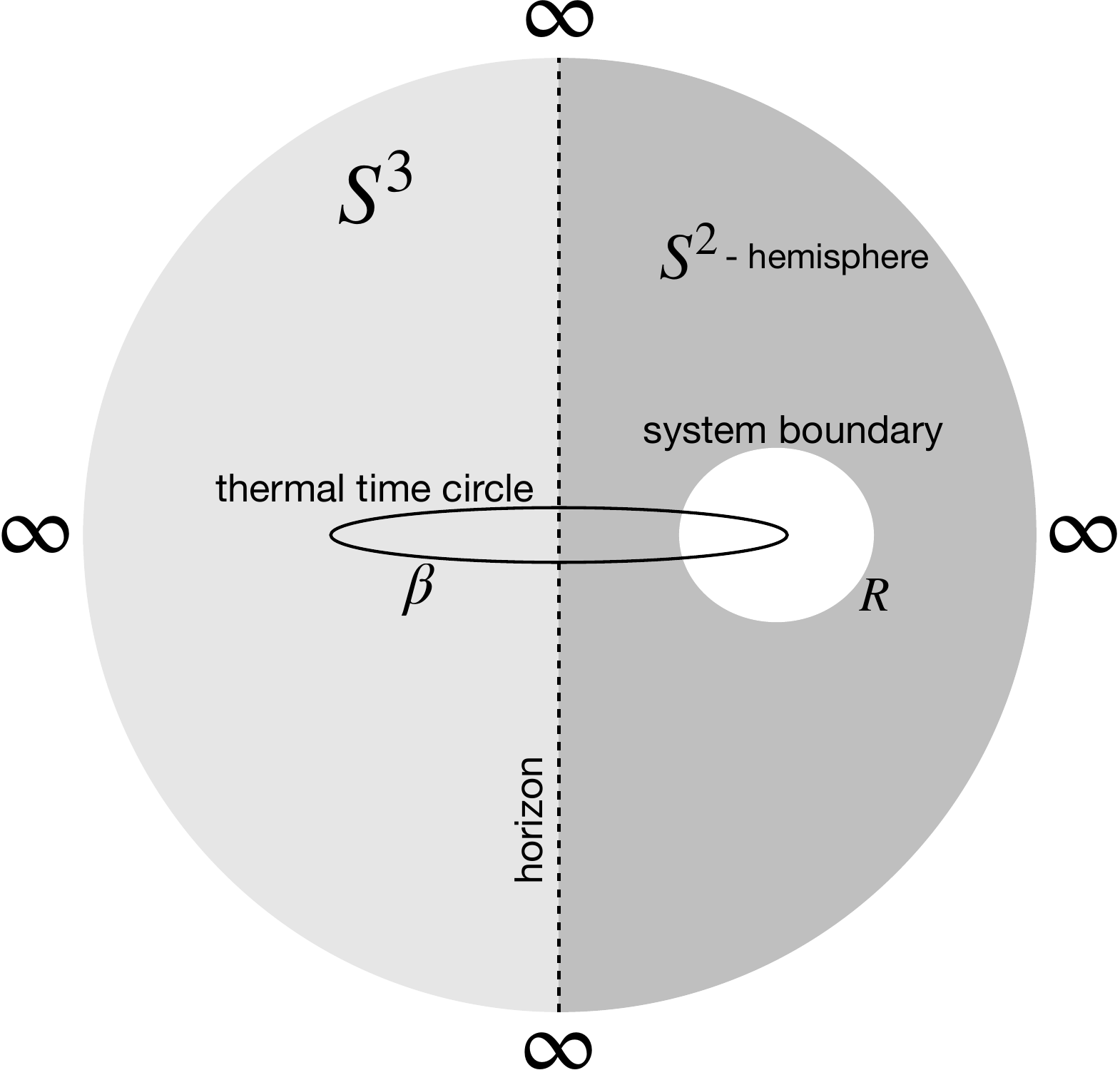}
\centering
\caption{A geometry in the Euclidean path integral for the 
partition function in three spacetime dimensions. 
The topology is $S^3$, represented as a 3-ball with all of its surface points identified (and labeled as ``$\infty$").
(See text for details.)
}
\label{dS3}
\end{figure}

To more easily visualize this limit, 
we describe the case of three spacetime dimensions, so the spatial system at each time is two dimensional.
Figure \ref{dS3} depicts the Euclidean spacetime as a 3-sphere,
presented as $\mathbb R^3$ compactified to a 
3-ball together with the point at infinity.
That is, all points on the surface of the ball are identified as one point, labeled here as ``$\infty$". 
The dashed line is a diameter of the 
3-ball whose endpoints are on the surface of the 3-ball and hence identified, so the dashed line is in fact a circle. 
The 3-ball is a figure of revolution about this dashed diameter, and
the Euclidean time 
is proportional to the angle of revolution. Space at one 
Euclidean time is indicated by the 
darker shaded region.
It contains the system boundary which is a circle of radius $R$, and is surrounded by a circular horizon consisting 
of the dashed line.
The partition function is (formally --- see discussion in section \ref{can}) a path integral over metrics on the manifold of topology $S^3$ minus a solid torus (corresponding to the white region ``outside" the system boundary) whose boundary is a flat torus $S^1\times S^1$ with circumferences $2\pi R$ for the spatial circle and $\beta$ for the time circle. 

As $R\rightarrow0$, this torus shrinks to a circle of circumference $\beta$. In that limit, the only restriction on the geometries in the path integral is that they contain a circle of circumference $\beta$, but this is no restriction at all. The partition function path integral thus becomes an integral
over {\it all} geometries on a manifold with topology $S^3$, weighted by 
the exponential of minus the Euclidean Einstein-Hilbert action. This integral has lost any dependence on system boundary conditions.
Since the system boundary has shrunk to zero, the Hamiltonian vanishes, 
so the partition function $Z={\rm Tr} \,e^{-\beta H}$ has become
${\rm Tr}\, \hat I_{\cal H}$, {\it where $\hat I_{\cal H}$ is the identity operator on the Hilbert space of 
states surrounded by a horizon}. That is, $Z$ has become
nothing but the dimension of the Hilbert space within a horizon, in the
presence of the cosmological constant $\Lambda$. 
Put differently, the canonical ensemble at any temperature is the maximally mixed state.

Since this  partition function has no boundary data to specify a temperature, the  de Sitter temperature and entropy --- which are determined by the size of the sphere --- can arise only from a stationary point approximation. 
A semiclassical (zero-loop) approximation to this $Z$ is given by $e^{-I_*}$, where $I_*$ is the action evaluated at the 
de Sitter solution with cosmological constant $\Lambda$. 
We have thus recovered --- and interpreted --- the GH proposal for the formal 
``partition function" of de Sitter space. 
Notice that
nearly the same reasoning 
implies that this 
``partition function" is 
also the $R\rightarrow0$ limit of the microcanonical 
path integral for the density of states, since the BY energy is always zero in that limit.\footnote{In $D=3$ spacetime dimensions the (unsubtracted) BY energy does not vanish in the limit $R \to 0$, because the integral of the extrinsic curvature over the boundary scales as $\sim R(1/R)=1$. Moreover, the solutions to the constraints do not all have the same BY energy in this limit, since the constraints allow for a conical singularity ``behind” a boundary of arbitrarily small $R$, outside the system. Different values of the conical deficit label different sectors of the Hilbert space. In ref.~\cite{Hikida:2022ltr} the contributions of such conical singularities to the partition function of dS$_3$ are discussed and matched with corresponding contributions in a dual CFT.}

The dS solution exists
in any spacetime dimension,
but is it the only solution? And
if it is {\it not} the only solution, does it nevertheless dominate the integral? In $D=4$ spacetime dimensions, the only known classical
solution to the Einstein equation with
positive cosmological constant is the round 4-sphere, i.e., \euc dS,
but it is not known whether this is 
the unique solution. 
In dimensions between 5 and 9
there are infinitely many solutions~\cite{akutagawa2019gap}.
The \euc action for a solution to the Einstein equation with cosmological constant $\Lambda$ is
proportional to $-\Lambda V$, where $V$ is the volume of the manifold. The de Sitter solution indeed dominates the ensemble
since it has the maximal volume among these solutions.\footnote{That the round sphere has the maximal volume among the Einstein metrics on $S^4$ topology follows from the 
Myers diameter bound and the Bishop volume comparison theorem. We thank Harish Seshadri for explaining this to us.}

\section{Discussion}
\label{Discussion}

In this paper we studied canonical and microcanonical ensembles for gravitational systems with positive cosmological constant,
with boundary conditions fixed on a system boundary which is chosen to 
be a round 2-sphere of fixed radius. Our approach and main results are
summarized in the Introduction, and foundational issues related to this 
study are discussed in section \ref{foundations}. We close the paper
in this discussion section with remarks on some outstanding questions and relation to other work. 

\paragraph{Negative temperature}
We reviewed in the Introduction the motivation for the
suggestion that the static patch of de Sitter space should be viewed as a system with negative temperature~\cite{KV, TJ&Manus,Jacobson:2019gco}; however, the microcanonical temperature obtained in (\ref{Tmic}) is evidently positive. Moreover our attempt to study the negative temperature canonical ensemble was inconclusive, since when using the stationary
point approximation to the \euc path integral we obtained a negative 
entropy, which indicated either a bad approximation or an invalid contour deformation. It would be interesting to understand precisely why a negative entropy results from the computation, and whether the negative temperature canonical ensemble actually exists --- which seems possible given the fact that the BY energy is bounded above. But even if neither of these ensembles exist at negative temperature, that does not necessarily mean that the previous negative temperature suggestion is incorrect, since that suggestion applies to a different system and Hamiltonian. 
For a causal diamond --- of which the de Sitter static patch is a special case --- there is no timelike boundary enclosing the system, hence there is no notion of Hamiltonian generating time evolution at the boundary.
Instead the ``first law", $\delta H_{\chi}=-\kappa \delta A /8\pi$,
derived in \cite{TJ&Manus}, relates the 
variation of the Hamiltonian $H_{\chi}$, which 
generates evolution along the flow of the 
timelike Killing (or conformal Killing) vector of the background diamond
being perturbed, to the variation of Bekenstein-Hawking entropy of the diamond. It is the minus sign in this relation that demands the assignment of negative temperature $-\hbar\kappa/2\pi$ to the diamond. 
In contrast to the ensembles considered in the present paper, the causal diamond has no timelike boundary, its edge area is not varied, and the Hamiltonian is a different operator, so it might be that both assignments of temperature are correct in their own context.

\paragraph{Lorentzian path integral for the dS partition function}

As carefully explained in section \ref{can}, the \euc path integral representation of the partition function should not be taken literally, since  it is essential that the integration contours for the lapse and shift are not purely Euclidean, although they may be Euclidean through a stationary point that dominates the integral. The contour deformation was needed in order to pass through the saddle,
but even if one were not concerned with the stationary point approximation, it would have been needed 
to obtain a real rather than a complex action
since otherwise the various boundary terms would not combine into one real GHY boundary term.
In view of this, it is interesting to note that 
the GH path integral for the dS
partition function is over metrics on a manifold without boundary, hence requires no contour deformation
in order to produce a path integral with a real action.
All of the boundary terms in \eqref{ZNh} 
vanish when the boundary shrinks to zero size, and one is left with a {\it Lorentzian} path integral.
On the other hand, the integral is over the manifold swept out by rotating a hemisphere of $S^3$
 through an angle $2\pi$ and identifying the initial and final hemispheres. A regular Lorentzian metric cannot be placed on this manifold, however. There would be a conical singularity at the hemisphere boundary, similar to
what is produced at the origin of
a 2-dimensional Minkowski cone if one identifies two Rindler time slices. Since the curvature is
regular everywhere except at the vertex of the cone where the metric is ill-defined, perhaps this
Lorentzian path integral could make sense. If so, some sort of regularity condition would need to be imposed if the de Sitter solution with its particular entropy and temperature were to emerge from such a calculation.\footnote{See also \cite{Marolf:2022ybi} for a recent study of the Lorentzian path integral for quantum gravity.}

\paragraph{Microstates}
Although formulated in principle at the level 
of the quantum mechanical ensembles, the calculations we have actually implemented here
within the spherical reduction of the theory
are able to capture only aspects of the 
thermodynamic limit, and so have nothing to say about the nature of the microstates beyond what might or might not be implied by the mere existence of the gravitational path integral. 
The same is of course true for the original Gibbons-Hawking path integral derivation of black hole entropy, but with the advent of AdS/CFT holography it became possible to relate black hole microstates to those of a dual quantum field theory that could be associated with the asymptotic boundary of spacetime~\cite{Maldacena:1997re,Witten:1998qj,Maldacena:2001kr}. 
It has been an ongoing aspiration to include de Sitter and other cosmological spacetimes within the realm of quantum gravity that can be understood via holographic duality, and some of the recent
progress in this direction has been achieved by introducing, in one way or another, a boundary 
at which the holographic degrees of freedom could reside.
In \cite{Susskind-matrix,Shaghoulian:2022fop} that boundary is located at the stretched horizon, while the boundary home of the dual theory in \cite{Coleman:2021nor} plays a role quite similar to the boundary we have introduced. 

Focusing on the 2+1 dimensional case, ref.~\cite{Coleman:2021nor}
succeeds in
exploiting the solvable $T\bar T +\Lambda_2$ deformation of a seed
CFT to count the states responsible for the horizon entropy of $dS_3$.
The bulk gravitational side of 
the holographic dictionary
discussed there is closely related to the one we developed here in 3+1 dimensions. Moreover,  
the systems on both sides of the boundary (the ``pole patch" and the ``cosmic horizon patch") 
are considered 
as configurations of the same ensemble there as well. 
The exchange of dominance between the two patches is characterized there
as a ``Hawking-page transition". This interpretation speaks to the question 
we raised in section \ref{stab},
as to whether what we called ``b-c transitions" between the thermal dS or black hole configurations and the cosmological horizon configuration are possible at all. This question arises 
since there is no continuous path through the (one-dimensional) space of spherical initial data, labeled by the horizon radius or mass parameter, that connects these configurations. (Although a transition can proceed by quantum  tunneling, we suppose there should nevertheless exist paths 
through the classical configuration space of 3-geometries that connect the initial and final configurations.) 
As we mentioned, it may be that nonspherical configurations connect them, but we have been unable
to identify  promising candidates for that role. Alternatively, 
perhaps complex metrics should be included, which would allow a continuous interpolation. 
But it also might be that only when going beyond the gravitational degrees of freedom 
can the transition be realized. Finally, 
it might be useful to point out that this potential blockade 
of the transition is absent in the special case that the boundary radius 
is equal to the Nariai radius $\ell/\sqrt{3}$. Then, configurations with vanishing extrinsic curvature 
of the boundary, and the same volume for the black hole or cosmological horizon system exist. 
The transition in this case was called the ``inside-out transition" in ref.~\cite{Susskind-matrix}.

\acknowledgments

We are grateful to 
Zhongshan An,
Michael Anderson,
Rodrigo Andrade e Silva,
Raphael Bousso,
David Brown,
William Donnelly,
Ruth Gregory,
Gary Horowitz,
Juan Maldacena,
Don Marolf, 
Ian Moss,
Alireza Parhizkar, 
Malcolm Perry,
Jorge Santos,
Harish Seshadri,
Sergey Solodukhin,
Douglas Stanford,
Manus Visser,
and
Bernard Whiting
for helpful discussions and correspondence.
B.B.\ gratefully acknowledges support through the Ruth Davis Fellowship.
This research was supported in part by the National Science Foundation 
under grants
PHY-1708139 and 
PHY-2012139 at UMD 
and PHY-1748958 at KITP.

\appendix

\section{Boundary-horizon distance in the maximal mass limit}
\label{NariaiApp}
\def\d{\delta}
\def\e{\epsilon}
\def\r{\rho}
Footnote \ref{failed} mentioned the curious
fact that,
if the radius of the system boundary in an SdS configuration
is fixed at $R=\ell/\sqrt{3}$, then in the limit
$M\rightarrow \ell/3\sqrt{3}$ the boundary winds up sitting at  equal proper distance from both horizons. To see this, we let $\ell=1$, introduce a new radial
coordinate $\r = r-1/\sqrt{3}$, and parametrize the cosmological horizon radius as $r_c = \d + 1/\sqrt{3}$. The corresponding 
mass \eqref{M-rh} then satisfies $2M = 2/3\sqrt{3}-\sqrt{3}\d^2 +O(\d^3)$.
The proper distance from the boundary to the cosmological horizon is then given, up to terms
that vanish in the limit $\d\rightarrow0$, by 
\begin{equation}\label{equator}
    \int_{\frac{1}{\sqrt{3}}}^{r_c} \frac{dr}{\sqrt{1-2M/r - r^2}}
    \approx \int_0^\d \frac{d\r\,\sqrt{\r+1/\sqrt{3}}}{\sqrt{\sqrt{3}(\d^2 -\r^2) -\r^3}} 
    \approx%\xrightarrow{\d\rightarrow0}
   \frac{1}{\sqrt{3}} \int_0^\d \frac{d\r}{\sqrt{\d^2 -\r^2}}
    =\frac{\pi}{2\sqrt{3}}\,.
\end{equation}
 The same result would be obtained for the limiting distance to the black hole horizon.
 The limiting configuration 
 can be smoothly embedded as a slice of 
 the \euc Nariai solution, in which
 case the boundary sits at the equator of the $\tau$-$\chi$ 2-sphere, and the proper distance
 \eqref{equator} is one quarter of the circumference of that  sphere.

\section{Accuracy of ``zero-loop" approximation at endpoints}
\label{endpoints}

As shown in section \ref{phasdiag}, for some values of the system boundary radius and temperature the action may be minimum at an endpoint; namely, at $r_b=0$ (thermal dS), $r_c=\ell$, $r_c=R$, or $r_c=\widehat R$ (where $\widehat R$ is defined in \eqref{Rhat}). 
When the action is expanded about a stationary local minimum, the next order term is quadratic, so that the leading order correction to the integral takes the form of an integral of a Gaussian whose width controls the size of contributing fluctuations. When expanded about an endpoint minimum of the action, by contrast, the next order term is typically linear, hence the fluctuations might be more important. To assess this possibility,
in the context of our spherical reduction of the 
path integral, we would like to find the next order correction to $\log Z \sim -I_*$ where $I_*$ is the action of such endpoints, and examine for what values of the ensemble parameters $R$ and $\beta$ (if any) it becomes comparable to the leading order term, invalidating the approximation. We find that, provided the ensemble parameters are much larger than the Planck length and the system volume remains finite, the leading order approximation is valid at any candidate dominant configuration.

The leading order approximation to an integral of the form $Z(x)=\int_a^b dr \rho(r) e^{-x Q(r)}$ for large $x$ is determined by the behavior of the integrand around the global minimum of $Q(r)$ in the interval $[a,b]$, which could be at one of the endpoints (cf.\ section 6.4 in \cite{Bender}). 
For estimating the partition function \eqref{Zc} in the symmetry reduced system with action \eqref{Igen}, we may factor out one of the macroscopic parameters  from the action, 
for example $\ell^2$, and
write $I/\hbar = (\ell/l_{\rm P})^2 Q(\beta,R,\ell,r_h)$, where $Q$ is generically much smaller than $(\ell/l_{\rm P})^2$. 
We then find corrections in powers of $(l_{\rm P}/\ell)$ in evaluating the partition function $Z\sim \int dr_h \exp \{ -\ell^2 Q(\beta,R,\ell,r_h) \}$, where now all lengths are in Planck units.

If $I'=dI/dr_h$ is finite at the minimum endpoint  (which is the case for endpoints at $r_b=0$ and $r_c=\ell$),  Taylor expanding the action around the minimum to linear order and using the Laplace method (which involves extending the integration limit to infinity \cite{Bender}) yields
\be
Z\sim e^{-\ell^2 Q_*} \int_0^\infty \! du \,e^{-\ell^2 |Q'_*|u}= \frac{e^{-\ell^2 Q_*}}{\ell^2 \, |Q'_*|}=\frac{e^{-I_*}}{|I'_*|} \, \Rightarrow \, \log Z \sim -I_* - \log  |I'_*| \, ,
\ee
where $(*)$ denotes evaluation at the endpoint.
The leading order approximation is valid as long as $|I_*|\gg |\log|I'_*||$. For thermal dS ($r_b=0$) this condition (in Planck units) becomes $\beta R \sqrt{1-R^2/\ell^2} \gg |\log \left ( \beta/\sqrt{1-R^2/\ell^2} \right )|$, which holds as long as $R$ does not approach $0$ or $\ell$.  When $R\rightarrow0$, 
the system volume goes to zero
at which point the semiclassical approximation is 
in any case not expected to be valid. When $R\rightarrow \ell$ only $r_b=0$ contributes, hence the action $I_b$ takes only one value, and the approximation method for an integral with only one contributing configuration does not apply.
For the $r_c=\ell$ endpoint 
to dominate
the validity condition is $|\beta R \sqrt{1-R^2/\ell^2}-\pi \ell^2| \gg |\log |2\pi \ell - \beta/\sqrt{1-R^2/\ell^2}||$.
As seen from figure \ref{c-phase}, $r_c=\ell$ can dominate only 
below the circular arc (at which the boundary temperature is equal to the blueshifted GH temperature), and in that region $\beta$ is too small for the left hand side of this inequality to vanish. 
Hence the inequality could fail only if the
argument of the log is either very large or very small. 
The log argument can be very large only for $R\rightarrow\ell$, 
which approaches the $r_c=\ell$ dominant region only in the lower right corner
of the phase diagram. 
The log argument vanishes on the
circular arc, which approaches the $r_c=\ell$ dominant region only in the lower right corner
of the phase diagram. Hence we may restrict attention to the
ensemble parameters  $(\beta \rightarrow 0, R\rightarrow \ell)$.  
Such parameters are not relevant for our analysis, however, since 
$R\rightarrow \ell$ corresponds to a zero volume configuration, 
and $\beta\rightarrow 0$ corresponds to the infinite temperature regime,
both of which take us beyond the low energy EFT framework that we employ 
in this work.

For the $r_c=R$ and $r_c=\widehat R$ endpoints $I'_*$ diverges, since the square root in \eqref{Igen} vanishes there. For $R\neq \ell/\sqrt{3}$, we change the integration variable to $u$ such that $r_c=r_c^*+u^2$, so the action near the endpoint $r_c^*$ becomes $I=-\pi r_c^{*2}+\beta R \sqrt{f'_*} \, u +O(u^2)$, with $f(r_c)=1-r_c(1-r_c^2/\ell^2)/R-R^2/\ell^2$. Thus we have
\be
Z\sim e^{-I_*}\int_0^\infty \! du \, u \, e^{-\beta R \sqrt{f'_*} \, u}= \frac{e^{-I_*}}{\beta^2 R^2 f'_*} \, \Rightarrow \, \log Z \sim -I_*-\log \left ( \beta^2  R^2 f'_* \right )\, .
\ee
The leading order approximation is valid when $|I_*|\gg |\log( \beta^2  R^2 f'_*)|$. For $R<\ell/\sqrt{3}$ when $r_c=\widehat R$ dominates, this condition becomes $\pi \widehat R^2 \gg |\log [\beta^2 R (3 \widehat R^2/\ell^2-1)]|$, which is violated when $R$ approaches $0$ or $\ell/\sqrt{3}$,  corresponding to cases where only one configuration contributes to the integral and where the geometry tends to Nariai, respectively (see below for the special case of Nariai). Moreover, for other values of $R$ at very small temperatures when $\beta \gtrsim \exp (\pi \widehat R^2/2)$ the approximation breaks down. At such exponentially low temperature, however, one does not
expect a semiclassical treatment to be accurate in any case.

For $R>\ell/\sqrt{3}$ when $r_c=R$ dominates,  the spatial volume vanishes, hence 
the validity of our semiclassical evaluation of the partition function is dubious. (As discussed in section \ref{c-instability} this configuration
is the endpoint of an instability in 
which the system seems to be destroyed.)
However, if one is interested in the mathematical estimation of the partition function as an integral, the above condition becomes $\pi R^2 \gg |\log [\beta^2 R (3R^2/\ell^2-1)]|$, which again does not hold when $R$ approaches $\ell/\sqrt{3}$. It also breaks down at very large temperatures $\beta \lesssim \exp(-\pi R^2/2)$ --- corresponding to the lower right corner of the phase diagram \ref{c-phase} --- where the low energy treatment is not supposed to apply, or at very small temperatures $\beta \gtrsim \exp(\pi R^2/2)$.

Finally, for the special case $R=\ell/\sqrt{3}$ when $r_c=\ell/\sqrt{3}$ dominates the c-configurations (see figure \ref{c-phase}) we note that $f'_*$ vanishes, so we include the next order. By a change of variable $r_c=\ell/\sqrt{3}+u$ the action near the endpoint becomes $I=-\pi \ell^2/3 +(\beta -2\pi \ell/\sqrt{3})u +O(u^2)$, hence we have $\log Z \sim \pi \ell^2/3+\log (\beta -2\pi \ell/\sqrt{3})$. Therefore the second term becomes comparable to the leading order when $\beta \gtrsim \exp(\pi \ell^2/3)$. 
However, this is not relevant since for the special value of the system boundary $R=\ell/\sqrt{3}$, the two branches of b- and c-horizon configurations join at $r_h=\ell/\sqrt{3}$, and there is a connected path in the phase space between the two, so a transition to the dominant b-configuration is possible. As a result, the dominant phase at large $\beta$ is the thermal dS (i.e.\ the $r_b=0$ configuration); see figure \ref{beta-R}, for which case the leading order approximation is valid, as discussed above.

\section{Dominance of cosmological horizon solutions over black holes}
\label{c-dominance}

We first prove that --- as figure \ref{actionsbc} shows --- the small (large) black hole solution is the local maximum (minimum) of the action $I_b$ (\ref{Igen}), thus the large black hole solution always has a lower action compared to the small black hole. 

For a given boundary size $R<\ell$ and a high enough temperature, i.e.\ when $\beta \! <\! \beta_{\text{B}}$ in plot \ref{beta-rh},
the equation $\partial I_b/\partial r_b = 0$ gives two extrema corresponding to the small and large black hole solutions, with horizon radius $r_{b1}^*$ and $r_{b2}^*$, respectively. 
It is straightforward to show that the second derivative of the action with respect to horizon radius at the stationary points become
\be 
I_b^{''  *}= (\text{a positive factor}) \left ( q(r_b^*)-\frac{R \beta^2}{8\pi^2} \right )\, ,
\ee
where $q(r)=r^3/(1+3r^2/\ell^2)$, and $r_b^*$ is either the small or the large black hole horizon radius.  
The horizon radii of the two black hole solutions satisfy $r_{b1}^*\! < \! r_{\text{B}} \! < \! r_{b2}^*$, where $r_{\text{B}}$ is the radius corresponding to $\beta_{\rm B}$, given by $q(r_{\text{B}})=R \beta^2_{\text{B}}/8\pi^2$.\footnote{This is obtained by finding the maximum of the function given at the right hand side of (\ref{temp}). That is a concave function of $r_b$, as its second derivative at the extremum point is $-4\beta_{\text{B}}/r_{\text{B}}^2(1-3r_{\text{B}}^2/\ell^2)^2<0$. Note also that the action has an inflection point at this temperature.}
The function $q(r)$ is a monotonically increasing function, thus 
\be 
\frac{R \beta^2}{8\pi^2} < \frac{R \beta^2_{\text{B}}}{8\pi^2}=q(r_{\text{B}})< q(r_{b2}^*)\, .
\ee
Therefore $I_b^{''}\!(r_{b2}^*)\! >\! 0$, hence the large black hole solution is a local minimum. Since there is only one other extremum point, the small black hole solution has to be a local maximum, thus its action is always greater compared to the large black hole.

We now show that for all values of $R\!<\!\ell$ and $0\!\leq\!\beta\!< \beta_{\text{B}}$ where the black hole stationary points exist, the value of $\Delta I(R,\beta)\equiv I_{b_2}^* - I_c^* $ is non-negative, where $I_{b_2}^*$ and $I_c^*$ are the \euc action for the large black hole and the cosmological horizon solutions, respectively.  Using the relations (\ref{Igen}) and (\ref{temp}), $\Delta I$ can be written as (dropping the $*$-superscript to reduce clutter)
\be
\begin{split}
\Delta I(R,\beta)&= \pi \left (r_c^2-r_b^2 \right )-\frac{R\beta^2}{4\pi} \left( \frac{1}{r_b}(1-3r_b^2/\ell^2)+\frac{1}{r_c} (3r_c^2/\ell^2-1) \right)\\
&=\frac{\pi}{ r_b r_c}(r_c-r_b)\Big[ r_b r_c (r_b+r_c)-\frac{R\beta^2}{4\pi^2}(1+3r_b r_c /\ell^2)\Big ]
\, .
\end{split}
\ee
At the Nariai limit where $r_b=r_c=R=\ell/\sqrt{3}$, $\Delta I$ vanishes. In all other cases, any value of $r_c$ is larger than $r_b$, thus proving $\Delta I\!>0$ is equivalent to showing
\be
\frac{R\beta^2}{4\pi^2}\stackrel{?}{<}\frac{r_b r_c(r_b+r_c)}{1+3r_br_c/\ell^2}\, .
\label{ineq}
\ee
For $r_b \! > \! r_{\text{B}}$ we have
\be 
\frac{R\beta^2}{4\pi^2}<\frac{R\beta_{\text{B}}^2}{4\pi^2}=2 q(r_{\text{B}})< 2q(r_b)\, ,
\ee
and using $r_b\!<\!r_c$ we get 
\be
 2q(r_b)=\frac{2r_b^3}{1+3r_b^2/\ell^2}<\frac{2r_b^2 r_c}{1+3r_br_c/\ell^2}<\frac{r_b r_c(r_b+r_c)}{1+3r_br_c/\ell^2}\, .
\ee
Therefore the inequality (\ref{ineq}) and hence $\Delta I\!>\!0$ holds, which means the cosmological horizon solution dominates over the black hole solutions in the path integral.

\bibliographystyle{JHEPmod}
\bibliography{SdS-ref}

\end{document}